\documentclass[12pt,english,aps,tightenlinesletterpaper,groupaddress,secnumarabic,nofootinbib]{revtex4}
\usepackage{mathptmx}

\usepackage[T1]{fontenc}
\usepackage[latin9]{inputenc}
\usepackage[letterpaper]{geometry}
\usepackage{babel}
\usepackage{verbatim}
\usepackage{textcomp}
\usepackage{amsmath}
\usepackage{amssymb}
\usepackage{cancel}
\usepackage{graphicx}
\usepackage[unicode=true,pdfusetitle,
 bookmarks=true,bookmarksnumbered=false,bookmarksopen=false,
 breaklinks=false,pdfborder={0 0 1},backref=false,colorlinks=false]
 {hyperref}

\makeatletter
\@ifundefined{textcolor}{}
{%
 \definecolor{BLACK}{gray}{0}
 \definecolor{WHITE}{gray}{1}
 \definecolor{RED}{rgb}{1,0,0}
 \definecolor{GREEN}{rgb}{0,1,0}
 \definecolor{BLUE}{rgb}{0,0,1}
 \definecolor{CYAN}{cmyk}{1,0,0,0}
 \definecolor{MAGENTA}{cmyk}{0,1,0,0}
 \definecolor{YELLOW}{cmyk}{0,0,1,0}
}

\usepackage{babel}
\usepackage[toc,page]{appendix}
\usepackage{pdfpages}

\usepackage{todonotes}
\setuptodonotes{color=green!20}
\newcommand{\M}{M}

\allowdisplaybreaks

\makeatother

\begin{document}
\title{A topological contribution to Bogoliubov coefficient for cosmological
particle production}
\author{Daniel J. H. Chung}
\email{danielchung@wisc.edu}

\affiliation{Department of Physics, University of Wisconsin-Madison, Madison, WI
53706, USA}
\author{Nidhi Sudhir}
\email{kandathpatin@wisc.edu}

\affiliation{Department of Physics, University of Wisconsin-Madison, Madison, WI
53706, USA}
\begin{abstract}
Particle production in cosmology is often efficiently computed in
terms of Bogoliubov transforms. Restricting to a particular class
of dispersion relationships, we identify a map between the number
of particles produced in a special kinematic limit and a Stokes phenomena
related topology of analytic continuation of the Bogoliubov coefficient
functions. Intuitively, this kinematic limit corresponds to the long
wavelength limit although a more precise description depends on the
nature of the curved spacetime. To identify the topology, we reformulate
the usual Bogoliubov computations as a type of SU(1,1) gauged differential
equation and utilize a special gauge together with a discrete symmetry
that naturally characterizes the dispersion relationship. Using a
dark matter model and a nonzero constant spatial curvature model,
we estimate how such topological contributions will arise in physical
applications.

\tableofcontents{}
\end{abstract}
\maketitle

\section{Introduction}

Stokes phenomena (see e.g.~\citep{Stokes:1864,FF:1965,Berry:1972na,meyer:1989,Boyd1999TheDI,Bender:1999box,ablowitz2003complex})
is mathematically striking because it leads to abrupt changes in the
coefficients of analytic continuations of asymptotic expansions, such
as those used in computing particle production through Bogoliubov
transformations. Relatively recently, there has been some interest
in applying Stokes phenomena in particle production in cosmology \citep{Li:2019ves,Hashiba:2020rsi,Enomoto:2020xlf,Enomoto:2021hfv,Yamada:2021kqw,Hashiba:2021npn,Hashiba:2022bzi,Enomoto:2022nuj,Kolb:2023ydq,Racco:2024aac}.\footnote{For earlier applications to dS space see for example \citep{Dabrowski:2014ica,Kim:2010xm}.
Also many other usage of Stokes phenomena to particle production exists
(for a sample of recent work, see e.g.~\citep{Dumlu:2010ua,Dumlu:2020wvd,Taya:2020dco,Taya:2021dcz,Fedotov:2022ely}).} These previous works focused mostly on providing a way to approximate
the large momentum $k$ limit of the Bogoliubov coefficients $\beta_{k}$
characterizing the spectrum of particles produced.

In this article, we use Stokes phenomenon to understand the opposite
limit in a class of dispersion relationships which may be approximated
as $\omega^{2}\approx C+A\eta^{n}$. Such effective dispersion relationships
can occur in cosmological scenarios when the effective mass passes
through a zero at some real conformal time. This limit, corresponding
to $C\rightarrow0$, involves merging zeroes of the effective dispersion
relation and therefore lies outside the range of validity of the previously
established approximation techniques using the Stokes data of well
separated simple zeroes \cite{Hashiba:2021npn} or saddle point approximations
\cite{Chung:1998bt}.

Here, we point out that for these effective dispersion relationships
with positive even integer $n$, we can relate the $C\rightarrow0$
limit (intuitively the small momentum limit) of the magnitude of the
Bogoliubov coefficient $\left|\beta_{k}\right|$ to the number of
Stokes sectors each of which supports approximately constant asymptotic
WKB expansion coefficients.\footnote{We restrict ourselves to the even $n$ because that corresponds to
adiabatic propagating particle vacua at $\eta=\pm\infty$ in contrast
with odd $n$ cases where there is an exponentially decaying particle
wave function in one of the end regions.} Since each of these Stokes sectors represent a region of functional
continuity (i.e.~approximately constant asymptotic expansion coefficients)
and their number is insensitive to variations in $A$ and $C$, the
number of Stokes sectors can be viewed as a topological index characterizing
the analytic continuation of the Bogoliubov coefficients which are
functions of time. In order to identify this topology, we reformulate
the usual Bogoliubov computation in terms of an $\mathrm{SU}(1,1)$
gauged differential equation. This will help us to use a technique
from the math literature \citep{FF:1965} and utilize an apparent
discrete symmetry to compute the topological index.

In physical applications, there will often be piecewise time regions
where the topological contribution will be relevant. We present couple
of such cosmological scenarios, where in one of the scenarios, $C\rightarrow0$
is achieved by taking $k\rightarrow0$ and in the other $C\rightarrow0$
is achieved by canceling $k^{2}$ against constant spatial curvature
in an Friedmann--Lema\^{i}tre--Robertson--Walker (FLRW) spacetime.
We can interpret the first scenario as a dark matter model embedded
in an inflationary cosmology, but the topological contribution will
be shown to be suppressed due to the phase space vanishing as $k\rightarrow0$.
This model is also a good illustration of how the approximate dispersion
relationship arises in physical applications where this paper's simple
analytic results can be applied. In the second scenario, the topological
contribution can be significant because of the fact that $k^{2}$
can be as large as the tunable background potential generated spatial
curvature.

It is interesting to note that there is an analogy of the Stokes sector
number to the Chern-Simons number carried by anomalous currents \citep{Harvey:2005it,Nakahara:2003nw}.
Both the gauge field description and the Bogoliubov analytic continuation
description employed in this paper are fictitious. The Chern-Simons
number change can be related to a physical charge production induced
by an instanton induced vacuum to vacuum transition. The Bogoliubov
coefficient related number of particles produced can also be viewed
as arising from a vacuum to vacuum transition. The Chern-Simons number
describes a topological characterization of the gauge field while
the Stokes sector number describes a topological character of the
analytically continued field mode function. It is also interesting
that unlike in typical steepest descent computations, the topological
number (partly owing to their integer nature) produces a large Bogoliubov
coefficient magnitude $|\beta_{k}|$.

From a mathematical side, one of what we have identified can be viewed
as a novel identity of a path ordered matrix integral which we will
make explicit. The topological nature can also be viewed as a special
conformal property of the Bessel functions together with their being
solutions to the mode equation in the special kinematic limit.

The order of presentation is as follows. In Sec.~\ref{sec:Particle-Production-scenario},
we define the class of models, standard particle production through
Bogoliubov transforms, and then define Stokes lines in this context.
In Sec.~\ref{sec:Gauge-Picture}, we show that the SU(1,1) based
complexification of the particle production in a first order formalism
naturally has a gauge symmetry. We will define the $F^{2}$-gauge
and the $0$-gauge in this section. In Sec.~\ref{sec:Topological-contribution-to},
we show that a certain soft limit of the Bogoliubov coefficient in
this class of models considered in this paper corresponds to measuring
the Stokes sector topology (defined explicitly in this section). In
Sec.~\ref{sec:Comparison-of-the}, we show how the topology can also
be viewed as a particular property of Bessel function mode functions.
The physical embedding of the topological contributions is investigated
in Sec.~\ref{sec:Illustrative-model}. We then conclude with a summary.
Appendix \ref{sec:1-loop-corrections-and} gives the details of the
1-loop correction to the tree-level potential and the dark matter
abundance computation used in Sec.~\ref{sec:Illustrative-model}.
Appendix \ref{sec:Covariance-of-linear} gives the details of the
discrete symmetry representation used in the paper.

\section{Particle Production scenario}

\label{sec:Particle-Production-scenario}Consider a non-minimally
coupled scalar field $\chi$ on flat FLRW spacetime 
\begin{equation}
\begin{split}S_{\chi}= & \frac{1}{2}\int\mathrm{d}\eta\,\mathrm{d}^{3}x\,\,\sqrt{-g}\left[\partial_{\mu}\chi\partial^{\mu}\chi-m_{\phi}^{2}(\eta)\chi^{2}+\xi R\chi^{2}\right]\\
 & ds^{2}=a^{2}(\eta)(d\eta^{2}-|d\vec{x}|^{2}).
\end{split}
\end{equation}
Expanding the $\chi$ field as 
\begin{equation}
\chi\left(\eta\right)=\int\frac{d^{3}k}{(2\pi)^{3}a(\eta)}\left(a_{k}\chi_{k}(\eta)e^{i\vec{k}\cdot\vec{x}}+h.c.\right)
\end{equation}
the Heisenberg equation of motion (E.O.M) yields the mode equation
\begin{equation}
\chi_{k}''(\eta)+\omega^{2}(\eta)\chi_{k}(\eta)=0\label{eq:modeeq}
\end{equation}
\begin{equation}
\omega^{2}(\eta)=k^{2}+a^{2}m_{\chi}^{2}(\eta)+(6\xi-1)\frac{a''}{a}\label{eq:dispersion}
\end{equation}
where $\xi=1/6$ corresponds to the conformal coupling and $m_{\chi}^{2}(\eta)$
is an effective time-dependent mass that can arise from couplings
to other fields. For example, if there exists another scalar field
$\phi$ which couples through the interaction 
\begin{equation}
\mathcal{L}_{I}=\frac{g}{2\Lambda^{2}}\phi^{4}\chi^{2}
\end{equation}
then if the $\phi$ has a time-dependent background $\phi(\eta)$,
the effective mass term contribution to $m_{\chi}^{2}(\eta)$ would
be $g\phi^{4}(\eta)/\Lambda^{2}$. Whenever this mode frequency becomes
time-dependent, some particle production occurs because time-translation
symmetry is broken. That usually leads to an ambiguity in the choice
of the vacuum.

One popular formalism to construct a vacuum is the adiabatic vacuum
\citep{Birrell:1982ix} relying on the WKB formalism. An adiabaticity
parameter $\delta_{k}$ can be defined as 
\begin{equation}
\delta_{k}(\eta)\equiv\frac{\omega_{k}'(\eta)}{4\omega_{k}^{2}(\eta)}\label{eq:nonad}
\end{equation}
which can be seen as counting the adiabatic order defined according
to the formal multiplication of every time derivative on $\omega$
with $1/T$ and the convention $T\rightarrow\infty$. For example
$\delta_{k}(\eta)$ defined above is an adiabatic order 1 quantity
while $\omega''/\omega^{3}$ is an adiabatic order 2 quantity. Adiabatic
time region is defined as when this formal adiabatic limit is an approximate
description of the time dependences of the frequencies, that is, $\eta$
intervals within which $\delta_{k}\left(\eta\right)$ and all higher
order adiabaticity parameters are of magnitudes $\ll1$.

Let $\chi_{k,1}(\eta)$ and $\chi_{k,2}(\eta)$ be two solutions of
the mode equation satisfying the following adiabatic quantization
boundary conditions 
\begin{equation}
\chi_{k,1}(\eta)\approx\frac{\exp\left(-i\int_{\eta_{-\infty}}^{\eta}d\eta'\omega(\eta')\right)}{\sqrt{2\omega(\eta)}}\hspace{1em}\mbox{as }\eta\rightarrow\eta_{-\infty}\label{eq:bc-1}
\end{equation}
and 
\begin{equation}
\chi_{k,2}(\eta)\approx\frac{\exp\left(-i\int_{\eta_{-\infty}}^{\eta}d\eta'\omega(\eta')\right)}{\sqrt{2\omega(\eta)}}\hspace{1em}\mbox{as }\eta\rightarrow\eta_{+\infty}\label{eq:bc-2}
\end{equation}
where $\eta_{\pm\infty}$ correspond to times in past and future adiabatic
regions.\footnote{In practice, we define $\eta_{\pm\infty}$ to be times when the nonadiabaticities
are sufficiently small for the desired accuracy of the computation.} Since $\left(\chi_{k,1},\chi_{k,1}^{*}\right)$ and $\left(\chi_{k,2},\chi_{k,2}^{*}\right)$
form two sets of complete basis for the solutions a second order differential
equation, we can write 
\begin{equation}
\chi_{k,1}(\eta)=\alpha_{k}(\eta_{+\infty})\chi_{k,2}(\eta)+\beta_{k}(\eta_{+\infty})\chi_{k,2}^{*}(\eta)\label{eq:bog_superpos}
\end{equation}
where $\{\alpha_{k}(\eta_{+\infty}),\beta_{k}(\eta_{+\infty})\}$
are two constant coefficients. Canonical quantization condition in
these basis requires the basis functions to be normalized to unity
under the Wronskian norm, implying 
\begin{equation}
|\alpha_{k}(\eta_{+\infty})|^{2}-|\beta_{k}(\eta_{+\infty})|^{2}=1.\label{eq:boundarynorm}
\end{equation}
The constant coefficients in Eq. (\ref{eq:bog_superpos}) are so named
since later, these will be understood as the limiting values taken
by time-dependent functions $\alpha_{k}(\eta)$ and $\beta_{k}(\eta)$
in adiabatic periods of time as $\delta\left(\eta\right)\rightarrow0$.
If the adiabatic regions are at finite time intervals, then the functions
$\alpha_{k}(\eta)$ and $\beta_{k}(\eta)$ will approximate to the
constants $\{\alpha_{k}(\eta_{+\infty}),\beta_{k}(\eta_{+\infty})\}$
to leading order in $\delta\left(\eta\right)$ in the adiabatic time
periods. Since at all such adiabatic time periods canonical quantization
implies Eq. (\ref{eq:boundarynorm}) (to leading order in $\delta\left(\eta\right)$
for finite adiabatic regions), later we will define $\alpha_{k}(\eta)$
and $\beta_{k}(\eta)$ functions such that this $SU(1,1)$ normalization
is always maintained. Particle production between the time $\eta_{-\infty}$
and $\eta_{+\infty}$ can be written in terms of 
\begin{equation}
\beta_{k}=-\left(\chi_{k,1},\chi_{k,2}^{*}\right)=i\left(\chi_{k_{1}}\partial_{\eta}\chi_{k,2}-\chi_{k,2}\partial_{\eta}\chi_{k,1}\right)\label{eq:exactbogo}
\end{equation}
where the number density in the comoving volume is 
\begin{equation}
n=\frac{1}{a^{3}}\int\frac{d^{3}k}{(2\pi)^{3}}|\beta_{k}|^{2}.\label{eq:NumDensity}
\end{equation}
These are thus far completely standard and well known.

In this paper, we focus on physical situations in which $\omega^{2}$
can be approximated as 
\begin{equation}
\omega^{2}\approx k^{2}+A(\eta-\eta_{0})^{n}+B\label{eq:scalingchoice}
\end{equation}
in the interval $(\eta_{-\infty},\eta_{\infty})$ containing $\eta_{0}$,
with $n$ a positive even integer and constants $A$ and $B$. For
example, with conformal coupling, the dispersion relationship Eq.~(\ref{eq:dispersion})
becomes 
\begin{equation}
\omega^{2}(\eta)=k^{2}+a^{2}(\eta)m_{\chi}^{2}(\eta)
\end{equation}
and in situations where $m_{\chi}^{2}(\eta)\propto\phi^{q}(\eta)$
dominates the time-dependence (i.e. $a(\eta)$ time-dependence being
subdominant) and goes through a zero, the approximate dispersion relationship
of Eq.~(\ref{eq:scalingchoice}) can be achieved with an appropriate
choice of the potential governing the homogeneous $\phi$ dynamics.
We will explicitly apply our general formalism to such a scenario
in Sec.~\ref{subsec:Tanh-model}. We can also find situations where
the dispersion relationship is approximately Eq.~(\ref{eq:scalingchoice})
in a much larger time range in very special cosmological periods.
For example, with minimal gravitational coupling and an engineered
potential for the background cosmology driving scalar field $\phi$
coupled to $\chi$, one can obtain dispersion relationships of the
form 
\begin{equation}
\omega^{2}(\eta)=k^{2}+a^{2}f(\phi)-\frac{a''}{a}\label{eq:withcurvature}
\end{equation}
for which $k^{2}$ term cancels a constant $a''/a$ term to yield
Eq.~(\ref{eq:scalingchoice}). This is studied in Sec.~\ref{subsec:Curvature-model}
and demonstrates that this special kinematic point corresponding to
the topological description that we define need not correspond to
the ultra-IR.

For these scenarios, we will find that in the limit $k^{2}+B\rightarrow0$
the $\beta_{k}$ coefficient has a simple relationship with the number
of asymptotic expansion sectors. By asymptotic expansion sectors,
we mean the number of contiguous regions in the analytically continued
$\eta$ plane coordinated by $z$ where the WKB basis functions have
either a uniform exponential suppression or a divergence in the asymptotic
radial limit. The boundaries of these regions can be defined as anti-Stokes
lines coming from the study of Stokes phenomena.

Stokes phenomena occurs when the basis of an \emph{asymptotic expansion}
has an analytic property that is mismatched with the analytic property
of the function that it resolves. For example suppose the analytic
continuation of $\chi_{k}(\eta)\rightarrow\chi_{k}(z)$ with $z\in\mathbb{C}$
is an entire function. Because of the approximate Lorentz group representation
properties reflected in the adiabatic boundary conditions above, the
mode function $\chi_{k}$ is decomposed in terms of WKB basis functions
as 
\begin{equation}
\chi_{k}(z)=\alpha_{k}(z)\frac{\exp\left(-i\theta(z,z_{0})\right)}{\sqrt{2\omega(z)}}+\beta_{k}(z)\frac{\exp\left(i\theta(z,z_{0})\right)}{\sqrt{2\omega(z)}}
\end{equation}
\begin{equation}
\theta(z,z_{0})\equiv\int_{z_{0}}^{z}dz'\omega(z')\label{eq:thetagen}
\end{equation}
where $\omega^{2}(z)$ is an analytic function of $z$ causing a branch
cut to appear in the WKB basis functions $\exp\left(\pm i\theta(z,z_{0})\right)/\sqrt{2\omega(z)}$.
The curve $z(s)$ where $i\theta(z(s),z_{0})$ is real (imaginary)
is called (an)a (anti-)Stokes line.\footnote{Here $s$ is parameterizes the curve. We will later discuss the subtleties
of the branch cut resolution when $k^{2}+B\rightarrow0$ limit is
taken.} The Stokes lines are further classified as ``+''(``-'') type
depending upon whether $i\theta(z,z_{0})$ increases (decreases) as
$|z(s)-z_{0}|$ is increasing.

Note that on the anti-Stokes lines, the magnitudes of each basis functions
are equal while on the Stokes lines, the ratio of the basis functions
can have a large hierarchy. This means when the function $\chi_{k}(z)$
is evaluated across a Stokes line, the coefficient of the suppressed
basis function can shift by a large number -unsuppressed in inverse
powers of $\vert z\vert$- without violating the smooth behavior of
$\chi_{k}$. In fact, with an exponential suppression, the suppressed
basis function has a representation of exactly zero in the asymptotic
expansion. This shifted coefficient of the (exponentially) suppressed
basis function can then become important in the asymptotic expansion
once an anti-Stokes line is crossed which exchanges the roles of suppressed
and unsuppressed basis functions before crossing. This then leads
to a shift in the asymptotic expansion representation. Such a shift
in the coefficients $\{\alpha_{k},\beta_{k}\}$ is called a Stokes
phenomena.

Our aim in this paper is to show that $\beta_{k}$ in this special
kinematic limit ($k^{2}+B\rightarrow0$) in a class of special models
may be dominated by topological information associated with the counting
of the Stokes sectors in the dispersion relationship. To demonstrate
this, we utilize the F-matrix formalism developed by \citep{FF:1965}.
Furthermore, we show how this F-matrix formalism is related to the
formalism used in the conventional cosmological literature by constructing
a unified matrix formalism which connects different formalisms through
a gauge transformation. It is important to note that the unified formalism
that we develop below is independent of Eq.~(\ref{eq:scalingchoice}).
However, to our current knowledge, the greatest utility of the unified
formalism will be to elucidate the topological nature of the Bogoliubov
transform in the special kinematic limit of interest in this paper.
One other byproduct of the gauge transformation formalism will be
a derivation of a novel integral identity Eq.~(\ref{eq:mathidentity}).

\section{Gauge Picture}

\label{sec:Gauge-Picture}Because the mode equations are second order
ordinary differential equations (ODEs), the equations can be rewritten
as a first order vector ODE as is typical in Hamiltonian dynamics.
This will help us to formulate a gauged set of ODEs that will allow
us to elucidate the relationship between the usual parameterization
found in typical physics literature (see e.g.~\citep{Kofman:1997yn,Chung:1998bt,Racco:2024aac})
and a parameterization that is useful to obtain bounds on the propagator
matrices as we will explain below.

As a first step, we write the first order formulation using the ansatz
\begin{equation}
\partial_{\eta}V_{k}(\eta)=M(\eta)V_{k}(\eta)\label{eq:maineq}
\end{equation}
where $V_{k}$ specifies the mode functions $\chi_{k}$ through a
projection onto a WKB basis $F_{k}(\eta)$, i.e. 
\begin{equation}
\begin{split}\chi_{k}(\eta)=\alpha_{k}(\eta)f_{-}(k,\eta)+\beta_{k}(\eta)f_{+}(k,\eta)=F_{k}(\eta)\cdot V_{k}(\eta)\end{split}
\label{eq:product}
\end{equation}
where 
\begin{equation}
f_{\pm}(k,\eta)=\frac{\exp\left(\pm i\int_{\eta_{(*)}}^{\eta}d\eta'\omega(\eta')\right)}{\sqrt{2\omega(\eta)}}\hspace{5mm}F_{k}(\eta)=(f_{-}(k,\eta),f_{+}(k,\eta))\hspace{5mm}V_{k}(\eta)=\left(\alpha_{k}(\eta),\beta_{k}(\eta)\right).\label{eq:Fdef-1}
\end{equation}

Here, $\eta_{(*)}$ is the origin with respect to which the WKB modes
are defined and is taken as a general point on the real line. The
Bogoliubov coefficients defined in Eqs.$\,$(\ref{eq:bc-2}) and (\ref{eq:bog_superpos})
correspond to taking $\eta_{(*)}=\eta_{-\infty}$. For dispersion
relations which are positive definite on the real line, as the ones
we will be interested in here, a general choice of $\eta_{(*)}$ results
in a phase shift in the functions $\alpha_{k}\left(\eta\right)$and
$\beta_{k}\left(\eta\right)$. We keep this general since taking $\eta_{(*)}=0$
will become convenient in sections \ref{subsec:Stokes-constants-from}
for symmetry reasons.

To take advantage of the WKB solutions $f_{\pm}(k,\eta)$ being an
approximate solution to the mode equations in the adiabatic region,
we impose the condition 
\begin{equation}
M(\eta)=O(\delta_{k})\label{eq:adiabatic}
\end{equation}
in the adiabatic region. Next, to specify $M(\eta)$, note that it
has 8 real functional degrees of freedom. Since we only require 2
real functional degrees of freedom to match a general mode function,
we have freedom to restrict $M(\eta)$. We choose $M(\eta)\in\text{Lie Algebra of }\,su(1,1)$
parameterized as 
\begin{equation}
M(\eta)=\begin{pmatrix}iM_{1}(\eta) & M_{2}^{*}(\eta)\\
M_{2}(\eta) & -iM_{1}(\eta)
\end{pmatrix}\label{eq:specifyM}
\end{equation}
where $M_{1}(\eta)$ and $M_{2}(\eta)$ are real and complex valued
functions, respectively. This ensures 
\begin{equation}
|\alpha_{k}(\eta)|^{2}-|\beta_{k}(\eta)|^{2}=1
\end{equation}
for all values of $\eta$, including Eq.$\,$(\ref{eq:boundarynorm})
at the boundaries $\eta_{\pm\infty}$. Furthermore, this choice implies
the existence of $3-2=1$ real functional gauge degree of freedom
in specifying $M(\eta)$ (as made explicit below).

The dynamical information governing $M(\eta)$ is provided by the
mode equation 
\begin{equation}
\chi_{k}''(\eta)+\omega^{2}(\eta)\chi_{k}(\eta)=0\label{eq:modeeq-1}
\end{equation}
where 
\begin{equation}
\omega^{2}(\eta)=k^{2}+a^{2}m^{2}+(6\xi-1)\frac{a''}{a}
\end{equation}
written in conformal time coordinates $ds^{2}=a^{2}(\eta)(d\eta^{2}-|d\vec{x}|^{2})$.
The mode equation in terms of $M(\eta)$ is then 
\begin{equation}
F_{k}(\eta)\left\{ M'+\left(\frac{3\omega'^{2}}{4\omega^{2}}-\frac{\omega''}{2\omega}\right)\mathbb{I}_{2\times2}+2\begin{pmatrix}-i\omega-\frac{\omega'}{2\omega} & 0\\
0 & +i\omega-\frac{\omega'}{2\omega}
\end{pmatrix}M+M^{2}\right\} V_{k}(\eta)=0.
\end{equation}
Since the above should be satisfied for arbitrary values of $V_{k}(\eta)$,
corresponding to arbitrary boundary conditions\footnote{Here, boundary conditions refer to the limiting values of the functions
$\left(\alpha_{k}\left(\eta\right),\beta_{k}\left(\eta\right)\right)=V_{k}\left(\eta\right)$,
in the limit $\eta\rightarrow\eta_{-\infty}$. For a general solution
of the mode equation Eq. (\ref{eq:modeeq-1}), the boundary conditions
on $\chi\left(\eta\right)$ given by $\lim_{\eta\rightarrow\eta_{-\infty}}\chi_{k}\left(\eta\right)$
and $\lim_{\eta\rightarrow\eta_{-\infty}}\chi_{k}'\left(\eta\right)$
can be described in terms of the limiting constants $\left(\alpha_{k}\left(\eta_{-\infty}\right),\beta_{k}\left(\eta_{-\infty}\right)\right)$
and can take arbitrary values such that $\chi_{k}\left(\eta\right)$
remains real on the real line.}, this equation implies 
\begin{equation}
F_{k}(\eta)\left\{ M'+\left(\frac{3\omega'^{2}}{4\omega^{2}}-\frac{\omega''}{2\omega}\right)\mathrm{I}_{2\times2}+2\begin{pmatrix}-i\omega-\frac{\omega'}{2\omega} & 0\\
0 & +i\omega-\frac{\omega'}{2\omega}
\end{pmatrix}M+M^{2}\right\} =0.\label{Eq:ModeEq-1}
\end{equation}
Since one of the two complex equations here is the conjugate of the
other, this equation provides two real constraints on the 3 real components
of $M(\eta)$. Eq.~(\ref{eq:adiabatic}) implies that for physical
applications, we are interested in the solutions of Eq.~(\ref{Eq:ModeEq-1})
with vanishing boundary conditions in the adiabatic region. At this
point, $M$ has been specified up to the gauge transformations that
we discussed below Eq.~(\ref{eq:specifyM}).

Let's construct the gauge transform explicitly. Let $\bar{V}\left(\eta\right)=\left(\bar{\alpha}(\eta),\bar{\beta}(\eta)\right)$
and $\tilde{V}\left(\eta\right)=\left(\tilde{\alpha}(\eta),\tilde{\beta}(\eta)\right)$
be related by a gauge transform (where we suppress the $k$ subscript
for brevity) 
\begin{equation}
\tilde{V}(\eta)=T(\eta)\bar{V}(\eta)
\end{equation}
where the matrix $T(\eta)$ belongs to a group element of the vector
representation of $SU(1,1)$ 
\begin{equation}
T=\begin{pmatrix}T_{1} & T_{2}\\
T_{2}^{*} & T_{1}^{*}
\end{pmatrix}
\end{equation}
with $|T_{1}|^{2}-|T_{2}|^{2}=1$. Since $T(\eta)$ should preserve
Eq.~(\ref{eq:adiabatic}) and leave the vectors $\bar{V}(\eta)$
and $\tilde{V}(\eta)$ unchanged in adiabatic regions, the matrix
components should be of the form 
\begin{equation}
T_{1}=1+g_{1}(\delta)\quad\text{and\ensuremath{\quad T_{2}=}}g_{2}(\delta)
\end{equation}
where $g_{i}(\delta)$ are functions of adiabatic order at least 1.
Now, since $\chi_{k}$ remains invariant under such a gauge transformation
\begin{equation}
\chi(\eta)=F(\eta)\bar{V}(\eta)=F(\eta)T(\eta)\bar{V}(\eta)
\end{equation}
this implies, for arbitrary values of $\bar{V}(\eta)$ 
\begin{equation}
F(\eta)[\mathbb{I}_{2\times2}-T(\eta)]=0.
\end{equation}

The above constraint along with $\det(T)=1$ implies $g_{1}(\delta)=ig(\delta)$,
$g_{2}(\delta)=ig(\delta)\exp(+2i\int_{\eta_{(*)}}^{\eta}\omega)$
with $g(\delta)$ real valued function of adiabatic order of at least
unity, allowing us to conclude 
\begin{equation}
T(\eta)=\begin{pmatrix}1+ig(\delta) & ig(\delta)\exp(2i\int_{\eta_{(*)}}^{\eta}d\eta'\omega(\eta'))\\
-ig(\delta)\exp(-2i\int_{\eta_{(*)}}^{\eta}d\eta'\omega(\eta')) & 1-ig(\delta)
\end{pmatrix}.
\end{equation}
Given our real functional gauge degree of freedom, we can choose a
gauge $\bar{M}_{1}(\eta)=0$ to solve Eq.~(\ref{Eq:ModeEq-1}) 
\begin{equation}
f_{-}(k,\eta)\left(\frac{3\omega'^{2}}{4\omega^{2}}-\frac{\omega''}{2\omega}+|\bar{M}_{2}|^{2}\right)+f_{+}(k,\eta)\left(\left(2i\omega-\frac{\omega'}{\omega}\right)\bar{M}_{2}+\bar{M}'_{2}\right)=0\quad\text{and\ensuremath{\quad} c.c.}\label{eq:MsimModEq}
\end{equation}
for $\bar{M}_{2}$. Combining these two equations give 
\begin{equation}
\frac{d}{dt}\left(f_{+}^{2}\bar{M}_{2}\right)=\frac{d}{dt}\left(f_{-}^{2}\bar{M}_{2}^{*}\right)
\end{equation}
whose solution is 
\begin{equation}
f_{+}^{2}\bar{M}_{2}=f_{-}^{2}\bar{M}_{2}^{*}+C
\end{equation}
and the unit %
{} modular nature of $\sqrt{2\omega}f_{\pm}$ on the real line allows
us to set $C=0$. Substituting this into Eq.$\,$(\ref{eq:MsimModEq})
and solving for $\bar{M}_{2}(\eta)$ gives 
\begin{equation}
M_{0}(\eta)\equiv\bar{M}(\eta)=\frac{\omega'}{2\omega}\begin{pmatrix}0 & \exp\left(2i\int_{\eta_{(*)}}^{\eta}d\eta'\omega(\eta')\right)\\
\exp\left(-2i\int_{\eta_{(*)}}^{\eta}d\eta'\omega(\eta')\right) & 0
\end{pmatrix}\label{eq:0gauge}
\end{equation}
which is what we will call the 0-gauge. The explicit gauge transformation
of $M$ can then be obtained from the invariance of Eq.(\ref{eq:maineq})
as 
\begin{equation}
M\rightarrow TMT^{-1}-T\partial_{\eta}T^{-1}.
\end{equation}
Given the above solution, the most general $M_{g}(\eta)$ which satisfies
Eq.$\,$(\ref{Eq:ModeEq-1}) can be obtained as a general gauge transform
of $\bar{M}(\eta)$ 
\begin{equation}
\begin{split}M_{g}(\eta) & =\left[-T(\eta)\partial_{\eta}T^{-1}(\eta)+T(\eta)\bar{M}(\eta)T^{-1}(\eta)\right]\\
 & =\begin{pmatrix}i\gamma_{1}(\eta) & e^{+2i\int_{\eta_{(*)}}^{\eta}d\eta'\,\omega\left(\eta'\right)}\left(\gamma_{2}(\eta)+i\gamma_{1}(\eta)\right)\\
e^{-2i\int_{\eta_{(*)}}^{\eta}d\eta'\,\omega\left(\eta'\right)}\left(\gamma_{2}(\eta)-i\gamma_{1}(\eta)\right) & -i\gamma_{1}(\eta)
\end{pmatrix}
\end{split}
\label{eq:Gengauge}
\end{equation}
where 
\begin{equation}
\gamma_{1}(\eta)=\left(-g(\delta)+\frac{\partial_{\eta}g(\delta)}{2g(\delta)\omega}+2\delta\right)2g(\delta)\omega,\hspace{3mm}\gamma_{2}(\eta)=\left(-g(\delta)+\delta\right)2\omega\label{eq:gamma_defs}
\end{equation}
and 
\begin{equation}
\delta(\eta)\equiv\frac{\omega'(\eta)}{4\omega^{2}(\eta)}.
\end{equation}
This together with Eq.~(\ref{eq:maineq}) represents the first order
formulation of any second order ODE of the mode function type with
a $SU(1,1)$ basis choice and adiabatic boundary conditions.

A particularly convenient gauge is the one used in the mathematical
formalism called the F-matrix formalism \citep{FF:1965} which in
our present gauge theory language corresponds to choosing $g(\delta)=\delta$
leading to 
\begin{equation}
M_{F}(\eta)=\frac{i\epsilon_{r}(\eta)\omega(\eta)}{2}\begin{pmatrix}-1 & -\exp\left(+2i\int_{\eta_{(*)}}^{\eta}d\eta'\,\omega(\eta')\right)\\
\exp\left(-2i\int_{\eta_{(*)}}^{\eta}d\eta'\,\omega(\eta')\right) & 1
\end{pmatrix}\label{Eq:FromanGauge}
\end{equation}
where 
\begin{equation}
\epsilon_{r}(\eta)\equiv\frac{3\omega'(\eta)^{2}}{4\omega^{4}(\eta)}-\frac{\omega''(\eta)}{2\omega^{3}(\eta)}.\label{eq:F2_error}
\end{equation}
We will call this the Fröman-Fröman gauge or the $F^{2}$-gauge.
It is interesting to note that $\det M_{F}=0$.

To summarize this section, we have constructed a gauged first order
ODE formalism (Eqs.~(\ref{eq:maineq}) and (\ref{eq:Gengauge}) where
$M=M_{g}$ for a gauge choice $g$) of computing Bogoliubov transformations
where the usual formalism found in the literature \citep{Kofman:1997yn,Chung:1998bt,Racco:2024aac}
is the 0-gauge obtained with Eq.~(\ref{eq:Gengauge}) with $g(\delta)=0$
and the F-matrix formalism found in \citep{FF:1965} is obtained with
$g(\delta)=\delta$ .

\section{Topological contribution to particle production}

\label{sec:Topological-contribution-to}In the previous section we
described a family of first order ODEs, which describe the time evolution
of the mode function $\chi_{k}\left(\eta\right)$ in terms of functions
$\left(\alpha\left(\eta\right),\beta\left(\eta\right)\right)$ and
the WKB basis functions. The functions $\left(\alpha\left(\eta\right),\beta\left(\eta\right)\right)$
are such that, in all gauges, their values converge to the Bogoliubov
coefficients of Eq. (\ref{eq:bog_superpos}) in the limit $\eta\rightarrow\eta_{+\infty}$
as $\delta(\eta)\rightarrow0$.\footnote{And in the limit $\eta\rightarrow\eta_{-\infty},$where again $\delta\rightarrow0$,
the functions asymptote to $\left(\alpha\left(\eta_{-\infty}\right),\beta\left(\eta_{-\infty}\right)\right)=\left(1,0\right)$
such that $\chi_{k}\left(\eta\right)$ satisfies the boundary conditions
in Eq. (\ref{eq:bc-1}).} Integrating Eq. (\ref{eq:maineq}) in a general gauge $M_{g}$ gives
\begin{equation}
V(\eta_{1})=U_{g}(\eta_{1},\eta_{0})V(\eta_{0})
\end{equation}
where 
\begin{equation}
U_{g}\left(\eta_{1},\eta_{0}\right)\equiv\mathbb{P}\left[\mathbb{I}_{2\times2}\,\exp\left(\int_{\eta_{0}}^{\eta_{1}}d\eta'\,M_{g}(\eta')\right)\right]
\end{equation}
relates the values of the functions $\left(\alpha(\eta),\beta(\eta)\right)=V\left(\eta\right)$
at two conformal times $\eta_{1}$ and $\eta_{0}$. In terms of these
matrices, which we henceforth refer to as the propagator matrices,
the problem of computing the Bogoliubov coefficients reduces to computing
the limiting matrix 
\begin{equation}
U\left(\eta_{-\infty},\eta_{+\infty}\right)\equiv\lim_{\eta_{1}\rightarrow\eta_{+\infty}}\lim_{\eta_{0}\rightarrow\eta_{-\infty}}U_{g}\left(\eta_{1},\eta_{0}\right)
\end{equation}
where the gauge index on the L.H.S of the above definition has been
dropped since the limit is a gauge invariant quantity. Having defined
these propagator matrices, we note that these are not any easier to
compute, being defined in terms of a series of nested oscillatory
integrals. Ref \citep{FF:1965}, makes great progress estimating these
matrices by analytically continuing the domain of the problem into
complex values, i.e. $\eta\rightarrow z$, and working with propagator
matrices in regions of the complex plane where $|\delta(z)|\ll1$-
that is, where the WKB modes present a good approximation to the analytically
continued mode functions.

In the following we review the mathematical ideas of \citep{FF:1965}
wherein certain propagator matrices in regions of $|\delta(z)|\ll1$
are shown to have a simplified structure when expressed in terms of
a small parameter $\mu$ quantifying the total ``non-adiabaticity''\footnote{As presented later, $\mu$ is defined as the integral of the analytically
continued $\epsilon_{r}\left(z\right)$ defined in Eq. (\ref{eq:F2_error}).
For real arguments $\epsilon_{r}\left(\eta\right)$, being adiabatic
order 2, quantifies the non-adiabaticity at conformal time $\eta$.} accrued along the contour of propagation. For propagator matrices
between two adjacent anti-Stokes lines and a Stokes line, this structure,
reviewed in \ref{subsec:Reduction-of-d.o.f}, implies that these can
be expressed in terms of the parameter $\mu$ and one of the cross
diagonal matrix elements. By itself, this is insufficient to determine
any matrix element without bounds on the relevant cross diagonal matrix
element. This is mitigated in section \ref{subsec:Stokes-constants-from}
for a class of models with ``dispersion relations'' $\omega^{2}=Az^{n}$,
for which a $\mathbb{Z}_{n+2}$ phase rotation symmetry in $z$ of
the differential equation satisfied by $\chi(z)$ bounds and fixes
the relevant cross diagonal element. We then use the derived propagator
matrices to compute $\beta_{k}$ in the kinematic limit described
at the end of Sec.~\ref{sec:Particle-Production-scenario} and show
that it has a simple relationship to the topological index counting
the number of asymptotic expansion sectors.

\subsection{\label{subsec:Reduction-of-d.o.f}Reduction of d.o.f in $F^{2}$-gauge}

In this section, we use the logic of \citep{FF:1965} to restrict
the form of the propagator matrix in the $F^{2}$-gauge. As we will
see, gauges (such as the $0$-gauge) in which $M_{g}$ is insufficiently
suppressed will not allow us to make similar restrictions.

After analytic continuation of the conformal time $\eta$ to the complex
plane parameterized by $z$, the integrated version of the propagation
Eq.~(\ref{eq:maineq}) is $V_{k}(z_{1})=U(z_{1},z_{0})V_{k}(z_{0})$
where 
\begin{equation}
U(z_{1},z_{0})\equiv\mathbb{P}\left[\mathbb{I}_{2\times2}\,\exp\left(\int_{\Gamma(z_{0},z_{1})}dz'\,M(z')\right)\right]\label{Eq:ConnecMat}
\end{equation}
which relates the values taken by $V_{k}(z)$ at two different points
$z_{0}$ and $z_{1}$, and $\Gamma(z_{0},z_{1})$ denotes a curve
connecting them. Even though $U$ is a functional of the curve $\Gamma$,
because we will mostly be interested in circular arcs, we will suppress
the $\Gamma$ dependence in the notation. Whenever we evaluate this
in a particular gauge, we will put a subscript such as $U_{F}$ (Eq.~(\ref{Eq:FromanGauge}))
or $U_{0}$ (Eq.~(\ref{eq:0gauge})).

Since $M(z)$ is proportional to $\delta(z)$ or it's derivatives
(Eq.$\,$(\ref{eq:Gengauge})) in all gauges, we may expect a simplification
in the form of $U_{g}(z_{1},z_{0})$ in the limit $\vert\delta\left(z\right)\vert\rightarrow0$.
The propagator matrix 
\begin{equation}
\begin{split}U_{g}(z_{1},z_{0}) & =\mathbb{P}\left(\mathbb{I}_{2\times2}\exp\left(\int_{z_{0},\Gamma}^{z_{1}}dz'\,M_{g}(z')\right)\right)\\
 & =\mathbb{I}_{2\times2}+\int_{z_{0},\Gamma}^{z_{1}}dz_{\bar{1}}\,{M_{g}}\left(z_{\bar{1}}\right)\,+\int_{z_{0},\Gamma}^{z_{1}}dz_{\bar{1}}\,{M_{g}}\left(z_{\bar{1}}\right)\int_{z_{0},\Gamma}^{z_{\bar{1}}}dz_{\bar{2}}\,{M_{g}}\left(z_{\bar{2}}\right)\\
 & +\int_{z_{0},\Gamma}^{z_{1}}dz_{\bar{1}}\,{M_{g}}\left(z_{\bar{1}}\right)\int_{z_{0},\Gamma}^{z_{\bar{1}}}dz_{\bar{2}}\,{M_{g}}\left(z_{\bar{2}}\right)\int_{z_{0},\Gamma}^{z_{\bar{2}}}dz_{\bar{3}}\,{M_{g}}\left(z_{\bar{3}}\right)+...
\end{split}
\label{eq:ConnecMat2}
\end{equation}
involves nested, oscillatory integrals which are difficult to estimate
on the complex plane. Moreover in a general gauge choice, all terms
in the above series make comparable contributions to the sum, adding
to the difficulty in estimating $U_{g}(z_{1},z_{0})$.

The $F^{2}$-gauge Eq.$\,$(\ref{Eq:FromanGauge}) gives a better
power series expansion in Eq.~(\ref{eq:ConnecMat2}) through its
proportionality to a second order complexified adiabatic parameter
$M_{F}(z)\propto\epsilon_{r}(z)$: i.e. $\int_{z_{0},\Gamma}^{z_{1}}dz_{a}\,{M_{F}}(z_{a})$
is still $\delta^{1}$ suppressed unlike $\int_{z_{0},\Gamma}^{z_{1}}dz_{a}\,{M_{0}}(z_{a})$
which is $\delta^{0}$ suppressed. For propagator matrices between
points connected by a path $\Gamma(z_{1},z_{0})$ along which $\lvert\exp\left(i\int_{z_{(*)}}^{z}dz'\,\omega(z')\right)\rvert$
increases monotonically (here $z_{(*)}$ is the origin with respect
to which the WKB basis functions are defined in the complex plane),
the infinite sum in Eq.$\,$(\ref{eq:ConnecMat2}) may be understood
as a perturbative series in terms of 
\begin{equation}
\mu(z,z_{0})=\int_{z_{0},\Gamma}^{z_{1}}\mid dz'\epsilon_{r}(z')\omega(z')\mid\ll1.\label{eq:mudef}
\end{equation}
For propagator matrices between two adjacent anti-Stokes lines (bounding
a region containing at least one Stokes line) a path connecting the
two end points can be constructed out of two monotonic paths, where
the two paths share one point on the Stokes line. The estimates for
propagator matrices on monotonic paths can then be used to make estimates
of the matrix elements of the propagator matrix relative to each other.
In the following we describe this in more detail.

\begin{figure}
\begin{centering}
\includegraphics[angle=15,scale=0.273]{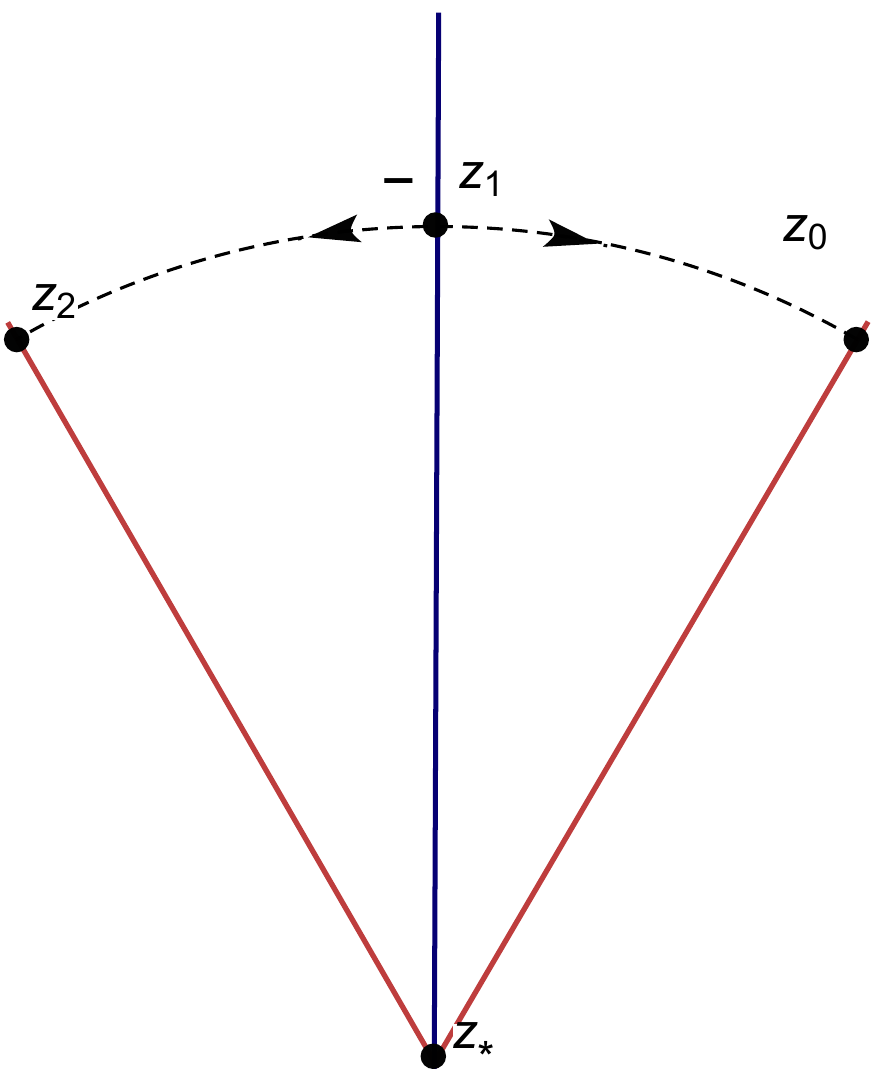}$\hspace{0.8cm}$\includegraphics[angle=15,scale=0.27]{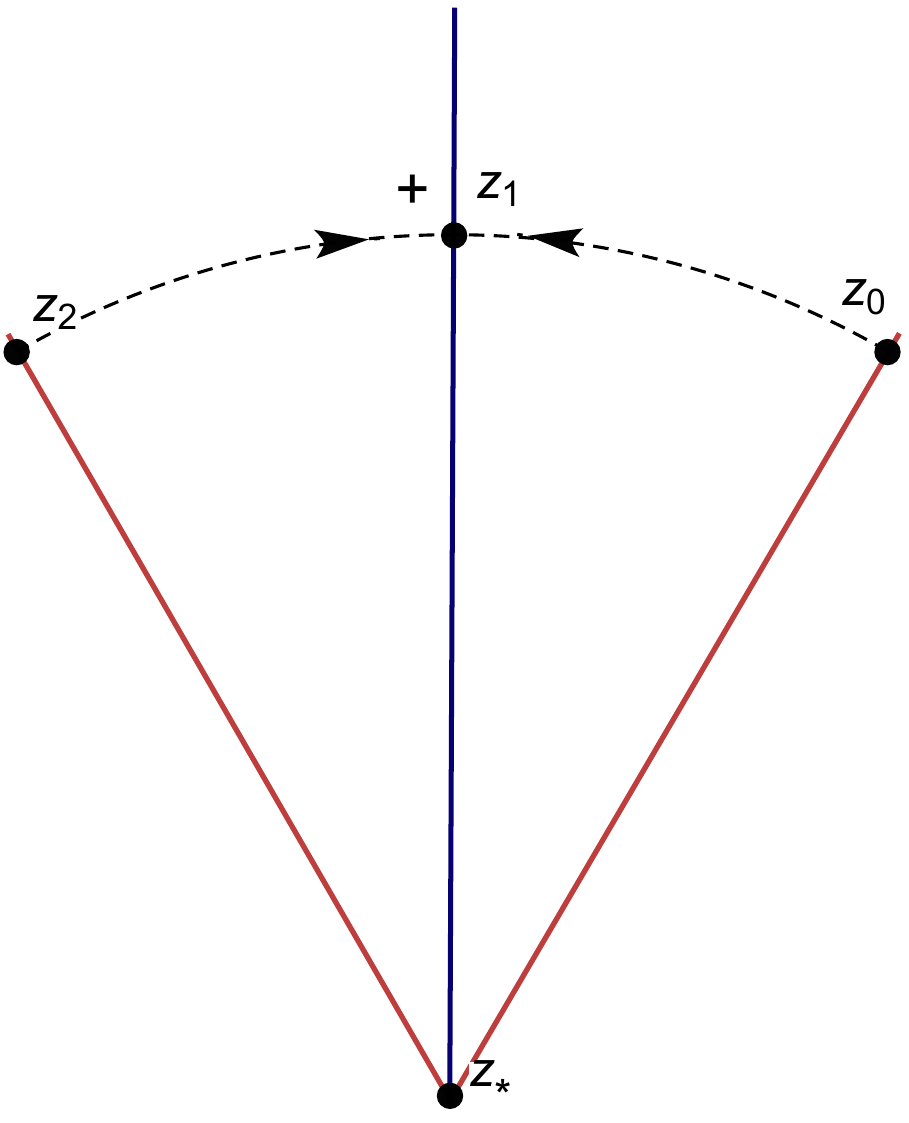}
\par\end{centering}
\caption{\label{fig:PM_SL} The red and blue lines denote the anti-Stokes and
Stokes lines respectively, arising from a zero $z_{*}$ of the dispersion
relation $\omega^{2}(z)$. The black dotted lines denote the integration
contour for the propagator matrix $U(z_{2},z_{0})$. Note, this contour
passes through the point $z_{1}$ on the Stokes line, and can be decomposed
into two monotonic contours between $\left(z_{1},z_{2}\right)$ and
$\left(z_{1},z_{0}\right).$ If the Stokes line crossed is the $(+)$
kind, then the point $z_{1}$ corresponds to a maximum of $\mid\exp\left(i\int_{z_{(*)}}^{z}dz'\,\omega(z')\right)\mid$,
whereas if the Stokes line is the $(-)$ kind, $z_{1}$ corresponds
to a minimum. The directions in which $\mid\exp\left(i\int_{z_{(*)}}^{z}dz'\,\omega(z')\right)\mid$
increases are denoted by black arrowheads.}
\end{figure}

Consider a path of integration $\Gamma$ along which $\mid\exp\left(i\int_{z_{(*)}}^{z}dz'\,\omega(z')\right)\mid$
increases monotonically from $z_{0}$ to $z_{1}$. Also, let the contour
be such that for $z\in\Gamma$ the function $\mu(z,z_{0})\ll1$. In
the $F^{2}$-gauge, a typical term in Eq.~(\ref{eq:ConnecMat2})
can be written as 
\begin{equation}
\begin{split}\M_{F}(z_{\bar{1}})\M_{F}(z_{\bar{2}})..\M_{F}(z_{\bar{n}}) & =\left(\frac{1}{2}i\right)^{n}\epsilon_{r}(z_{\bar{1}})\omega(z_{\bar{1}})\epsilon_{r}(z_{\bar{2}})q(z_{\bar{2}})...\epsilon_{r}(z_{\bar{n}})\omega(z_{\bar{n}})\\
 & \times\left(1-\exp\left[-2i\int_{z_{\bar{2}}}^{z_{\bar{1}}}dz_{a_{1}}\omega\left(z_{a_{1}}\right)\right]\right)\left(1-\exp\left[-2i\int_{z_{\bar{3}}}^{z_{\bar{2}}}dz_{a_{2}}\omega\left(z_{a_{2}}\right)\right]\right)...\\
 & \times\left(1-\exp\left[-2i\int_{z_{\bar{n}}}^{z_{\bar{n}-\bar{1}}}dz_{a_{n-1}}\omega\left(z_{a_{n-1}}\right)\right]\right)\\
 & \times\begin{pmatrix}-\exp\left[2i\int_{z_{\bar{n}}}^{z_{\bar{1}}}dz_{a_{n}}\omega\left(z_{a_{n}}\right)\right] & -\exp\left[2i\int_{z_{(*)}}^{z_{\bar{1}}}dz_{a_{n}}\omega\left(z_{a_{n}}\right)\right]\\
\exp\left[-2i\int_{z_{(*)}}^{z_{\bar{n}}}dz_{a_{n}}\omega\left(z_{a_{n}}\right)\right] & 1
\end{pmatrix}.
\end{split}
\end{equation}
Due to monotonicity in $\left|\exp\left(i\int_{z_{(*)}}^{z}dz'\,\omega(z')\right)\right|$
exponential factors which contribute to the integrals in Eq.$\,$(\ref{eq:ConnecMat2})
may be bounded as 
\begin{equation}
\frac{1}{2}\left|1-\exp\left[-2i\int_{z_{\bar{i}+\bar{1}}}^{z_{\bar{i}}}dz'\,\omega\left(z'\right)\right]\right|\leq1
\end{equation}
where $z_{\bar{i}}$ lies in between $z_{\bar{i}+\bar{1}}$ and the
end point $z_{1}$ (as seen from the integration limits in Eq.(\ref{eq:ConnecMat2})).
This implies that the absolute value of the $(n+1)\text{th}$ contribution
in Eq.$\,$(\ref{eq:ConnecMat2}) is bounded by $\leq\mu^{n}\left(z_{1},z_{0}\right)A_{ij}(z_{1},z_{0})/\left(2n!\right)$,
where $A_{ij}(z_{1},z_{0})$ depends on the matrix component considered:\footnote{Without going through a general gauged framework, \citep{FF:1965}
uses the vector $\left(\beta,\alpha\right)$ ($\left(a_{+},a_{-}\right)$
in their notation) to define the propagator matrix (which they call
the $F$-matrix) instead of the vector $V_{k}=\left(\alpha,\beta\right)$
used here. Hence the expressions derived in this subsection are the
same as those in \citep{FF:1965} up to a $1\leftrightarrow2$ switch
in the matrix element indices.} 
\begin{align}
 & \left|U_{F22}(z_{1},z_{0})-1\right|\leq\frac{\mu}{2}+O(\mu^{2})\label{eq:starteq}\\
 & \left|U_{F21}(z_{1},z_{0})\right|\leq\left|\exp\left[-2i\int_{z_{(*)}}^{z_{0}}dz'\omega(z')\right]\right|\left(\frac{\mu}{2}+O(\mu^{2})\right)\\
 & \left|U_{F12}(z_{1},z_{0})\right|\leq\left|\exp\left[2i\int_{z_{(*)}}^{z_{1}}dz'\omega(z')\right]\right|\left(\frac{\mu}{2}+O(\mu^{2})\right)\\
 & \left|U_{F11}(z_{1},z_{0})-1\right|\leq\frac{\mu}{2}+\left|\exp\left[2i\int_{z_{0}}^{z_{1}}dz'\omega(z')\right]\right|\left(\frac{\mu^{2}}{4}+O(\mu^{3})\right)\label{eq:asymptotic-constraint}
\end{align}
which gives asymptotic expansion constraints for propagator matrices
across monotonic paths. For example, if one can give an upper bound
on $\left|\exp\left[2i\int_{z_{0}}^{z_{1}}dz'\omega(z')\right]\right|$
then one can conclude $\mu^{2}$ times this will vanish as fast as
$\mu^{2}$ in the ``adiabatic'' regions\footnote{These are regions of the complex plane where the magnitudes of the
complexified adiabaticity parameters are $\ll1$.} of the complex plane.

The above can now be used to constrain the propagator matrices between
two adjacent anti-Stokes lines (bounding a region that contains at
least one Stokes line). Let, $z_{0}\text{(anti-Stokes)},\,z_{1}\text{(Stokes)},$
and $z_{2}\text{(anti-Stokes)}$ be points on the anti-Stokes and
Stokes lines as in Fig. \ref{fig:PM_SL}. Now, the path connecting
these points can be divided into two monotonic paths $\Gamma_{01}(z_{0},z_{1})$
and $\Gamma_{12}(z_{1},z_{2})$. Depending on whether the Stokes line
is the ``$-$'' or ``$+$'' kind, the point $z_{1}$ is a minimum
or a maximum of the two monotonic paths.

Consider first the case of a ``$-$'' Stokes line corresponding
to a minimum at $z_{1}$. The $22-$ and $12-$components of the composition
property 
\begin{equation}
U_{F}(z_{2},z_{1})=U_{F}(z_{2},z_{0})U_{F}(z_{0},z_{1})
\end{equation}
and su(1,1) property of $M$ 
\begin{equation}
\text{det}\left[U_{F}(z_{2},z_{0})\right]=1
\end{equation}
can be used to get the following relations 
\begin{equation}
\begin{split}U_{F}{}_{22}(2,0) & =\frac{U_{F}{}_{22}(2,1)}{U_{F}{}_{22}(0,1)}-\frac{U_{F}{}_{12}(0,1)}{U_{F}{}_{22}(0,1)}U_{F}{}_{21}(2,0)\\
U_{F}{}_{11}(2,0) & =\frac{U_{F}{}_{22}(0,1)}{U_{F}{}_{22}(2,1)}+\frac{U_{F}{}_{12}(2,1)}{U_{F}{}_{22}(2,1)}U_{F}{}_{21}(2,0)\\
U_{F}{}_{12}(2,0) & =\frac{U_{F}{}_{12}(2,1)}{U_{F}{}_{22}(0,1)}-\frac{U_{F}{}_{12}(0,1)}{U_{F}{}_{22}(2,1)}\\
 & -\frac{U_{F}{}_{12}(0,1)U_{F}{}_{12}(2,1)}{U_{F}{}_{22}(0,1)U_{F}{}_{22}(2,1)}U_{F}{}_{21}(2,0)
\end{split}
\end{equation}
where $U_{F}{}_{ab}(i,j)\equiv U_{F}{}_{ab}(z_{i},z_{j})$. Eqs.~(\ref{eq:starteq})-(\ref{eq:asymptotic-constraint})
then imply 
\begin{equation}
\begin{split}U_{F}{}_{11}(2,0) & =1+O(\mu)+O(\mu)U_{F}{}_{21}(2,0)\\
U_{F}{}_{22}(2,0) & =1+O(\mu)+O(\mu)U_{F}{}_{21}(2,0)\\
U_{F}{}_{12}(2,0) & =O(\mu)+O(\mu^{2})U_{F}{}_{21}(2,0).
\end{split}
\label{MyEq:50}
\end{equation}
A similar set of equations and estimates can be found for the second
case of a maximum at $z_{1}$, using the 22- and 21-components of
$U_{F}(1,0)=U_{F}(1,2)U_{F}(2,0)$ and the determinant condition.
These are 
\begin{equation}
\begin{split}U_{F}{}_{22}(2,0) & =\frac{U_{F}{}_{22}(1,0)}{U_{F}{}_{22}(1,2)}-\frac{U_{F}{}_{21}(1,2)}{U_{F}{}_{22}(1,2)}U_{F}{}_{12}(2,0)\\
U_{F}{}_{11}(2,0) & =\frac{U_{F}{}_{22}(1,2)}{U_{F}{}_{22}(1,0)}+\frac{U_{F}{}_{21}(1,0)}{U_{F}{}_{22}(1,0)}U_{F}{}_{12}(2,0)\\
U_{F}{}_{21}(2,0) & =\frac{U_{F}{}_{21}(1,0)}{U_{F}{}_{22}(1,2)}-\frac{U_{F}{}_{21}(1,2)}{U_{F}{}_{22}(1,0)}\\
 & -\frac{U_{F}{}_{21}(1,0)U_{F}{}_{21}(1,2)}{U_{F}{}_{22}(1,0)U_{F}{}_{22}(1,2)}U_{F}{}_{12}(2,0).
\end{split}
\end{equation}
Eqs.~(\ref{eq:starteq})-(\ref{eq:asymptotic-constraint}) in this
case then imply 
\begin{equation}
\begin{split}U_{F}{}_{11}(2,0) & =1+O(\mu)+O(\mu)U_{F}{}_{12}(2,0)\\
U_{F}{}_{22}(2,0) & =1+O(\mu)+O(\mu)U_{F}{}_{12}(2,0)\\
U_{F}{}_{21}(2,0) & =O(\mu)+O(\mu^{2})U_{F}{}_{12}(2,0).
\end{split}
\label{MyEq:51}
\end{equation}
Here, we see that the perturbative structure in Eqs.~(\ref{eq:starteq})-(\ref{eq:asymptotic-constraint})
helps us make estimates of the relative magnitudes of the elements
of the F-matrix. Note that because the above discussion leaves the
off-diagonal elements such as $U_{F21}$ and $U_{F12}$ unconstrained,
without further analysis, Eqs.~(\ref{MyEq:50}) and (\ref{MyEq:51})
do not actually state that the diagonal terms have a leading magnitude
of unity and one of the off-diagonal elements is suppressed. On the
other hand, if one can give an additional condition that these off-diagonal
elements are at most order unity, then the number of $U_{F}$ matrix
elements that need to be determined to leading order in $\mu$ become
one. This is the main advantage of using the $F^{2}$-gauge.

The additional constraint on the propagator elements $U_{F21}$ and
$U_{F12}$ (across a $-$ and a $+$ Stokes lines respectively), fixing
them to be $O(\mu^{0})$ to leading order, is obtained from the leading
order WKB approximation of solutions to the mode equation. In the
annulus and away from the Stokes lines, WKB approximation fixes $V(z)$
to be approximately constant (up to higher orders in $\mu$). This
in turn bounds the propagator elements. A stepwise derivation of this
result, i.e. 
\begin{equation}
\mu U_{F21}\rightarrow0,\qquad\text{and}\qquad\mu U_{F12}\rightarrow0
\end{equation}
in the limit $\mu\rightarrow0$ is detailed in appendix \ref{sec:An-asymptotic-property}.
Now, taking the limit $\mu\rightarrow0$ of Eq.~(\ref{MyEq:50})
and Eq.~(\ref{MyEq:51}), the propagator matrices reduce to 
\begin{equation}
U_{F\text{sgn}(s)}=\begin{pmatrix}1 & \left(\frac{1+s}{2}\right)\lim_{\mu\rightarrow0}U_{F12}\\
\left(\frac{1-s}{2}\right)\lim_{\mu\rightarrow0}U_{F21} & 1
\end{pmatrix}\label{eq:reduced_prop}
\end{equation}
where $s=\pm1$ corresponds to the positive and negative Stokes lines
respectively. Interestingly, $U_{F+}\left(U_{F-}\right)$ is an element
of an additive unipotent group. To compute the Bogoliubov coefficient
$\beta$ the cross diagonal elements of Eq.$\,$(\ref{eq:reduced_prop})
need to be determined. We will see in Sec.$\,$(\ref{subsec:Stokes-constants-from})
that in the case of dispersion relations of the form $\omega^{2}=Az^{n}$
these coefficients are fixed by symmetries of the mode equation.

\subsection{\label{subsec:Stokes-constants-from}Stokes constants from symmetries}

In this section, we will work in the $F^{2}$-gauge to compute the
$O\left(\mu^{0}\right)$ propagator matrices $U(z_{1},z_{2})$ across
two adjacent anti-Stokes lines in systems where $\omega^{2}=Az^{n}$.
These systems as we will see are simply related to the long wavelength
limit $\left(k\rightarrow0\right)$ of dispersion relations of the
form $\omega^{2}\approx k^{2}+Az^{n}$. For systems with $\omega^{2}=Az^{n}$,
the mode equation is symmetric under a discrete set of coordinate
rotations which in turn induces a symmetry representation in the propagators.
Combining this with the reduced form of the propagator matrices described
in the previous section and analyticity properties of solutions to
the mode equation will allow us to determine the unknown cross diagonal
propagators - also known as Stokes constants.

The analytic structure of the solutions of the mode equation and that
of the (WKB) basis functions play an important role in constraining
these constants. This may be understood as follows. Note, that the
analytically continued mode function $\chi_{k}\left(z\right)$ is
single valued on the complex plane whereas the WKB basis functions,
being proportional to $\omega^{-1/2}\left(z\right)$, are multivalued
for general polynomial dispersion relations. Choosing a specific branch
as our choice of the WKB basis functions the transformation/discontinuity
of these functions across their chosen branch cuts provides a measure
of the mismatch of these basis functions with the single valued function
$\chi_{k}\left(z\right)$. This in turn imposes a constraint on the
Stokes constants of the system.

\begin{figure}
\centering{} \includegraphics[scale=0.22]{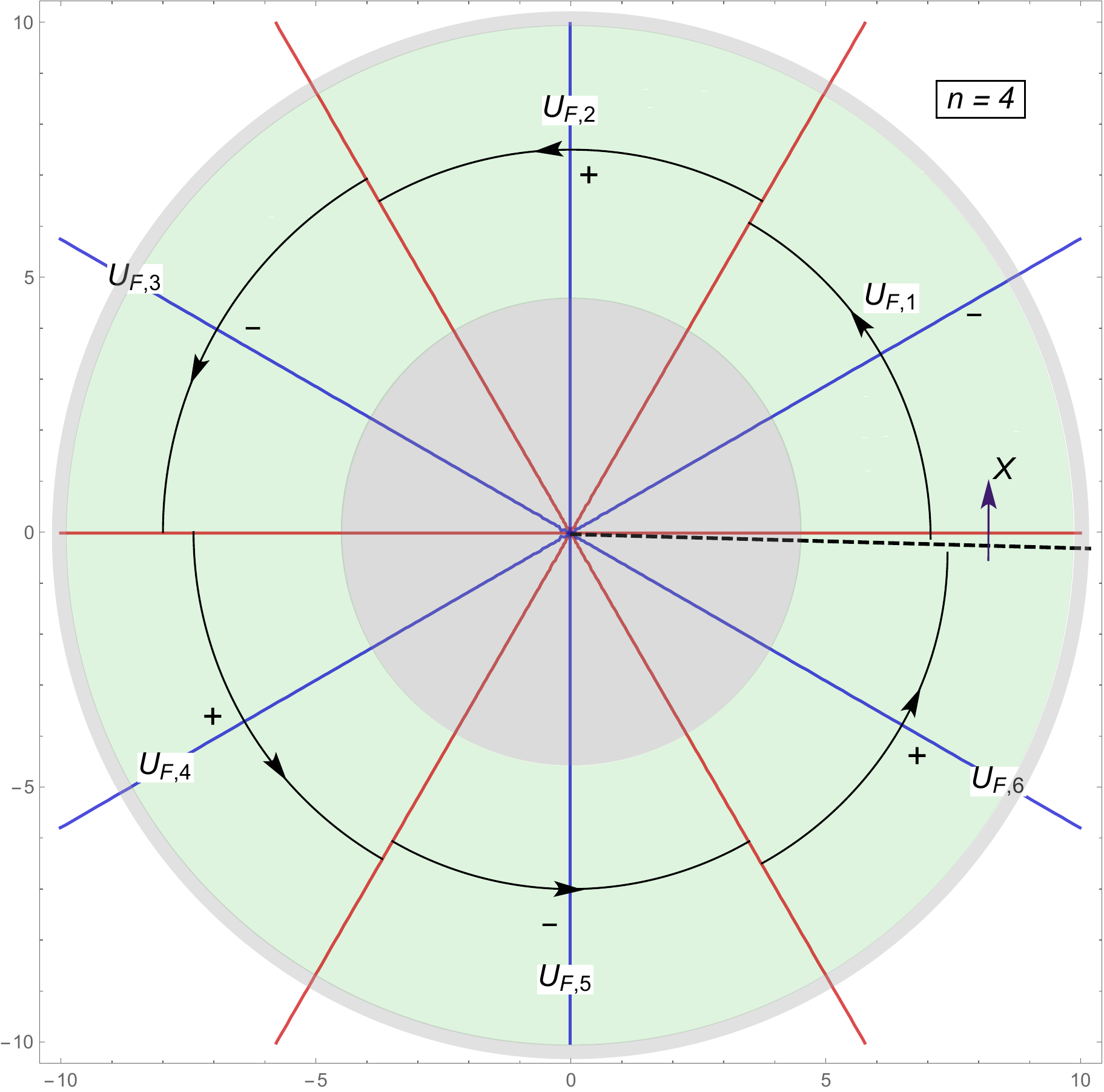}\caption{\label{fig:n=00003D00003D4Ex}This figure shows the Stokes (blue)
lines and anti-Stokes (red) lines in the case of the \textquotedblright dispersion
relation\textquotedblright{} $\omega^{2}=Az^{4}$ , and a branch cut
(black dashed) describing the general scenario of a positive even
integer $n$. $X$ denotes the transformation of $V(z)=\left(\alpha(z),\beta\left(z\right)\right)$
across the branch cut. The branch cut, chosen below the positive real
line, is such that $\omega$ is positive definite on $\mathbb{R}_{>0}$.
For $n\mod4=0$, the branch cut vanishes, i.e. $X=\mathbb{I}.$ For
$n\mod4=2$ the branch cut choice is such that $\omega<0$ on $\mathbb{R}_{<0}$
and the transformation $X$ is the inverse of the transformation in
Eq. (\ref{eq:basis_trans}). The label \textquotedblleft$+$\textquotedblright{}
(\textquotedblleft$-$\textquotedblright ) indicates Stokes contours
from $z_{0}=0$ the branch point along which $i\int_{0}^{z}dz'\omega(z')$
is real and increases (decreases). The propagator matrices across
a Stokes sector is denoted by $U_{F}\left(j\right)$ where$j\in\mathbb{Z}_{\protect\leq n+2=6}$.
The green annulus corresponds to the region in the complex plane where
$\vert\delta\left(z\right)\vert\gtrsim1$ and $z<z_{max}$ , where
the latter is imposed to avoid the essential singularity in the WKB
basis functions.}
\end{figure}

Specifically, for ``dispersion'' relations of the form

\begin{equation}
\omega^{2}(z)=Az^{n}
\end{equation}
where $n$ is a positive even integer, we may choose $\omega^{-1/2}\left(z\right)$
to be defined by a single branch cut below the positive real line
(Fig. \ref{fig:n=00003D00003D4Ex}). \footnote{This choice of branch cut does not satisfy the physical requirement
of positive definite particle energy $\omega$, since $\omega<0$
on the negative real line for $n\mod4=2$.} Under this choice, the WKB basis functions defined as 
\begin{equation}
f_{\pm}\left(z\right)=\frac{\exp\left(\pm i\int_{0}^{z}dz'\:\omega\left(z'\right)\right)}{\sqrt{2\omega\left(z\right)}}
\end{equation}
with origin taken as $0$ for simplicity, transform as 
\begin{equation}
F_{k}\left(ze^{2\pi i}\right)=\left(-iB\right)^{n}F_{k}\left(z\right)\label{eq:basis_trans}
\end{equation}
with 
\begin{equation}
B\equiv\left(\begin{array}{cc}
0 & 1\\
1 & 0
\end{array}\right)\label{eq:Bdef}
\end{equation}
under $z\rightarrow ze^{2\pi i}$ across the branch cut. \footnote{Note, for $n\mod4=0$, there is no branch cut - that is the transformation
of the WKB basis functions under $z\rightarrow ze^{2\pi i}$ is identity.} This implies that the functions $\left(\alpha\left(z\right),\beta\left(z\right)\right)$
should transform inversely under $z\rightarrow ze^{2\pi i}$ to maintain
the single valuedness of the modefunction $\chi_{k}(z)$, implying
an analytic structure as in Fig. \ref{fig:n=00003D00003D4Ex}. Restricting
to propagator matrices in the green annular region, which avoids the
essential singularity as $\vert z\vert\rightarrow\infty$ and regions
of $\vert\delta\left(z\right)\vert\gtrsim1$ near the zero of $\omega^{2}\left(z\right)$,
the propagator matrices between any two anti-Stokes lines should be
such that propagation in a full circle returns identity. This constrains
the Stokes constants in the system, as in Eq. (\ref{eq:SingleVal}).
In fact we will see that symmetries in the system equate all the Stokes
constants, and the above constraint fixes them in terms of the branch
cut transformation of $\left(\alpha\left(z\right),\beta\left(z\right)\right)$.

\begin{figure}
\centering{}a) \includegraphics[scale=0.21]{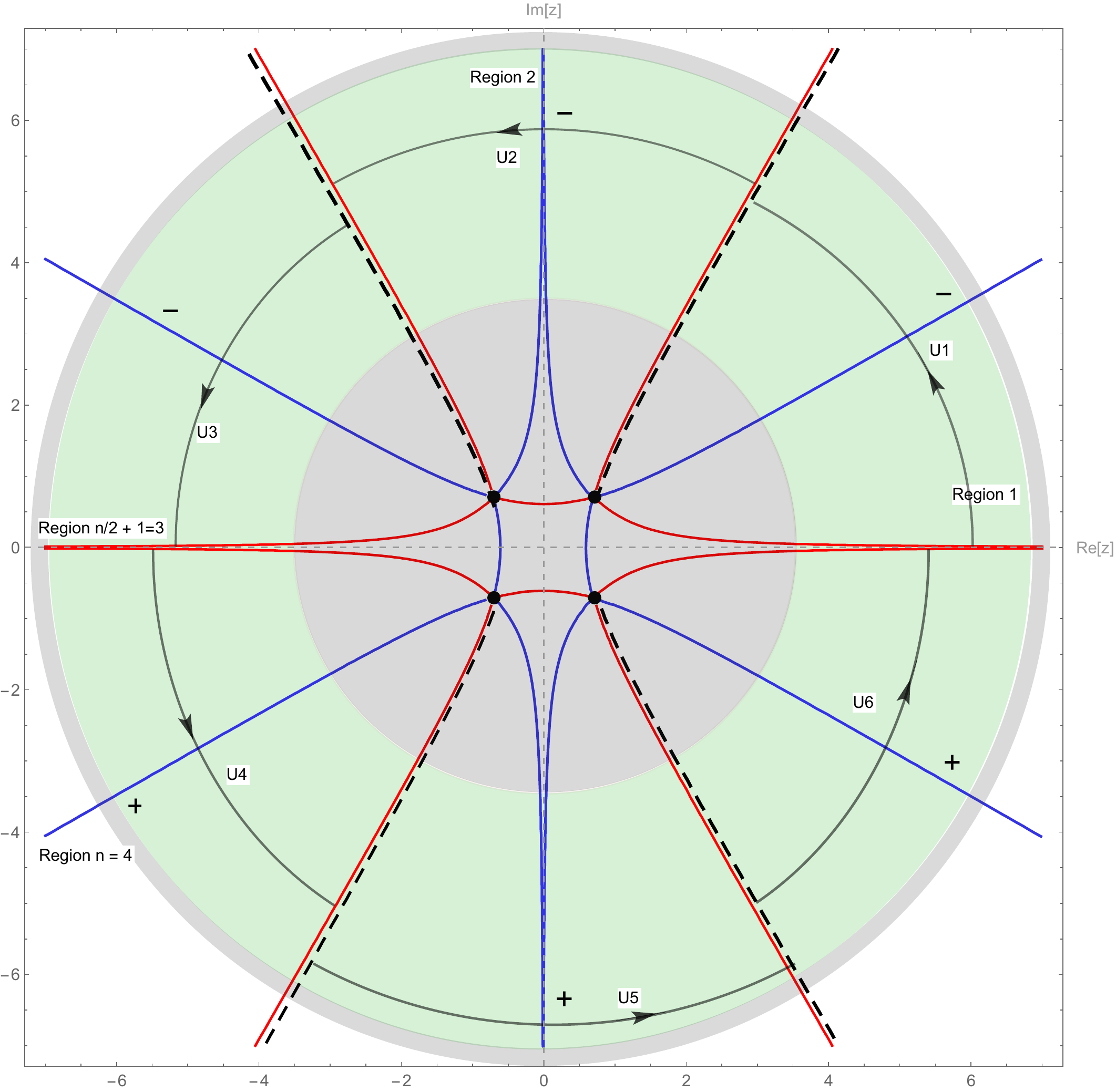}b)\includegraphics[scale=0.237]{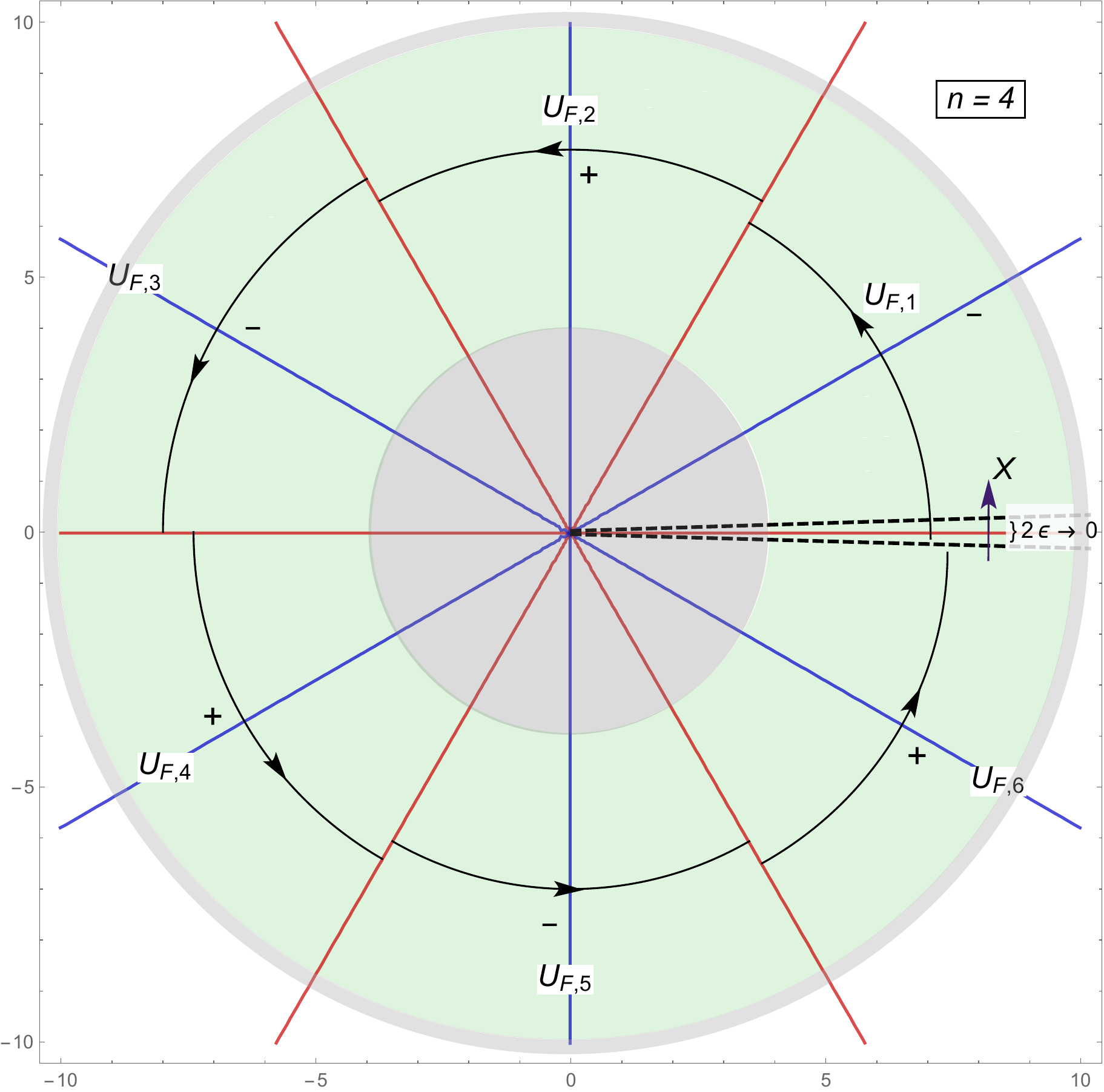}\caption{\label{fig:n=00003D00003D00003D4Ex}This figure shows the Stokes (blue)
lines, anti-Stokes (red) lines and the branch cut (black dashed) choices
for the dispersion relation $\omega^{2}=k^{2}+Az^{4}$ (left) and
$\lim_{k\rightarrow0}\omega^{2}=k^{2}+Az^{4}$ (right). The label
\textquotedblleft$+$\textquotedblright{} (\textquotedblleft$-$\textquotedblright )
indicates Stokes lines where the value of $i\int_{z_{b}}^{z}dz'\omega(z')$
increases (decreases) from the branch point. The corresponding propagator
matrices across a Stokes sector have been denoted by $U_{F,j}$ where$j\in\mathbb{Z}_{\protect\leq6}$.
Left figure: The branch cuts are such that the dispersion relation
$\omega$ remains positive definite and continuous on the real line.
Some of the (anti-)Stokes lines merge (separation decreasing as $|z|^{-2}$)
in the green region which represents the region in which the asymptotic
expansion is a good approximation. Right figure: A different branch
cut structure is used that is technically a different analytic continuation
compared to the figure on the left (see Fig.~\ref{fig:Deformation-of-the}
and text). For either analytic continuation, we define a Stokes sector
in a $k/A^{1/6}$ independent manner as a region in the green annulus
bounded by two anti-Stokes lines containing at least one Stokes line.}
\end{figure}

The analogous picture for dispersion relations of the form 
\begin{equation}
\omega^{2}=k^{2}+z^{n}\label{eq:dispe_rel1}
\end{equation}
is as follows. For finite $k$, defining $\omega^{-1/2}\left(z\right)$
with branch cuts extending from the $n$ simple zeroes to infinity
such that $\omega\left(z\right)$ is positive definite on the real
line, we see that the WKB basis functions 
\begin{equation}
f_{\pm}\left(z\right)=\frac{\exp\left(\pm i\int_{0}^{z}dz'\:\omega\left(z'\right)\right)}{\sqrt{2\omega\left(z\right)}}\label{eq:WKB_basis_fink}
\end{equation}
transform discontinuously across each of these branch cuts. In the
limit $k\rightarrow0$, the transformation across each of these branch
cuts reduces to 
\begin{equation}
F_{k}\left(ze^{2\pi i}\right)=-iBF_{k}\left(z\right)\label{eq:basistrans_n=00003D00003D1}
\end{equation}
corresponding to $n=1$ in Eq. (\ref{eq:basis_trans}), whereas the
combined transformation across all the $n$ branch cuts reproduces
the transformation in Eq. (\ref{eq:basis_trans}). Once again, single
valuedness of $\chi_{k}\left(z\right)$ imposes a dual analytic structure
on the functions $\left(\alpha\left(z\right),\beta\left(z\right)\right)$,
such that across each of the above branch cuts they transform inversely
to the WKB basis functions. This results in Fig. \ref{fig:n=00003D00003D00003D4Ex}a
for finite $k$. In the $k\rightarrow0$ limit the transformations
across each of these branch cuts are inverses of the transformations
of Eq. (\ref{eq:basistrans_n=00003D00003D1}). Merging the branch
cuts in the upper and lower half planes in this limit, Fig. \ref{fig:n=00003D00003D00003D4Ex}b
, shows the similarity and distinction from Fig. \ref{fig:n=00003D00003D4Ex}.
We see that in the $k\rightarrow0$ limit of Eq. (\ref{eq:dispe_rel1}),
the functions $\left(\alpha\left(z\right),\beta\left(z\right)\right)$
correspond to a different analytic continuation than in Fig. \ref{fig:n=00003D00003D4Ex}.
When computing Bogoliubov coefficients, this difference must be taken
into account (rf. Eq. (\ref{eq:complication})).

The figure also includes the Stokes and anti-Stokes lines drawn from
each of the zeroes.\footnote{The Stokes and anti-Stokes lines from each zero are drawn with respect
to WKB basis functions defined with origin at the respective zeroes.
These asymptote to the (anti-)Stokes lines from Eq. (\ref{eq:WKB_basis_fink})
for $\left|Az^{n}\right|\gg k^{2}$.} In the region $\left|Az^{n}\right|\gg k^{2}$ as in the green annulus
we see that the Stokes lines from the zeroes merge and asymptote to
those in the scenario of $\omega^{2}=Az^{n}$. The Stokes constants
across these Stokes lines are nontrivial functions of $k$, but it
may be shown that in the $k\rightarrow0$ limit these reduce to those
of the previous scenario with $\omega^{2}=Az^{n}$.\footnote{This is true up to factors of $i$ which arise from the differences
in the analytic continuation.}.

To compute these Stokes constants, note after the analytic continuation
the operator governing the mode equation in Eq.~(\ref{eq:modeeq-1})
becomes $\mathcal{O}_{z}\equiv\partial_{z}^{2}+\omega^{2}(z)$. For
dispersion relationships 
\begin{equation}
\omega^{2}(z)=Az^{n}\hspace{5mm}\text{where}\,\,n\in\mathbb{Z}\label{eq:power}
\end{equation}
the mode equation operator $\mathcal{O}_{z}$ is symmetric under the
$\mathbb{Z}_{n+2}$ discrete symmetry 
\begin{equation}
z\rightarrow R^{q}z\equiv\exp\left(\frac{2\pi iq}{n+2}\right)z.\label{eq:Symm_Rep}
\end{equation}
This can be used to find a $\mathbb{Z}_{n+2}$ symmetry representation
of the propagator as discussed in Appendix \ref{sec:Covariance-of-linear}.
The propagator connecting anti-Stokes boundaries take the form 
\begin{equation}
V\left(z_{1}\right)=U_{F}\left(z_{1},z_{0}\right)V\left(z_{0}\right)
\end{equation}
where $V\left(z_{0}\right)=\left(\alpha(z_{0}),\beta(z_{0})\right)$
and $V\left(z_{1}\right)=\left(\alpha(z_{1}),\beta(z_{1})\right)$.
Note that $U_{F}(z_{1},z_{0})$ and $F_{k}(z)$ now depend on $n$
because of their $\omega$ dependences. Here again, we restrict to
propagator matrices in the green annulus region (as in Fig. \ref{fig:n=00003D00003D00003D4Ex}a
for $n=4$ ) which avoids regions of $\vert\delta\left(z\right)\vert\gtrsim1$
at the center of the complex plane and the essential singularity as
$\vert z\vert\rightarrow\infty$. That is, we will focus on $\left|Az^{n}\right|\gg k^{2}$
which is satisfied for 
\begin{equation}
\left|\left(zA^{\frac{1}{n+2}}\right)^{\frac{n}{2}}\right|\gg\left(\frac{n}{2}\right)^{\frac{n}{2+n}}\gg\frac{k}{A^{\frac{1}{n+2}}}
\end{equation}
where the $(n/2)^{n/(2+n)}$ related condition comes from condition
$4\delta=\omega'/\omega^{2}\ll1$, and $|z|<|z_{\max}|$ where $|z_{\mathrm{max}}|$
can be chosen to be arbitrarily large.

To see how the $\mathcal{O}_{z}$ symmetry representation in $U_{F}$
can be used to compute the Bogoliubov coefficient, consider the identity
obtained from single valuedness of the mode functions, i.e. $\chi(z)=\chi(e^{2\pi i}z)$.
Single valuedness implies that propagating the vector $V(z)$ in a
full circle should return the same value i.e $V(z)=V(e^{2\pi i}z)$
for any $z\in\mathbb{C}$. Within the green annulus, propagating around
a closed path implies 
\begin{equation}
XU_{F}(n+2)...U_{F}(2)U_{F}(1)=\mathbb{I}\label{eq:singlevalued}
\end{equation}
where $U_{F}\left(j\right)$ is an abbreviation for the propagator
matrices across the $j$th Stokes line and $X$ is the transformation
associated with the branch cut in Fig. \ref{fig:n=00003D00003D00003D4Ex},
that is 
\begin{equation}
V\left(ze^{2\pi i}\right)=X^{-1}V(z)\quad\text{where}\quad X^{-1}=(-iB)^{-n}
\end{equation}
obtained as the inverse of the WKB basis transformation in Eq. (\ref{eq:basis_trans}).
Note, Fig. \ref{fig:n=00003D00003D00003D4Ex}b implies the same single
valuedness constraint as in Eq. (\ref{eq:singlevalued})since the
combined transformation across the branch cuts equals $X$.

\begin{figure}
\begin{centering}
a)\includegraphics[scale=0.36]{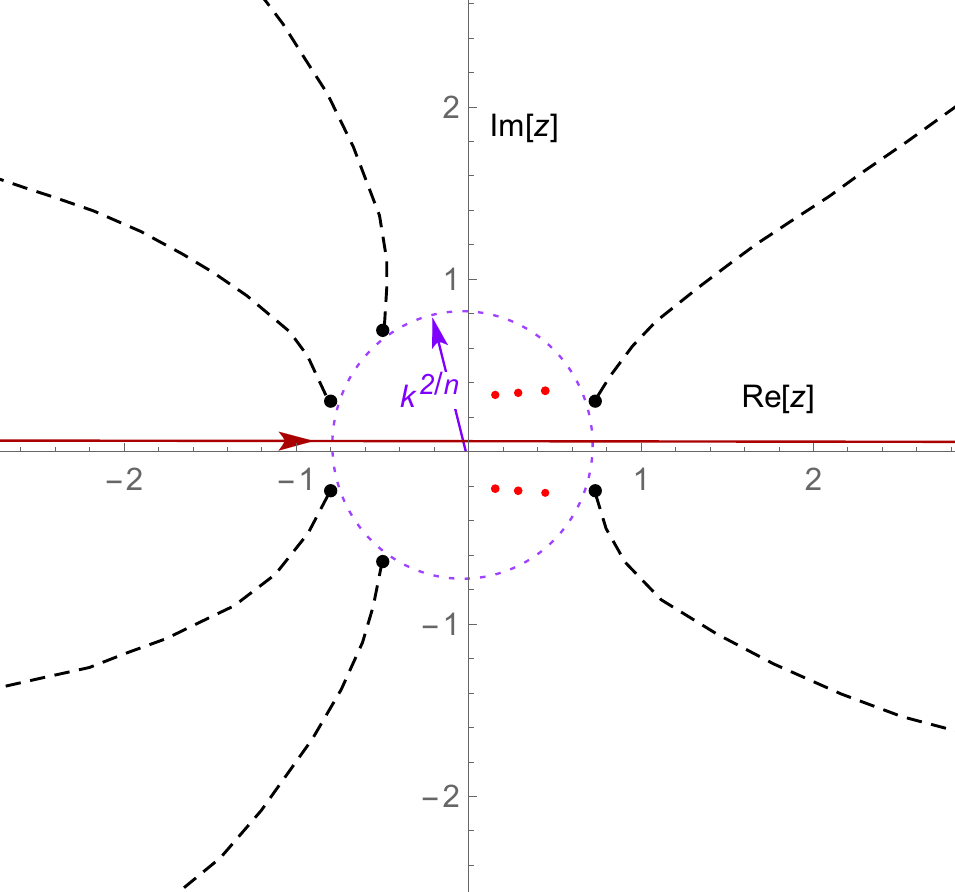}\hspace{1.3cm} b)\includegraphics[scale=0.35]{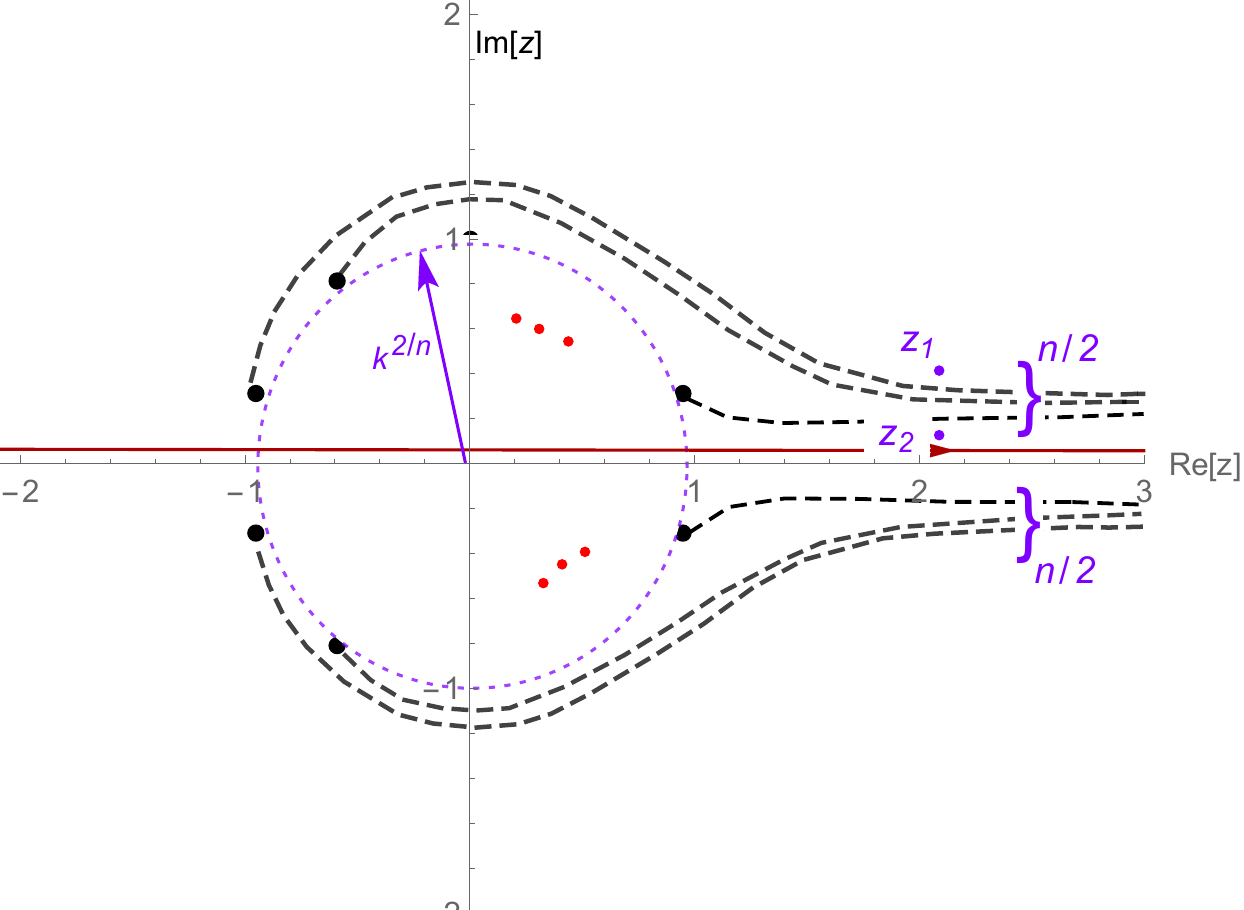}
\par\end{centering}
\begin{centering}
c)\includegraphics[scale=0.35]{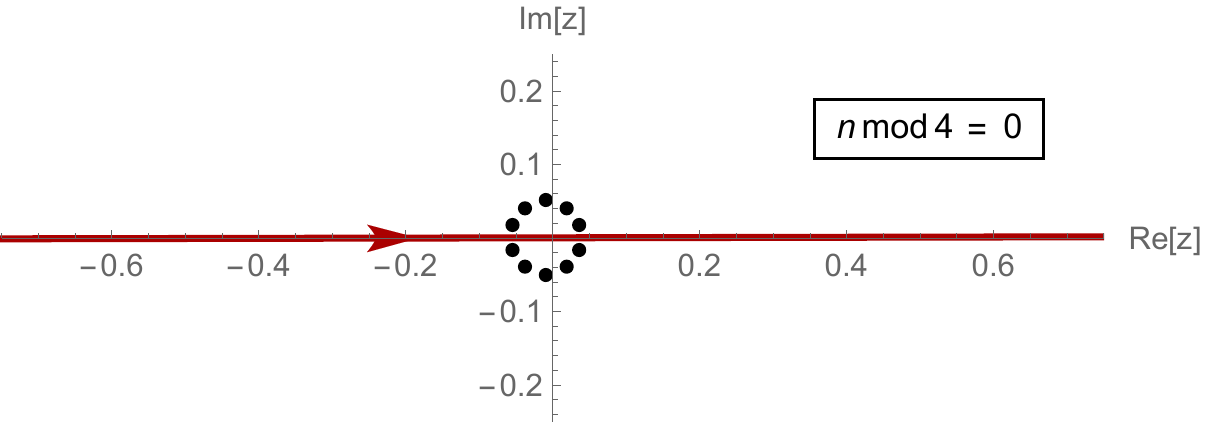}~ d)\includegraphics[scale=0.35]{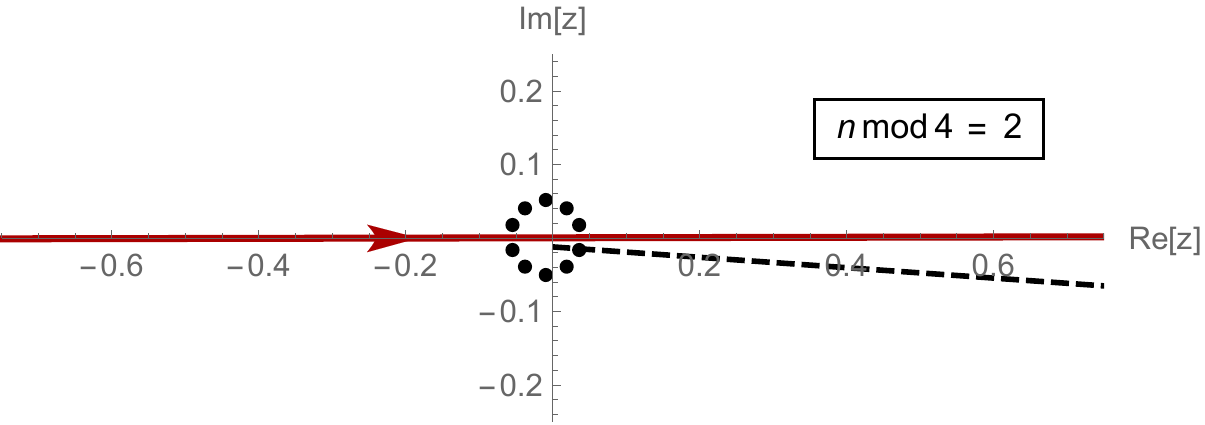}
\par\end{centering}
\caption{\label{fig:Deformation-of-the}a) Shown is the choice of the original
branch cuts with a total of $n$ branch points, half in the upper
half plane and the other half symmetrically in the lower half plane.
b) The branch cuts are deformed such that there are $n/2$ branch
cuts overlapping above and below the real axis. This gives a different
analytic continuation. The $V_{k}$ between points $z_{1}$ and $z_{2}$
are related by a factor of $(i^{-1}B)^{n/2}$. In the limit $k/A^{1/6}\rightarrow0$,
this corresponds to the physical choice since the cuts are taken to
not cross the real line and limit to two sets of $n/2$ merged branch
cuts approaching the real line from above and below. c,d) In contrast,
when $k/A^{1/6}=0$, the branch points and branch cuts are completely
merged making the branch cuts disappear when $n\equiv0\mod4$ and
leaving a single branch cut when $n\equiv2\mod4$.}
\end{figure}

In Appendix \ref{sec:Covariance-of-linear}, we motivate the following
symmetry representation in the propagators induced from the symmetries
of $\mathcal{O}_{z}$ 
\begin{equation}
U_{F}(\bar{z},\bar{z}_{0})=B^{-1}U_{F}(z,z_{0})B\label{eq:transformB}
\end{equation}
where the bar is not a complex conjugation but the discrete rotation
defined as 
\begin{equation}
\left(\bar{z},\bar{z}_{0}\right)\equiv\left(\exp\left(\frac{2\pi i}{n+2}\right)z,\exp\left(\frac{2\pi i}{n+2}\right)z_{0}\right).
\end{equation}
To understand the $U_{F}$ representation, it is useful to note that
each region divided by the anti-Stokes lines has a Stokes line on
which the absolute value of the WKB phase $\left|\exp\left(i\int_{z_{(*)}}^{z}dz'\,\omega(z')\right)\right|$
increases or decreases, labeled by $"+"$ and $"-"$ Stokes lines,
respectively (introduced in Eq.~(\ref{eq:thetagen})). For the propagator
as $U_{F}\left(j\right)$ across the $j$th Stokes line Eq.~(\ref{eq:transformB})
can be used to write 
\begin{equation}
U_{F}\left(j+1\right)=B^{-1}U_{F}\left(j\right)B
\end{equation}
\begin{equation}
U_{F}\left(j+2\right)=B^{-2}U_{F}\left(j\right)B^{2}=U_{F}\left(j\right).
\end{equation}
Therefore, along a circular contour, the propagator matrices between
any two adjacent anti-Stokes lines where the propagation crosses a
``$+(-)$'' Stokes line are the same. Now defining the propagation
matrix $U_{F}$ across a ``$\pm$'' Stokes lines as $U_{F\pm}$,
we can write 
\begin{equation}
U_{F\pm}=B^{-1}U_{F\mp}B.\label{eq:relateplusminus}
\end{equation}
More specifically, if there are $n+2$ anti-Stokes lines then the
number of $U_{F}$ matrix multiplications required to return to identity
is $n+2$. The explicit form of the $U_{F\pm}$ matrices is given
by Eqs.~(\ref{eq:reduced_prop}). Defining $\lim_{\mu\rightarrow0}U_{F12}=S$
this can be written as 
\begin{equation}
U_{F+}=\left(\begin{array}{cc}
1 & S\\
0 & 1
\end{array}\right)\label{eq:UFplusparam}
\end{equation}
and an analogous expression for $U_{F-}$determined by Eq.$\,$(\ref{eq:relateplusminus}).

In the appendix \ref{sec:An-asymptotic-property}, we show that $\mu^{n>0}U_{F21}\rightarrow0$
(and similarly for $\mu^{n>0}U_{F12})$ in the annulus.

To obtain $S$, we use single valuedness of the mode function rewritten
as a propagation around a closed contour as written in Eq.~(\ref{eq:singlevalued}).
Now, traversing a closed loop in the counter-clockwise direction,
Eq.~(\ref{eq:singlevalued}) implies %
\begin{equation}
\left(U_{F+}U_{F-}\right)^{\frac{n}{2}+1}\left(-iB\right)^{n}=\mathbb{I}.
\end{equation}
Using Eq.~(\ref{eq:relateplusminus}), we have the condition 
\begin{equation}
\left(U_{F+}B\right)^{n+2}=(-iB)^{-n}=i^{n}\label{eq:SingleVal}
\end{equation}
for $n=2m$, $m\in\mathbb{Z}_{>0}$. To solve this equation for $S$
(valid for $\mu=0$), we diagonalize 
\begin{equation}
\Lambda^{-1}\left(U_{F+}B\right)\Lambda=\left(\begin{array}{cc}
\frac{S-\sqrt{4+S^{2}}}{2} & 0\\
0 & \frac{S+\sqrt{4+S^{2}}}{2}
\end{array}\right)\label{eq:UFplus}
\end{equation}
where 
\begin{equation}
\ensuremath{\Lambda=\left(\begin{array}{cc}
\frac{S-\sqrt{4+S^{2}}}{2} & \frac{S-\sqrt{4+S^{2}}}{2}\\
1 & 1
\end{array}\right)}.
\end{equation}
This and Eq.$\,$(\ref{eq:SingleVal}) can be used to obtain the following
condition: 
\begin{equation}
\left(\frac{S\pm\sqrt{4+S^{2}}}{2}\right)^{2m+2}=(-1)^{m}\label{eq:EigenValRel}
\end{equation}
which can be solved to obtain 
\begin{equation}
S=2i\cos\left(\frac{\pi}{n+2}\right).\label{eq:s-cosine}
\end{equation}

Eq.~(\ref{eq:s-cosine}) is derived in Eq.~(7.11) of \citep{FF:1965}
using mostly the same ideas except that reference constructs and uses
the explicit power series general solution information whereas we
here have not made any reference to an explicit solution construction.
Note that the appearance of $n+2$ which counts the number of regions
bounded by two anti-Stokes lines with at least one Stokes line in
between. We emphasize that this $n+2$ counting identification is
manifest in $S^{n+2}$ that arises from the $n+2$ factors of $U_{F}$
in Eq.~(\ref{eq:singlevalued}). Eq.~(\ref{eq:UFplusparam}) thus
gives $U_{F+}$ to $O(\mu^{0})$ and $U_{F-}$ as 
\begin{equation}
U_{F-}=B^{-1}U_{F+}B=\left(\begin{array}{cc}
1 & 0\\
S & 1
\end{array}\right)\label{eq:matrixrep}
\end{equation}
from Eq.~(\ref{eq:transformB}). Note that even though we derived
this result using the $F^{2}$-gauge, this propagator result is also
valid in the 0-gauge due to the form of the gauge transformation.

One way to view this result is as a mathematical identity when comparing
the $U_{0}$ and $U_{F}$ to zeroth order in $\mu$. In this limit,
we can write $U_{0+}$ as 
\begin{equation}
U_{0+}=\lim_{R\rightarrow\infty}\mathbb{P}\left[\mathbb{I}_{2\times2}\,\exp\left(\frac{n}{4}\int_{\frac{2\pi}{n+2}}^{\frac{4\pi}{n+2}}d\theta\begin{pmatrix}0 & e^{i\frac{4}{2+n}R^{1+\frac{n}{2}}e^{i\left(1+\frac{n}{2}\right)\theta}}\\
e^{-i\frac{4}{2+n}R^{1+\frac{n}{2}}e^{i\left(1+\frac{n}{2}\right)\theta}} & 0
\end{pmatrix}\right)\right]
\end{equation}
where we have taken the integral from an anti-Stokes to anti-Stokes
line with at least one ``$+$'' Stokes line in between. Hence, the
mathematical identity is 
\begin{equation}
\lim_{R\rightarrow\infty}\mathbb{P}\left[\mathbb{I}_{2\times2}\,\exp\left(\frac{n}{4}\int_{\frac{2\pi}{n+2}}^{\frac{4\pi}{n+2}}d\theta\begin{pmatrix}0 & e^{i\frac{4}{2+n}R^{1+\frac{n}{2}}e^{i\left(1+\frac{n}{2}\right)\theta}}\\
e^{-i\frac{4}{2+n}R^{1+\frac{n}{2}}e^{i\left(1+\frac{n}{2}\right)\theta}} & 0
\end{pmatrix}\right)\right]=\left(\begin{array}{cc}
1 & 2i\cos\left(\frac{\pi}{n+2}\right)\\
0 & 1
\end{array}\right).\label{eq:mathidentity}
\end{equation}
This is one of the key non-trivialities that the $F^{2}$-gauge affords
us compared to the 0 gauge of Eq.~(\ref{eq:0gauge}). As we will
see, this will play a role in the computation of particle production
in cosmology in a particular kinematic limit.

Assuming $n\in$even, we identify the discrete group $\mathbb{Z}_{q}$
that $U_{F}$ belongs to by imposing the condition that the phase
\begin{equation}
\left(\left(U_{F+}U_{F-}\right)^{\frac{n}{2}+1}\right)^{q}=B^{nq}\exp(-in\pi q/2)\hspace{1em}n\mbox{ even}
\end{equation}
returns to unity. This means 
\begin{equation}
q=\mathrm{min}_{p\in\mathbb{N}}\left(\frac{4p}{n}\right)\in\mathbb{Z}.\label{eq:integer}
\end{equation}
Hence, we can summarize the representation as $U_{F}$ belonging to
$\mathbb{Z}_{q}$ and $X$ belonging to $\mathbb{Z}_{4}$.

In summary, we have computed the propagator matrix $U$ in the $F^{2}$-gauge
for general $z^{n}$ using the methods of \citep{FF:1965} and used
our gauge formalism to attribute it also to the 0-gauge to leading
order in the complexified nonadiabaticity parameter $\mu$ in the
$\omega^{2}\sim z^{n}$ model. The computation the reduced form of
the propagators Eq.$\,$(\ref{eq:reduced_prop}) derived in the $F^{2}$-gauge
and the discrete symmetries explained in this section and Appendix
\ref{sec:Covariance-of-linear}. One way to view the non-triviality
of having a gauge picture which allows us to compare the propagator
expressions in different gauges is that it allows us to derive a mathematical
identity of Eq.~(\ref{eq:mathidentity}).

\subsection{Computing particle production with $k\rightarrow0$}

We will use Eqs.~(\ref{eq:UFplusparam}), (\ref{eq:matrixrep}),
and (\ref{eq:s-cosine}) to compute the Bogoliubov coefficients for
a special class of kinematic context introduced below Eq.~(\ref{eq:withcurvature}).

Although the $V_{k}(\eta)$ in Eq.~(\ref{eq:maineq}) can be approximately
interpreted directly as Bogoliubov coefficients in the adiabatic region,
their map to particle production acquires additional permutation structure
for cases when $n/2$ in Eq.~(\ref{eq:scalingchoice}) is odd. This
is because on the real axis, the positive frequency modes are defined
with respect to $\sim\exp\left(-i\int^{\eta}d\eta'\,\omega\left(\eta'\right)\right)$
where $\omega(\eta)>0$ (positivity of energy) whereas the analytic
continuation of the frequency by itself does not contain the positivity
constraint. For example, when we write $\int_{0}^{\eta}d\eta'\,\omega\left(\eta'\right)=\int_{0}^{\eta}d\eta'\,\sqrt{A}\left(\eta'\right)^{n/2}$
for odd $n/2$ (which is implicit in the analytic continuation), we
are treating $\omega$ to be negative on the negative real axis, thereby
effectively flipping the definition of negative and positive frequencies.
Hence, the map to the physical particle production requires an additional
permutation that can be written as 
\begin{equation}
(\alpha(\eta),\beta(\eta))=P(\eta)V\left(\eta\right)\label{eq:complication}
\end{equation}
\begin{equation}
P(\eta)\equiv\Theta(-\eta)B^{n/2}+\Theta(\eta)\mathbb{I}
\end{equation}
where $B$ is the matrix defined in Eq.~(\ref{eq:Bdef}). This permutation
is accounted for in the branch cut choice of Fig.~\ref{fig:Deformation-of-the}.b
using an effective branch matrix in the upper half plane given by
\begin{equation}
X_{1}\equiv(-iB)^{-n/2}.\label{eq:eff}
\end{equation}
This gives 
\begin{equation}
\beta\left(z_{+\infty}\right)=-i\cot\left[\frac{\pi}{n+2}\right].\label{eq:betafinresult}
\end{equation}

Eq.~(\ref{eq:betafinresult}) shows that the particle production
at the special kinematic point depends only on $\cot(\pi/(n+2))$
and not on $\sqrt{A}$. The factor $n+2$ counts the number of Stokes
sectors on the complex plane for Eq.~(\ref{eq:power}) where we define
a Stokes sector to be the region in the annulus bounded by two anti-Stokes
lines with at least one Stokes line in the region. For example, one
can see that there are six Stokes sectors in Fig.~\ref{fig:n=00003D00003D00003D4Ex}
which illustrates the case of $n=4$. We also see in Fig.~\ref{fig:n=00003D00003D00003D4Ex}
that $k\neq0$ case sometimes has two Stokes lines in a single Stokes
sector. That is why the it more convenient to define the topology
as counting the Stokes sectors which is invariant under the $k$ deformations
away from zero, unlike the number of Stokes lines in the annulus.
Note that since the number of Stokes sectors depend on the basis choice
of Eq.~(\ref{eq:product}) (weakly since the $k$ deformations do
not change this quantity) which is left invariant under gauge transformations
by construction, this definition of number of Stokes sectors is also
gauge invariant.

Furthermore, since these sectors partition the annulus into regions
where each region has a common characteristic (of having an approximately
fixed asymptotic expansion), one can view each sector as a connected
region with respect to the asymptotic expansions, and thus a natural
notion of topology characterized by Stokes sectors exist.\footnote{Although related, this is distinct from the notion of change in the
topology of steepest descent paths discussed in \citep{Boyd1999TheDI}.} Moreover, we have already emphasized below Eq.~(\ref{eq:s-cosine})
that $n+2$ comes from the number of $U_{F}$ matrices in the single
valuedness condition. Hence, the count is not directly about the number
of branch points (or equivalently the zeroes of $\omega^{2}$) but
directly about the number of Stokes sectors. In fact, the branch cut
factor $X_{1}$ of Eq.~(\ref{eq:eff}) does not change the results
for $|\beta_{k}|$.

To intuitively understand the notion of connectedness provided by
the Stokes sector (and Stokes phenomena in general), consider what
happens to the smooth function $\chi_{k}(z)=F_{k}(z)\cdot V_{k}(z)$
as any one given Stokes line is crossed. There is a jump in the coefficient
component in $V_{k}(z)$ of the exponentially suppressed component
of the mode function in WKB basis $F_{k}(z)$. However, that component
does not contribute to the asymptotic expansion in any given Stokes
sector (since as an asymptotic expansion, $e^{-1/|\delta|}$ is exactly
zero as $\delta\rightarrow0$) but manifests itself later (as one
continues towards the next anti-Stokes boundary) as the boundary anti-Stokes
lines are crossed. In that sense, the asymptotic expansion has a connected
character in any given Stokes sector, and the number of Stokes sectors
correspond to a topological characterization of the asymptotic expansion.
Furthermore, each sector is insensitive to the continuous deformation
of the $A$ parameter. Finally, as one can see in the left Fig.~\ref{fig:n=00003D00003D00003D4Ex},
the number of Stokes sectors is insensitive to the changes in $k$.

One can also see from the last Fig.~\ref{fig:Deformation-of-the}
that there is a pinched singularity of the original integration of
Eq.~(\ref{eq:maineq}) along the real axis for $V_{k}(\eta)$. Unfortunately,
that only indicates some type of derivative singularity because of
the pinching comes from branch points. In reality, at least in the
$F^{2}$-gauge and $0$-gauge, $M$ is singular at the origin with
$k=0$ while it is not singular for any $k>0$. This means our computation
of $U_{F}$ effectively evaluated the Cauchy principle value of the
integral but indirectly through the analytic continuation and symmetries.
The branch point information (from which the Stokes lines emanate)
is still felt by the original contour integral from the pinched singularity.
That is why intuitively the $k/A^{1/(n+2)}\rightarrow0$ limit is
the topological limit (as the influence on the integral is maximal).

\section{Comparison of the F-matrix method with the exact solution}

\label{sec:Comparison-of-the}Since the differential equation of the
form 
\begin{equation}
\chi''(z)+z^{n}\chi(z)=0\label{eq:diffeq}
\end{equation}
is exactly solvable in terms of Bessel functions, we can compare the
F-matrix results with the exact solutions. What this means is that
the F-matrix methods are not necessary to compute the $k\rightarrow0$
limit of $\beta_{k}$ variables in this paper, but it does elucidate
the topological structure hidden in the Bessel function solutions.

For specificity and intuitive clarity, we focus on $n=4$. Exact solutions
to the mode Eq.~(\ref{eq:diffeq}) take the form 
\begin{equation}
\chi(z)=C_{1/6}z^{1/2}J_{1/6}\left(\frac{z^{3}}{3}\right)+C_{-1/6}z^{1/2}J_{-1/6}\left(\frac{z^{3}}{3}\right)\label{eq:ExacSolQuart}
\end{equation}
where $C_{\pm1/6}$ is determined by the boundary conditions. The
Bogoliubov coefficient $\beta$ may be computed using Eq.$\,$(\ref{eq:exactbogo})
and the exact solutions $\chi_{1}(z)$ and $\chi_{2}(z)$ satisfying
the asymptotic boundary conditions on the real axis 
\begin{equation}
\chi_{1}(z)\sim f_{-}(z)=\frac{1}{\sqrt{2z^{2}}}\exp\left(-i\frac{z^{3}}{3}\right)\quad\text{as\ensuremath{\quad}}z\rightarrow z_{-\infty}\label{eq:BCNegInfQuart}
\end{equation}
\begin{equation}
\chi_{2}(z)\sim f_{-}(z)=\frac{1}{\sqrt{2z^{2}}}\exp\left(-i\frac{z^{3}}{3}\right)\quad\text{as\ensuremath{\quad}}z\rightarrow z_{+\infty}\label{eq:BCPosInfQuart}
\end{equation}
respectively which matches Eqs.~(\ref{eq:Fdef-1}) up to a phase.
Here $z_{-\infty}<0$ and $z_{+\infty}>0$ correspond to times when
the $\vert\delta\left(z\right)\vert\ll1$. To obtain $\chi_{1}(z)$
and $\chi_{2}(z)$, we match the lowest order terms of the asymptotic
series expansion of Eq.$\,$(\ref{eq:ExacSolQuart}) to Eq.$\,$(\ref{eq:BCNegInfQuart})
and Eq.$\,$(\ref{eq:BCPosInfQuart}) respectively. Being dependent
on the $J_{\pm1/6}(\kappa)$ Bessel functions, the lowest order terms
of the asymptotic series of the exact solutions are determined by
\begin{equation}
\begin{split}\left(3\kappa\right)^{1/6}J_{\nu}(\kappa) & \sim b_{\nu,+}\left(\frac{24}{\pi^{3}\kappa^{2}}\right)^{1/6}\exp\left(i\left(\kappa-\frac{\nu\pi}{2}-\frac{\pi}{4}\right)\right)\\
 & +b_{\nu,-}\left(\frac{24}{\pi^{3}\kappa^{2}}\right)^{1/6}\exp\left(-i\left(\kappa-\frac{\nu\pi}{2}-\frac{\pi}{4}\right)\right)
\end{split}
\end{equation}
where $\kappa=\frac{z^{3}}{3}$ and $\nu=\pm\frac{1}{6}$. The coefficients
$b_{\nu,\pm}$ in different sectors of the complex $\kappa$ plane
are 
\begin{equation}
\begin{split} & b_{\nu,+}=\frac{1}{2}\exp\left(2p\left(\nu+1/2\right)\pi i\right)\\
 & b_{\nu,-}=\frac{1}{2}\exp\left(2p\left(\nu+1/2\right)\pi i\right)\quad\quad\text{for}\,\,\,\,(2p-1)\pi<\arg(\kappa)<(2p+1)\pi\label{eq:Coeff1-1}
\end{split}
\end{equation}

\begin{equation}
\begin{split} & b_{\nu,+}=\frac{1}{2}\exp\left(2(p+1)\left(\nu+1/2\right)\pi i\right)\\
 & b_{\nu,-}=\frac{1}{2}\exp\left(2p\left(\nu+1/2\right)\pi i\right)\quad\quad\qquad\text{for}\,\,\,\,2p\pi<\arg(\kappa)<(2p+2)\pi\label{eq:Coeff2-1}
\end{split}
\end{equation}
where $p\in\mathbb{Z}.$

For $\chi_{1}(z)$, the relevant approximation corresponds to the
sector to which $z_{-\infty}$ belongs. The fact that $\arg(z_{-\infty})=\pi$
implies $\arg(\kappa_{-\infty})=3\pi\in(2\pi,4\pi)$. Thus, the correct
asymptotic series approximation is obtained from Eq.$\,$(\ref{eq:Coeff1-1})
with $p=1.$ Matching with the boundary condition gives 
\begin{equation}
\chi_{1}(z)=C_{1,1/6}z^{1/2}J_{1/6}\left(\frac{z^{3}}{3}\right)+C_{1,-1/6}z^{1/2}J_{-1/6}\left(\frac{z^{3}}{3}\right)
\end{equation}
where 
\begin{equation}
C_{1,1/6}=D_{1}\exp\left(-7\pi i\left(\frac{1}{12}+\frac{1}{4}\right)\right),\quad\text{\ensuremath{\quad C_{1,-1/6}=}}-D_{1}\exp\left(-7\pi i\left(-\frac{1}{12}+\frac{1}{4}\right)\right)
\end{equation}
and 
\begin{equation}
D_{1}=\left(\frac{\exp\left(-\pi i\left(\frac{1}{6}+\frac{1}{2}\right)\right)-\exp\left(-\pi i\left(-\frac{1}{6}+\frac{1}{2}\right)\right)}{2}\right)^{-1}\left(\frac{12}{\pi}\right)^{-1/2}.
\end{equation}
Similarly for $\chi_{2}(z)$, the sector to which $z_{+\infty}$ belongs
is $\arg(\kappa_{+\infty})\in(-\pi,\pi)$ because $\arg(z_{+\infty})=0=\arg(\kappa_{+\infty})$.
Matching the asymptotic series obtained from Eq.$\,$(\ref{eq:Coeff2-1})
with $p=0$ to Eq.$\,$(\ref{eq:BCPosInfQuart}) gives 
\begin{equation}
\chi_{2}(z)=C_{2,1/6}z^{1/2}J_{1/6}\left(\frac{z^{3}}{3}\right)+C_{2,-1/6}z^{1/2}J_{-1/6}\left(\frac{z^{3}}{3}\right)
\end{equation}
where 
\begin{equation}
C_{2,1/6}=D_{2}\exp\left(i\pi\left(\frac{1}{12}+\frac{1}{4}\right)\right),\quad\text{\ensuremath{\quad C_{2,-1/6}=}}-D_{2}\exp\left(-i\pi\left(\frac{1}{12}-\frac{1}{4}\right)\right)
\end{equation}
and 
\begin{equation}
D_{2}=\left(\frac{12}{\pi}\right)^{-1/2}\left(\frac{\exp\left(i\pi\left(\frac{1}{6}+\frac{1}{2}\right)\right)-\exp\left(-i\pi\left(\frac{1}{6}-\frac{1}{2}\right)\right)}{2}\right)^{-1}.
\end{equation}

Therefore, Eq.$\,$(\ref{eq:exactbogo}) with the use of the Wronskian
identity 
\begin{equation}
W\equiv J_{1/6}\left(Az^{\xi}\right)\overleftrightarrow{\partial}_{z}J_{-1/6}\left(Az^{\xi}\right)=\frac{-\xi}{\pi z}
\end{equation}
where $A$ is a constant gives the $\beta$ coefficient as 
\begin{equation}
\beta_{\bar{k}=0}=\frac{3i\left(-C_{2,1/6}C_{1,-1/6}+C_{2,-1/6}C_{1,1/6}\right)}{\pi}=i\sqrt{3}.
\end{equation}
Because of the physical branch cut resolution discussed in Subsec.~\ref{subsec:Stokes-constants-from},
we need to map this to $\beta_{\bar{k}\rightarrow0}$ using Eq.~(\ref{eq:eff}).
The result is 
\begin{equation}
\beta_{\bar{k}\rightarrow0}=i^{n/2}\beta_{\bar{k}=0}
\end{equation}
with $n=4$. Here we obtain the topological result of Eq.$\,$(\ref{eq:betafinresult})
in the $\bar{k}\rightarrow0$ limit in terms of the Wronskian identities
satisfied by the Bessel functions. In other words, in the $\bar{k}\rightarrow0$
limit, the Wronskian of the Bessel function solutions to the mode
equation count the number of Stokes sectors defined below Eq.~(\ref{eq:betafinresult}).
One way to understand this link between the Bessel solution and the
topology is that the Wronskian of the the Bessel function satisfies
a differential equation 
\begin{equation}
\frac{dW}{dz}=\frac{-1}{z}W\label{eq:besselwronsk}
\end{equation}
which is invariant under scaling $z$. In other words, conformal invariance
naturally erases geometrical information, which happens to leave the
topological information left in Eq.~(\ref{eq:betafinresult}).

\section{Illustrative model}

\label{sec:Illustrative-model} In this section we present couple
of physical models for which the topological limit of the particle
production computation is manifestly relevant. These models are motivated
from scenarios where dark matter production results from a time dependent
effective mass (\cite{Ema:2018ucl,Kaneta:2022gug,Kolb_2021,gross2021gravitationalvectordarkmatter,Ahmed_2020,Ling:2021zlj,Garcia_2023,Yu_2023})
or weak couplings to fields with similar non-adiabatic dispersion
relations. Examples of the latter include dark matter production from
the thermal bath produced from inflaton decay during reheating (\cite{Garcia_2021,arcadi2024zprimemediateddarkmatterfreezein,Becker_2024,Kaneta_2019,Ellis:2015jpg,Bernal:2019mhf,Chen:2017kvz,Zhang:2023xcd})
and bubble dynamics in first order phase transitions (\cite{azatov2021darkmatterproductionrelativistic,giudice2024nonthermalheavydarkmatter,baldes2023bubbletrons,Chung:1998ua,Baker_2020}).

The first model considered here corresponds to the mass of the dark
matter $\chi$ conformally coupled to gravity being controlled by
a dimension 6 coupling to a spectator field $\phi$ whose time evolution
causes the frequency squared $\omega^{2}$ of $\chi$ to go through
a zero (approximately) analytically. This first model is easily embeddable
in the context of inflationary cosmology during the reheating phase.
Our second model is presented as a purer mathematical match of the
cosmology and the topological production scenario. It involves three
scalar fields and a background FLRW spacetime with constant spatial
curvature. The second model will have a large fraction of the particle
production coming from the topological contribution unlike the first
model.

\subsection{\label{subsec:Tanh-model}Tanh model}

In this subsection, we will consider a scenario in which the dark
matter field $\chi$ obtains its mass through a dimension 6 coupling
to a spectator scalar $\phi$ which is rolling down a $\tanh$ potential
in a post-quasi-dS phase of inflation. When the mass of the $\chi$
field goes through a zero, there will be an approximately topological
contribution to the $\chi$ particle production because of the nonadiabaticity
of the dispersion relations. The boundary conditions and parameters
are chosen to separate other sources of nonadiabaticities such that
Eq.~(\ref{eq:betafinresult}) approximately applies. Given that the
vacua change is responsible for the particle production here, and
given that the production amplitude has a topological character, there
is a semblance to the usual anomalous current equation 
\begin{equation}
\partial_{\mu}j_{A}^{\mu}=\frac{-g^{2}}{8\pi^{2}}\mathrm{Tr}F\tilde{F}
\end{equation}
where $j_{A}^{\mu}$ is the current anomalous with respect to the
gauge group whose field strength is $F$.

Consider a spectator scalar field $\phi$ governed by the following
potential 
\begin{equation}
V(\phi)=\rho_{0}\left[1-\tanh\left(\phi/M\right)\right]
\end{equation}
and the $\chi$ field coupling 
\begin{equation}
\mathcal{L}\supset\frac{g}{2\Lambda^{2}}\phi^{4}\chi^{2}
\end{equation}
where $\chi$ particles will be produced through the classical field
motion of $\phi$. The classical equation of motion for $\phi$ is
\begin{equation}
\ddot{\phi}+3H\dot{\phi}-\frac{\rho_{0}}{M}\mathrm{sech}^{2}\left(\frac{\phi}{M}\right)=0\label{eq:orig}
\end{equation}
which depends on 3 scales $H,\,M,$ and $\rho_{0}/M$. We can use
the freedom to scale time and the field to define a single dynamical
parameter 
\begin{equation}
\frac{\rho_{0}}{M}=10^{-6}MH_{I}^{2}\label{eq:rhobar}
\end{equation}
where we will explain later that $10^{-6}$ ultimately comes from
ensuring that there is at least an order of magnitude separation between
when $k/(a_{e}H_{I})\sim O(1)$ becomes nonadiabatic and $k/(a_{e}H_{I})\ll O(1)$
becomes nonadiabatic owing to the fact that $\phi\propto\eta^{6}$
during the $\tanh\left(\phi/M\right)\sim\phi/M$ phase. For example,
if we had chosen $10^{-3}$ here, the corresponding number $(10^{-3})^{1/6}\sim O(1)$
will not lead to a hierarchy. Note that Eq.~(\ref{eq:rhobar}) automatically
ensures that $\phi$ is a spectator field since 
\begin{equation}
\frac{\rho_{0}}{3M_{P}^{2}H_{I}^{2}}=\frac{10^{-6}M^{2}}{3M_{P}^{2}}\ll1
\end{equation}
(where $M<M_{P}$ owing to the EFT validity condition).

The field value at the beginning and the end of inflation are 
\begin{equation}
\phi_{p}<-0.50016M\label{eq:initialvalue}
\end{equation}
\begin{equation}
\phi_{e}=-0.500M\label{eq:phi_end_of_infl}
\end{equation}
respectively. The first condition ensures at least 60 e-folds of inflation.
The expansion rate is parameterized as 
\begin{equation}
\begin{split}H(t) & =\begin{cases}
H_{I} & t<t_{e}\\
\frac{H_{I}}{1+\frac{3H_{I}}{2}(t-t_{e})} & t\geq t_{e}
\end{cases}\end{split}
\end{equation}
where $t_{e}$ is the time at the end of inflation.

During inflation, the field trajectory of $\Delta\phi$ obeys slow
roll 
\begin{equation}
3H_{I}\dot{\phi}-\frac{\rho_{0}}{M}\mathrm{sech}{}^{2}\left(\phi/M\right)=0
\end{equation}
whose solution is 
\begin{equation}
\frac{\phi}{2}+\frac{M\sinh\left(2\phi/M\right)}{4}-\left(\frac{\phi_{p}}{2}+\frac{M\sinh\left(2\phi_{p}/M\right)}{4}\right)=\frac{\rho_{0}}{3MH_{I}}(t-t_{p}).\label{eq:slow-roll-sol}
\end{equation}
If inflation ends in the linear section of the potential, as in Eq.$\,$(\ref{eq:phi_end_of_infl})
then the field solution for $t>t_{e}$ to Eq.~(\ref{eq:orig}) is
\begin{equation}
\frac{\phi}{M}=\frac{1}{9}\left(\frac{2\rho_{0}}{3M^{2}H_{I}^{2}}\left[1+\frac{3}{2}H_{I}(t-t_{e})\right]^{2}-\frac{3c_{1}/2}{1+\frac{3}{2}H_{I}(t-t_{e})}\right)+c_{2}\label{eq:phisol}
\end{equation}
where 
\begin{equation}
c_{1}\equiv\frac{8}{9}\frac{\rho_{0}}{M^{2}H_{I}^{2}}\left(\frac{2-\cosh\left(2\phi_{e}/M\right)}{1+\cosh\left(2\phi_{e}/M\right)}\right)
\end{equation}
\begin{equation}
c_{2}\equiv\frac{\phi_{e}}{M}-\frac{2}{9}\frac{\rho_{0}}{M^{2}H_{I}^{2}}\left(\frac{\cosh\left(2\phi_{e}/M\right)-1}{1+\cosh\left(2\phi_{e}/M\right)}\right).
\end{equation}
Note that $c_{2}$ should be interpreted as the field displacement
needed to reach the nonadiabatic point where $\phi=0$.

The conformal time is related to the comoving observer's proper time
as 
\begin{equation}
1+\frac{3}{2}\left(H_{I}t-H_{I}t_{e}\right)=\left(\frac{H_{I}a_{e}\eta}{2}+1\right)^{3}
\end{equation}
where $\eta=0$ corresponds to the end of inflation. The effective
mode frequency of Eq.~(\ref{eq:modeeq}) in conformal time is 
\begin{equation}
\omega^{2}(\eta)=k^{2}+\frac{g}{\Lambda^{2}}a^{2}(\eta)\phi^{4}(\eta).
\end{equation}
For long wavelengths characterized by 
\begin{equation}
k\ll\frac{g}{\Lambda^{2}}a^{2}(\eta)\phi^{4}(\eta),
\end{equation}
for generic values of $\eta$, nonadiabaticity occurs near $\phi=0$
in this model corresponding to the time $\eta=\eta_{0}$. This nonadiabaticity
is taken to be well separated from that at the end of inflation by
choosing the parameter 
\begin{equation}
\bar{\rho}\equiv\frac{\rho_{0}}{M^{2}H_{I}^{2}}\ll1,\label{eq:rhobardef}
\end{equation}
and maximizing $|\phi_{e}|$ while still lying within the linear approximation
range of the potential $V(\phi).$

\begin{figure}
\begin{centering}
\includegraphics[scale=0.35]{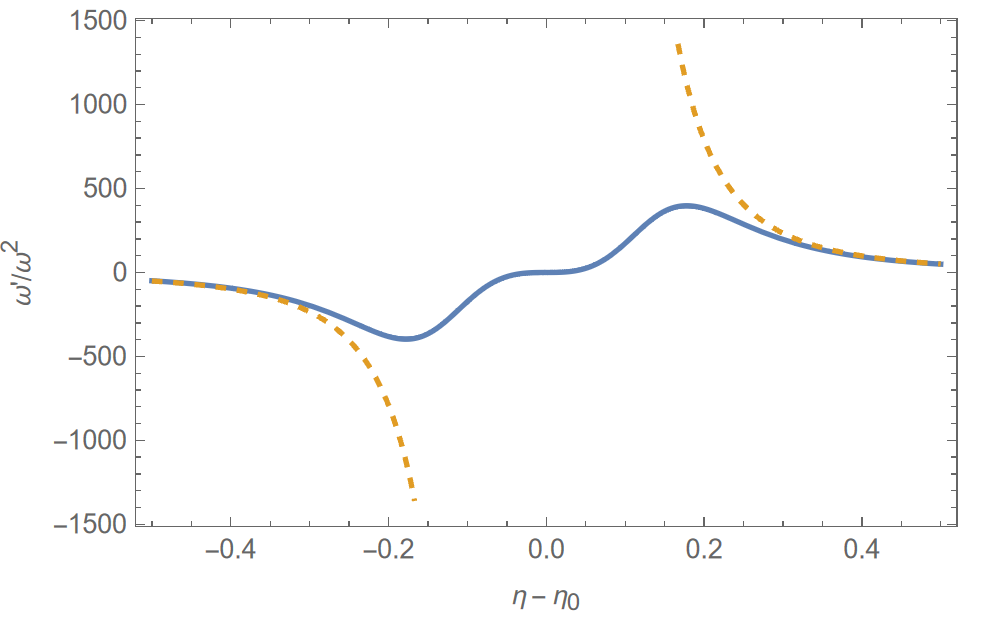}
\par\end{centering}
\caption{\label{fig:Nonadiabaticity-estimates-}Nonadiabaticity estimates $\omega'/\omega^{2}$
for the toy function $\omega^{2}=k^{2}+10^{-1}(\eta-\eta_{0})^{4}$
for $k=10^{-2}$ (solid blue) and $k=0$ (dashed orange) illustrate
that although both share the double peaked structure, the $k=0$ case
has an infinitely larger peak nonadiabaticity. Nonetheless, the particle
production for the two situations will be similar, tending to the
topological limit.}
\end{figure}

The nonadiabaticity for this frequency can be defined through Eq.~(\ref{eq:nonad}).
Parameterizing 
\begin{equation}
\frac{g}{\Lambda^{2}}a^{2}(\eta)\phi^{4}(\eta)=A(\eta-\eta_{0})^{4}\label{eq:QuarticApprox}
\end{equation}
near $\eta=\eta_{0}$ where $A$ now contains all the scales we can
write 
\begin{equation}
\frac{\omega'}{\omega^{2}}=\frac{1}{\omega}\frac{1}{\frac{k^{2}}{Aa^{2}(\eta)(\eta-\eta_{0})^{n-1}}+(\eta-\eta_{0})}\left(\frac{n}{2}+\frac{a'}{a}(\eta-\eta_{0})\right)\label{eq:nonadexample}
\end{equation}
with $n=4$ and the corresponding time region for large nonadiabaticity
might be defined to have the width $\Delta\eta$ satisfying 
\begin{equation}
\left[\frac{\omega'}{\omega^{2}}\right]_{\eta_{\max}+\Delta\eta}=0.1\times\left[\frac{\omega'}{\omega^{2}}\right]_{\eta_{\max}}\label{eq:wdith-naive}
\end{equation}
In the $k\rightarrow0$, this expression would formally give $\Delta\eta\rightarrow0$.
One of the main points of this paper is the topological nature of
the Bogoliubov coefficient in this parametric limit. Such situations
generically cannot be characterized by a scale $a(\eta_{0})\Delta\eta$
unlike for $k\gtrsim O(aH)$ for which this width does capture the
qualitative aspects of the nonadiabatic physics.\footnote{The cases with modes with $k\gtrsim aH$ will not be well approximated
by the topological production computation. This constraint will play
role in our discussion in Sec.~\ref{subsec:Curvature-model}.} Fig.~\ref{fig:Nonadiabaticity-estimates-} explicitly illustrates
the qualitative behavior of the frequency time dependence we are trying
to model with the current physical scenario. Note that $|\omega'/\omega^{2}|$
has a double peak structure surrounding the zero-crossing time $\eta_{0}$.

For the $k\rightarrow0$ case, let us define a time region during
which the system be considered nonadiabatic differently. One criteria
that can be chosen is to define $\eta_{c}$ where 
\begin{equation}
\left(\frac{\omega'}{\omega^{2}}\right)_{\eta_{c}}=\pm1\label{eq:widthk0}
\end{equation}
which has a solution 
\begin{equation}
a(\eta_{0})\eta_{c}=a(\eta_{0})\eta_{0}+\left[\frac{2\Lambda}{\sqrt{g}\left[\frac{1}{a(\eta_{0})}\partial_{\eta}\phi(\eta_{0})\right]^{2}}\right]^{\frac{1}{3}}.\label{eq:time-scale}
\end{equation}
Hence, one of the remarkable simplification that occurs in the Bogoliubov
coefficient computations at the complex frequency threshold is that
the time scale associated with the nonadiabaticity in Eq.~(\ref{eq:time-scale})
disappears. It is this topological character that the F-matrix formalism
allows us to make precise. The topological index will be associated
with the counting the number of Stokes regions in this limit.

\subsubsection{Bogoliubov coefficient from $U_{F}$ propagators in $F^{2}$-gauge}

The non-adiabaticity $\delta=\omega'/(4\omega^{2})$ of the dispersion
relation 
\begin{equation}
\omega^{2}(\eta)=k^{2}+\frac{g}{\Lambda^{2}}a^{2}(\eta)\phi^{4}(\eta)
\end{equation}
peaks in the neighborhood of the critical point $\eta_{0}$ in the
long wavelength limit. The width of this non-adiabatic region, defined
as the interval outside which $|\delta(\eta)|\ll|\delta|_{\mathrm{max}}$,
can be seen to be directly proportional to $k^{\nu}$ where $\nu$
is a positive power. Therefore, for small values of $k$, we may approximate
the dispersion relation as 
\begin{equation}
\omega^{2}(\eta)\approx k^{2}+\frac{g}{\Lambda^{2}}a^{2}(\eta_{0})\left(\phi'(\eta_{0})\right)^{4}(\eta-\eta_{0})^{4}\label{eq:OmegaQuartAppr}
\end{equation}
if the non-adiabatic width lies within the width of the linear approximation.
More explicitly, the time region $\Delta\eta_{l}$ of the linear approximation
satisfies 
\begin{equation}
\frac{1}{4}\times\frac{2f'(\eta_{0})}{f''(\eta_{0})}\approx\frac{1}{7}\left(\frac{27|c_{2}|}{2\bar{\rho}}\right)^{1/6}\gg a_{e}H_{I}\Delta\eta_{l}\label{eq:delz_bound}
\end{equation}
where 
\begin{equation}
f(\eta)\equiv a^{1/2}(\eta)\frac{\phi(\eta)}{M}
\end{equation}
and the factor of $1/4$ is related to the power of $f(\eta)$ in
the dispersion relation. In deriving Eq.$\,$(\ref{eq:delz_bound}),
we have used Eq.$\,$(\ref{eq:phisol}) in conformal time $\eta$
and 
\begin{equation}
\frac{2\rho_{0}}{3M^{2}H_{I}^{2}}\left[\frac{H_{I}a_{e}\eta}{2}+1\right]^{6}\gg\frac{3c_{1}/2}{\left(\frac{H_{I}a_{e}\eta}{2}+1\right)^{3}}\,.
\end{equation}
The above is justified for cases where the $\phi$ energy is sufficiently
suppressed (see Eq. (\ref{eq:rhobardef})), $\phi_{e}\approx-0.5M$,
$c_{1}\sim\bar{\rho}$, and $H_{I}a_{e}\eta/2\gg1$. In terms of $\phi(\eta)-\phi(\eta_{0})$
around $\eta_{0},$ Eq.$\,$(\ref{eq:delz_bound}) implies 
\begin{equation}
\frac{\phi(\eta)-\phi(\eta_{0})}{M}\ll\frac{3}{7}|c_{2}|\approx\frac{3\phi_{e}}{7M}.
\end{equation}
Within the approximation Eq.$\,$(\ref{eq:OmegaQuartAppr}), the width
of the non-adiabatic region for a particular value of $k$ may be
estimated as 
\begin{equation}
\Delta\eta_{w}\sim3\left(\frac{k^{2}\Lambda^{2}}{ga^{2}(\eta_{0})\left(\phi'(\eta_{0})\right)^{4}}\right)^{1/4}\label{eq:quarticwidth}
\end{equation}
by considering when $\omega^{2}$ that controls the denominator of
$\delta(\eta)$ is dominated by $\eta-\eta_{0}$.\footnote{The factor of 3 here is an ansatz that works well near the fiducial
parametric point. In other words, with the fiducial parametric choices,
the non-adiabaticity function has become $\delta\lesssim O(0.1)$
at a time $\eta=\eta_{0}+\Delta\eta_{w}$.}

\begin{figure}
\begin{centering}
\includegraphics[scale=0.4]{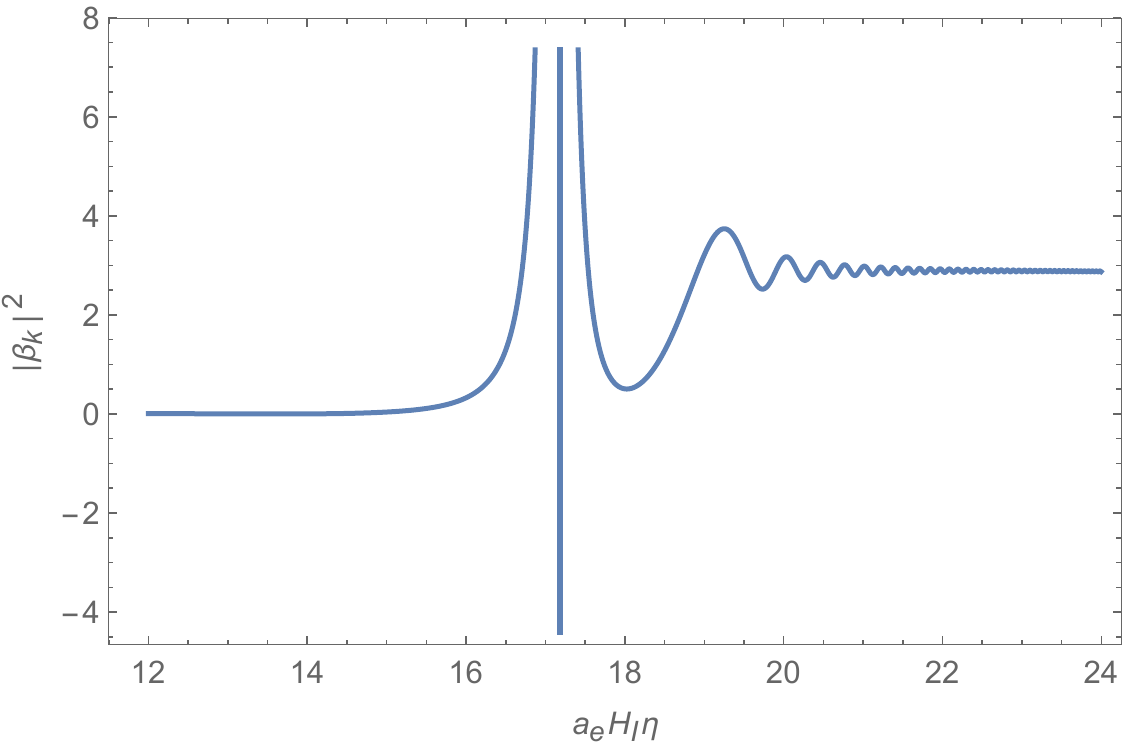}
\par\end{centering}
\caption{\label{fig:Numerical-computation-of}Numerical computation of $|\beta_{k}|^{2}$
for $k=10^{-6}a_{e}H_{I}$ in the Tanh-model.}
\end{figure}

Requiring $\Delta\eta_{w}\ll\Delta\eta_{l}$, we find 
\begin{equation}
\frac{k}{a_{e}H_{I}}\lesssim4\times10^{-4}\left(\frac{\frac{gM^{4}}{\Lambda^{2}H_{I}^{2}}}{10^{-2}}\right)^{1/2}\left(\frac{\left|\phi_{e}\right|/M}{0.5}\right)^{7/3}\left(\frac{\mathcal{A}}{10^{-1}}\right)^{2}\label{eq:linearversuswidth}
\end{equation}
where $\mathcal{A}$ is the desired accuracy and $\phi_{e}$ measures
the field distance from the end of inflation to zero of $\omega^{2}$.
The fiducial value of $10^{-2}$ for $gM^{4}/(\Lambda^{2}H_{I}^{2})$
can be interpreted as the square of time ratio measuring the isolation
of the nonadiabatic time region compared to the linear $\phi$ approximation
time region. This is a model-dependent limitation to the topological
contribution. Within this range of $k$ values, Eq.$\,$(\ref{eq:betafinresult})
estimates the Bogoliubov coefficient as $|\beta_{k}|^{2}\approx3$.
Numerical computation of $|\beta_{k}|^{2}$ for $k=10^{-6}a_{e}H_{I}$,
shown in Fig.~\ref{fig:Numerical-computation-of} shows agreement
with this estimate. Although the modes satisfying Eq.~(\ref{eq:linearversuswidth})
are superhorizon at the end of inflation, they are subhorizon at a
time 8 efolds before the end of inflation. This establishes that these
topological contributions can be physical.

There is a constraint on the $M$ parameter from the resolution of
the classical field $\phi$ being $\Delta\phi\sim H/(2\pi)$. Since
Eqs.~(\ref{eq:initialvalue}) and (\ref{eq:phi_end_of_infl}) require
a resolution of 
\begin{equation}
\frac{\Delta\phi}{M}\sim20\frac{\rho_{0}}{M^{2}H_{I}^{2}}\left(\frac{N}{60}\right)
\end{equation}
we require 
\begin{equation}
\frac{H_{I}}{M}\ll O(100)\frac{\rho_{0}}{M^{2}H_{I}^{2}}\left(\frac{N}{60}\right)\sim10^{-4}\left(\frac{N}{60}\right).\label{eq:quant_resolution}
\end{equation}
Although this condition can be violated without drastically upsetting
the phenomenology, we impose this here for illustrative convenience.

Next, let's compute the total cosmological dark matter density produced
from the tanh model. We will see that the topological production contribution
makes up a negligible part of the total production density. Before
we discuss the detailed computation, let's see what the main non-triviality
of the analysis will be. We generically expect that the number density
contribution to the dark matter density will take the form of an integral
that is cutoff at $\Lambda_{2}$: 
\begin{equation}
\int_{0}^{\Lambda_{2}}\frac{dk\,k^{2}}{(2\pi)^{3}}\mid\beta_{k}\mid^{2}\sim f\Lambda_{2}^{3}\label{eq:kintegral}
\end{equation}
where $f$ represents the strength of non-adiabaticity. For the non-topological
contribution, the cutoff $\Lambda_{2}$ is expected to be distinguishable
from $k=0$, unlike the $k$ values for which 
\begin{equation}
\omega^{2}=k^{2}+m^{2}\sim m^{2}
\end{equation}
(i.e. $k$ values for which the topological approximation will be
valid). We will compute $\Lambda_{2}$ numerically and find that it
is much larger than the bound given by Eq.~(\ref{eq:linearversuswidth}).

The cosmological dark matter energy density $\rho_{\chi}$ is obtained
through the $k$ integral 
\begin{equation}
\rho_{\chi}=\frac{1}{2\pi^{2}}\frac{1}{a^{3}}\int_{0}^{\infty}dkk^{2}\frac{\omega}{a}\mid\beta_{k}\mid^{2}\label{eq:rhochi}
\end{equation}
where the effective cutoff in Eq.~(\ref{eq:rhochi}) occurs at $k/a_{e}\sim H_{I}$.
The observable relic abundance depends on the details of the $\phi$
evolution and cosmology after the particle production. As worked out
in Appendix \ref{sec:1-loop-corrections-and}, the final relic abundance
is 
\begin{equation}
\Omega_{\chi}h^{2}=0.27\left(\frac{T_{rh}}{10^{7}\mathrm{GeV}}\right)\left(\frac{H_{I}}{10^{2}\mathrm{GeV}}\right)\left(\frac{M}{10^{9}\mathrm{GeV}}\right)\left(1+0.21\log\left(\frac{\rho_{0}/(M^{2}H_{I}^{2})}{10^{-6}}\right)\right)\label{eq:omega}
\end{equation}
and the mass of this dark matter is 
\begin{equation}
m=1.6\times10^{11}\left(\frac{M}{10^{9}\mathrm{GeV}}\right)\,\mathrm{GeV}
\end{equation}
where the mass $M$ can be easily increased from this fiducial value
without upsetting the assumptions of the computation. Note that the
chosen fiducial parametric values of $H_{I}=10^{2}$ GeV and $M=10^{9}$
GeV, the cutoff scale is of the order 
\begin{equation}
\frac{\Lambda}{\sqrt{g}}\sim10^{17}\mathrm{GeV}
\end{equation}
as explained more in Appendix \ref{sec:1-loop-corrections-and}.

To understand this result qualitatively (and from comparison with
numerical exploration), the dominant contribution to $\rho_{\chi}$
in Eq.~(\ref{eq:rhochi}) comes from the upper part of the integration
with 
\begin{equation}
f\sim O(1)
\end{equation}
in the notation of Eq.~(\ref{eq:kintegral}) and $\Lambda_{2}$ that
is determined by approximately the exponential cutoff controlled by
\begin{equation}
\exp\left(-C\int^{\Delta\tilde{\eta}}d\eta k\right)\sim\exp\left(-Ck\Delta\tilde{\eta}\right)
\end{equation}
where $C$ is presumably an order unity coefficient and $\Delta\tilde{\eta}$
corresponds to the $k$-dependent time period when the given $k$
mode is most nonadiabatic. For the fiducial parameters shown in Eq.~(\ref{eq:omega}),
we expect 
\begin{align*}
\Delta\tilde{\eta} & \sim\left(\frac{k}{a_{e}}\right)^{1/2}\left(\frac{gM^{4}}{\Lambda^{2}}\right)^{-1/4}\frac{1}{c_{2}a_{e}}\left(\frac{1}{H_{I}}\right)
\end{align*}
Hence, the non-topological contribution to $|\beta_{k}|^{2}$ integral
is expected to be 
\begin{equation}
\int\frac{d^{3}k}{(2\pi)^{3}}|\beta_{k}|^{2}\sim O\left(\frac{c_{2}^{2}a_{e}^{3}H_{I}^{2}}{2\pi^{2}}\left(\frac{gM^{4}}{\Lambda^{2}}\right)^{1/2}\right).
\end{equation}

\begin{figure}
\begin{centering}
\includegraphics[scale=0.17]{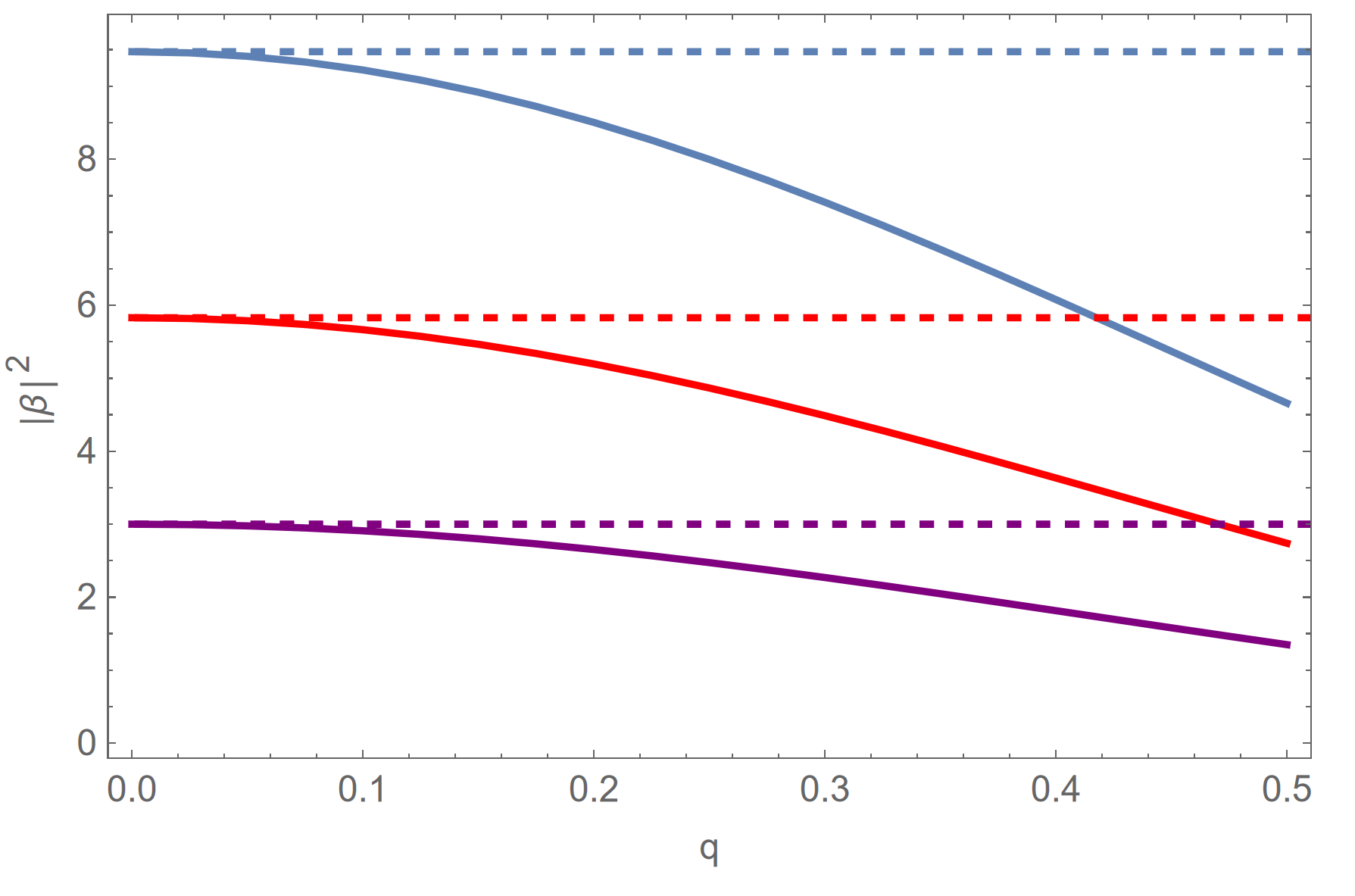}
\par\end{centering}
\caption{\label{fig:neq4_numer}The numerical value of the $\left|\beta\right|^{2}$
as a function of $q\equiv k/A^{1/(n+2)}$ in $\omega^{2}=k^{2}+A\eta^{n}$
is plotted for $n=4\,(\mathrm{purple),6\,(\mathrm{red),8\,(\mathrm{blue)}}}$
in solid. The topological formula given by Eq.~(\ref{eq:betafinresult})
is plotted in horizontal dashed lines. An $O(10\%)$ deviation occurs
when $k/A^{1/(n+2)}$ changes by $O(0.1)$ from zero. Note that these
are non-perturbatively large Bogoliubov coefficients.}
\end{figure}

Let's now compare this explicitly with the topological contribution
which we parameterize as 
\begin{equation}
\int\frac{d^{3}k}{(2\pi)^{3}}|\beta_{k}|^{2}\sim O(1)f\Lambda_{3}^{3}
\end{equation}
where $\Lambda_{3}$ corresponds to the cutoff of the integral associated
with the topological contribution. There are two length scales which
determine this cut-off: i) the length scale $\Lambda_{3}^{\text{lin}}$
within which the linear approximation is good, determined by Eq.$\,$(\ref{eq:linearversuswidth})
and ii) the length scale $\Lambda_{3}^{\text{corr}}$ within which
the small $\bar{k}$ corrections to $\beta_{\bar{k}}$ is negligible,
determined by Fig.~\ref{fig:neq4_numer}.. These length scales are
given by 
\begin{equation}
\frac{\Lambda_{3}^{\text{lin}}}{a_{e}H_{I}}\lesssim4\times10^{-4}\left(\frac{\frac{gM^{4}}{\Lambda^{2}H_{I}^{2}}}{10^{-2}}\right)^{1/2}\left(\frac{\left|\phi_{e}\right|/M}{0.5}\right)^{7/3}\left(\frac{\mathcal{A}}{10^{-1}}\right)^{2}
\end{equation}
and given that the effective $A$ parameter in $\omega^{2}\approx k^{2}+A\eta^{4}$
is $ga^{2}(\eta_{0})\left(\phi'(\eta_{0})\right)^{4}/\Lambda^{2}$,
we can evaluate 
\begin{equation}
\frac{\Lambda_{3}^{\text{corr}}}{a_{e}H_{I}}\lesssim0.6\times O\left(10^{-1}\right)\left(\frac{\mathcal{A}}{10^{-1}}\right)\left(\frac{\frac{gM^{4}}{H_{I}^{2}\Lambda^{2}}}{10^{-2}}\right)^{1/6}
\end{equation}
using similar steps as Eq.~(\ref{eq:linearversuswidth}). From the
above we see that $\Lambda_{3}^{\text{corr}}\gg\Lambda_{3}^{\text{lin}}$,
and therefore, the topological scale determined by the minimum of
the above two is set by $\Lambda_{3}^{\text{lin}}$ 
\begin{equation}
\Lambda_{3}=\text{min}\left(\Lambda_{3}^{\text{corr}},\Lambda_{3}^{\text{lin}}\right)=\Lambda_{3}^{\text{lin}}\sim4\times10^{-5}\left(\frac{\frac{gM^{4}}{\Lambda^{2}H_{I}^{2}}}{10^{-2}}\right)^{1/2}\left(\frac{\left|\phi_{e}\right|/M}{0.5}\right)^{7/3}\left(\frac{\mathcal{A}}{10^{-1}}\right)^{2}
\end{equation}
Hence, the topological contribution for this model is estimated to
be less than $O(10^{-10})$ fraction of the total particles produced.
On the other hand, it is interesting that $\Lambda_{3}^{\text{corr}}$
is larger which indicates that in a different scenario in which the
linear approximation can be extended, the topological contribution
can be more significant. We present an extreme version of this in
the next model.

\subsection{\label{subsec:Curvature-model}Curvature model}

The tanh model discussed above has the topological production as an
approximation for $k$ satisfying Eq.~(\ref{eq:linearversuswidth})
coming partially from the equation governing the time range for which
$\tanh\left(\phi/M\right)\sim\phi/M$ during which $\phi-\phi_{e}\sim(t-t_{e})^{2}$
with $\phi_{e}<0$. One can instead also set up a nonlinear potential
and cosmology for which $\phi=c_{1}(t-t_{0})^{2}$ nearly exactly.
Furthermore, in this scenario the exact topological limit will nearly
correspond to an arbitrarily large $k$ that is matched to the spatial
curvature scale of cosmology.

Consider the Lagrangian of 3 real scalar fields $\phi,\chi,$ and
$\psi$ minimally coupled to gravity: 
\begin{equation}
\mathcal{L}=\frac{1}{2}\left(\partial\phi\right)^{2}-V(\phi)+\frac{1}{2}\left(\partial\chi\right)^{2}-\frac{1}{2}f(\phi)\chi^{2}+\frac{1}{2}\left(\partial\psi\right)^{2}-2h^{2}M_{P}^{2}e^{-\sqrt{2}\psi/M_{P}}
\end{equation}
\begin{equation}
V(\phi)=-8c_{1}\phi
\end{equation}
\begin{equation}
f(\phi)=\frac{c_{1}\left(E/a_{0}^{2+n_{2}}\right)}{\phi\left(K/a_{0}\right)^{2+n_{2}}}\left(\ln\left[\frac{K\sqrt{\frac{\phi}{c_{1}}}}{a_{0}}\right]\right)^{n_{2}}\label{eq:potential}
\end{equation}
where both $\chi$ and $\phi$ are spectators in a universe driven
by $\psi$. The free parameters in this model are $\{c_{1},K/a_{0},h,E/a_{0}^{2+n_{2}},n_{2}\}$
where when $\phi=\phi_{0}\equiv c_{1}\left(a_{0}/K\right)^{2}$ the
mode frequency 
\begin{equation}
\omega^{2}=k^{2}+a^{2}(\eta)f(\phi)-a''(\eta)/a(\eta)\label{eq:curvaturedispersion}
\end{equation}
vanishes for the one particular $k$-mode of 
\begin{equation}
k=K.
\end{equation}
This is the main attractive feature of this model since one can produce
physical non-vanishing $k$-mode particles with arbitrarily large
momentum matched to the parameter $K$ which we will see below is
the spatial curvature of this cosmology. Note $c_{1}$ in Eq.~(\ref{eq:potential})
has units of mass cubed and 
\begin{equation}
[E]=[k]^{2+n_{2}}
\end{equation}
such that the units of $f(\phi)$ are determined by $c_{1}/\phi$.
We will later see that $h$ is a parameter that determines the origin
of $\psi$.

The background equations are 
\begin{equation}
\ddot{\phi}+3H\dot{\phi}-8c_{1}+\frac{1}{2}f'(\phi)\chi^{2}=0
\end{equation}
\begin{equation}
\ddot{\chi}+3H\dot{\chi}+gf(\phi)\chi=0
\end{equation}
\begin{equation}
\ddot{\psi}+3H\dot{\psi}-2\sqrt{2}h^{2}M_{P}e^{-\sqrt{2}\psi/M_{P}}=0
\end{equation}
\begin{equation}
3M_{P}^{2}\left(\frac{\dot{a}(t)}{a(t)}\right)^{2}\approx\frac{1}{2}\dot{\psi}^{2}+2h^{2}M_{P}^{2}e^{-\sqrt{2}\psi/M_{P}}\label{eq:spectatorapprox}
\end{equation}
to which there exists an explicit solution 
\begin{equation}
\chi=0
\end{equation}
\begin{equation}
\phi=c_{1}(t-t_{1})^{2}
\end{equation}
\begin{equation}
\psi=\sqrt{2}M_{P}\ln\left[h(t-t_{1})\right]
\end{equation}
\begin{equation}
a(t)=K(t-t_{1}).
\end{equation}
The approximation made in Eq.~(\ref{eq:spectatorapprox}) is that
$\chi$ and $\phi$ fields do not contribute significantly to the
background energy density. This yields in conformal coordinates 
\begin{equation}
K\eta+C_{2}=\ln\left(K(t-t_{1})\right)
\end{equation}
yielding 
\begin{equation}
a=a_{0}e^{K\eta}\label{eq:scale-factor}
\end{equation}
corresponding to an open universe dominated by the spatial curvature
\begin{equation}
H^{2}=\left(\frac{K}{a}\right)^{2}.
\end{equation}
Hence, if one were to embed this model into a realistic cosmology
some work needs to be done, but we will not pursue that here since
our point in this section is to illustrate how the topological particle
production is not limited to the $k=0$ approximation. This spatial
curvature will cancel the $k^{2}$ in Eq.~(\ref{eq:curvaturedispersion})
to yield an analytic dispersion relationship 
\begin{equation}
\omega^{2}=E\eta^{n_{2}}
\end{equation}
going through a zero at time $\eta=0$.\footnote{It is clear from Eq.~(\ref{eq:scale-factor}) that the spacetime
is regular at $\eta=0$ in this coordinate system.} In light of our interest topological particle production, we will
assume $n_{2}$ is an even positive integer. The nonadiabaticity corresponding
to this dispersion relationship is 
\begin{equation}
\frac{\omega'}{\omega^{2}}=\frac{n_{2}}{2\sqrt{E}}\eta^{-1-\frac{n_{2}}{2}}
\end{equation}
indicating that a smaller $E$ and larger $n_{2}$ generates a steeper
approach to the non-adiabatic singularity. If we define the width
of this region to be where the absolute value of this nonadiabaticity
reaches $\frac{1}{2}$, we find 
\begin{equation}
\eta_{1/2}=\left(\frac{n_{2}^{2}}{E}\right)^{\frac{1}{2+n_{2}}}
\end{equation}
which shows that the width decreases only modestly with larger $E$
and is largely insensitive to $n_{2}$.

Let's estimate the topological contribution to the total dark matter
production. The dispersion relation of the $\chi_{k}$ modes for general
$k$ values is 
\begin{equation}
\omega^{2}=k^{2}-K^{2}+E\eta^{n_{2}}.
\end{equation}
The corresponding Bogoliubov coefficient $\beta_{k}$ takes the topological
value 
\begin{equation}
|\beta_{K}|=\cot\left[\frac{\pi}{n_{2}+2}\right]
\end{equation}
at $k=K$. For $k>K$, the dispersion relation is positive definite
and $\beta_{k}$ may be estimated as 
\begin{equation}
|\beta_{k}|\sim|\exp\left(O(1)i\int_{0}^{\eta_{*}}d\eta\,\omega\right)|\sim\exp\left(-O(1)\left(\frac{k^{2}-K^{2}}{E^{2/(2+n_{2})}}\right)^{(n_{2}+2)/2n_{2}}\right)
\end{equation}
whereas for $k<K$, the dispersion relation is tachyonic for a finite
interval of $\eta$, and we expect exponentially enhanced particle
production 
\begin{equation}
\beta_{k}\sim\exp\left(\int_{-\eta_{*}}^{\eta_{*}}d\eta\,|\omega|\right)\sim\exp\left(2\left(\frac{K^{2}-k^{2}}{E^{2/(2+n_{2})}}\right)^{(n_{2}+2)/2n_{2}}\right).
\end{equation}
Now including the phase space contribution, the integrand in Eq.$\,$(\ref{eq:NumDensity})
defining $n_{\chi}$, i.e., $k^{2}|\beta_{k}|^{2}$ has a peak around
$k_{p}$ given by 
\begin{align}
\left(\frac{2(2+n_{2})}{n_{2}}\right)^{\frac{2n_{2}}{n_{2}-2}} & \left(\frac{k_{p}}{E^{1/(2+n_{2})}}\right)^{\frac{4n_{2}}{n_{2}-2}}+\left(\frac{k_{p}}{E^{1/(2+n_{2})}}\right)^{2}=\left(\frac{K}{E^{1/(2+n_{2})}}\right)^{2}
\end{align}
Hence, for $n_{2}>2$, this implies $k_{p}\approx K$, for $k_{p}/E^{1/(2+n_{2})}<1$,
i.e. the integrand peaks at its topological value $K^{2}|\beta_{K}|^{2}.$
The width of the integrand around this peak may be estimated to be
\begin{equation}
\Delta k_{w}\sim O(1)E^{1/(2+n_{2})}.\label{eq:kwidth}
\end{equation}
We then expect significant contributions to the total number density
$n_{\chi}$ from the topological quantity $\beta_{K}$. Testing this
numerically for $n_{2}=6$, we see that $\bar{K}=K/E^{1/(2+n_{2})}=0.6$
corresponds to a peak $\bar{k}_{p}=k_{p}/E^{1/(2+n_{2})}\approx\bar{K}$.The
$k$ width within which the topological contribution becomes significant
can then be estimated from Eq.~(\ref{eq:kwidth}) as 
\begin{equation}
\Delta\bar{k}_{\text{topo}}\sim O(0.1).
\end{equation}
With this, the fraction of topological contribution to the total particle
production (the latter found numerically) is 
\begin{equation}
\frac{n_{\chi,\text{topo}}}{n_{\chi}}\approx0.2
\end{equation}
which indicates that cosmological models where the zero of the dispersion
relationship can occur at large $k$ can have a large topological
contribution. Even though this model has not been fully embedded into
a realistic cosmological setting, it is encouraging that the topological
production can give a physically significant contribution. We defer
the the embedding of this type of model into a realistic cosmology
to a future work.

\section{Summary}

Previous literature using Stokes phenomena to compute particle production
in the cosmological context focused mostly on considering large $k/(ma(t_{c}))$
limit at the time $t_{c}$ of particle production. In the present
work, we considered the $k/(ma(t_{c}))\rightarrow0$ region using
a Stokes phenomena inspired method of computation and showed that
one can relate the topology in the form of Stokes sectors of the analytic
continuation of the $(\alpha_{k}(\eta),\beta_{k}(\eta))$ to the non-perturbatively
large $\left|\beta(z_{+\infty})\right|$ as given in Eq.~(\ref{eq:betafinresult}).
Since the WKB \emph{asymptotic} expansion in each of the Stokes sectors
can be viewed as a choice of vacuum, this is analogous to the Chern-Simons
number separating different gauge vacua. From the perspective of a
topological quantity being rigid in the presence of continuous deformations,
the $n+2$ count of the number of Stokes sectors (defined below Eq.~(\ref{eq:betafinresult}))
is insensitive to continuous variations of the strength of the time
dependence characterized by the parameters $A$ and $C$ in $\omega^{2}=C+Az^{n}$.
Note, the above result, and those in the previous literature apply
to dispersion relations which may be approximated as $\omega^{2}\approx C+A\eta^{n}$
in intervals of conformal time wherein the dispersion relation becomes
non-adiabatic. For such dispersion relations, Stokes constants which
are otherwise difficult to compute, may be approximated in certain
limits using symmetries of the mode equation. This allows one to then
use the very simple exact solution of Eq.~(\ref{eq:betafinresult})
in the limit $C\rightarrow0$ as an approximation

The key mathematical ingredients that determine the topology in the
$C\rightarrow0$ limit are the single-valuedness and the nature of
Stokes phenomena (reviewed above Eq.~(\ref{eq:thetagen})). One of
the key technological apparatus to derive the concrete results was
a mathematical technique of \citep{FF:1965} which constrains the
form of the propagator matrix (reviewed in subsection \ref{subsec:Reduction-of-d.o.f}).
To use that technique and relate that to the standard complexification
basis \citep{Kofman:1997yn,Chung:1998bt,Racco:2024aac}, we developed
a gauge formulation of the equations governing $(\alpha_{k}(\eta),\beta_{k}(\eta))$.
From a purely mathematical perspective, our result can be viewed as
a novel identity of Eq.~(\ref{eq:mathidentity}). An intuitive understanding
of why a topological limit exists in the $C\rightarrow0$ limit is
the special conformal property of the Bessel function Wronskian Eq.~(\ref{eq:besselwronsk}).

We presented two cosmological scenarios illustrating the topological
contribution to $\beta_{k}$. One scenario involves the $\beta_{k}$
amplitude describing a dark matter $\chi$ number spectrum where the
time dependence of the dark matter mass is controlled by a scalar
field $\phi$ rolling in a $\tanh$ potential during the inflationary
coherent oscillations period. During a time interval surrounding $\eta_{0}$
when $\phi$ is in the linear part of the $\tanh$ potential, the
dispersion relationship of $\chi$ takes the form of $\omega^{2}\approx k^{2}+A(\eta-\eta_{0})^{4}$.
The 1-loop correction to the potential generates a global minimum
of the potential at a finite $\phi$ value, determining the final
post-inflationary heavy mass of the $\chi$ particle in this scenario.
In this scenario, the topological contribution is naturally suppressed
since the phase space is proportional to $k^{3}$ whereas the topological
contribution is in the kinematic region $k/(ma(\eta_{0}))\rightarrow0$.
In a second illustrative scenario, we constructed an FLRW solution
with a nonzero constant spatial curvature which enters the dispersion
relationship of $\chi$. In such cases, the kinematic point corresponding
to the topological contribution can be at a large value of the momentum
$k$. We have shown that the fractional contribution to the particle
production in such scenarios can be $O(0.2)$.

There are many future directions to consider in extending the present
work. One can extend the discrete symmetry representation to nonvanishing
$k$ values in this class of models leading to constraints on the
Bogoliubov coefficients. Given that the S-matrices in the background
field driven vacuum transitions can be expressed in terms of $(\alpha_{k},\beta_{k})$,
and given that the background fields can be resolved in terms of quantum
fields, it would interesting to identify the full quantum S-matrix
interpretation of the Stokes sector topological charges. An interesting
direction would be to embed the finite constant curvature FLRW scenario
into a phenomenologically viable cosmological scenario. Yet another
interesting direction to explore is to understand what constraints
can be imposed on the $(\alpha_{k},\beta_{k})$ in the intermediate
$k$ ranges based on the fact that the functional dependence on $k$
is constrained in the $k\rightarrow0$ by our present work and $k\rightarrow\infty$
by exactly solvable model of \citep{Kofman:1997yn}.
\begin{acknowledgments}
DJHC acknowledges partial support from DOE grant DE-SC0017647. NS
acknowledges support from the Ray MacDonald Fund and Wisconsin Alumni
Research Fund at the University of Wisconsin-Madison.
\end{acknowledgments}

\appendix

\section{\label{sec:1-loop-corrections-and}1-loop corrections and dark matter
relic abundance}

The 1-loop effective potential seen by $\phi$ around the background
fields $\left(\phi_{cl},\chi_{cl}=0\right)$ 
\begin{align}
V_{\mathrm{eff}}\left(\phi,\cancel{\chi}\right) & =\rho_{0}\left(1-\tanh\left(\frac{\phi}{M}\right)\right)+\frac{g^{2}}{4(4\pi)^{2}}\left(\ln\left(\frac{g}{\bar{\Lambda}^{2}\Lambda^{2}}\phi^{4}\right)-\frac{3}{2}\right)\frac{\phi^{8}}{\Lambda^{4}}\nonumber \\
 & +\frac{1}{(4\pi)^{2}}\frac{\rho_{0}}{M^{4}}\left(\ln\left(\frac{2\rho_{0}}{\bar{\Lambda}^{2}M^{2}}\text{sech}^{2}\left(\frac{\phi}{M}\right)\tanh\left(\frac{\phi}{M}\right)\right)-\frac{3}{2}\right)\rho_{0}\left(\text{sech}^{2}\left(\frac{\phi}{M}\right)\tanh\left(\frac{\phi}{M}\right)\right)^{2}\label{eq:1-loop-1}
\end{align}
where $\bar{\Lambda}$ is the scale at which the coupling constants
are defined. Since evolution in the tree level potential implies $\phi\left(a_{p}\right)\sim100M$,
we will set $\bar{\Lambda}=100M$. The above potential may be understood
as the loop corrections generating $\phi^{8}/\Lambda^{4}$ and $\rho_{0}\left(\text{sech}^{2}\left(\frac{\phi}{M}\right)\tanh\left(\frac{\phi}{M}\right)\right)^{2}$
non-renormalizable interaction terms. To ensure perturbativity, their
respective (running) couplings should be $\lesssim1$: 
\begin{align}
\frac{g^{2}}{64\pi^{2}}\left(\ln\left(\frac{g}{\bar{\Lambda}^{2}\Lambda^{2}}\phi^{4}\right)-\frac{3}{2}\right) & \lesssim1\\
\frac{1}{64\pi^{2}}\left(\frac{4\rho_{0}}{M^{4}}\right)\left(\ln\left(\frac{2\rho_{0}}{\bar{\Lambda}^{2}M^{2}}\text{sech}^{2}\left(\frac{\phi}{M}\right)\tanh\left(\frac{\phi}{M}\right)\right)-\frac{3}{2}\right) & \lesssim1.
\end{align}
For our choice of parameters $\bar{\rho}=10^{-6}$ and $\bar{g}=10^{-2}$,
the above are satisfied for $g\lesssim1$ and $\frac{H_{I}}{M}\ll1$
(which is also required by the resolution condition Eq.$\,$(\ref{eq:quant_resolution})),
for the range of values of $\phi$ during it's evolution in the 1-loop
corrected potential Eq.$\,$(\ref{eq:1-loop-1}). As we will see below,
evolution in this potential implies $\phi(a_{p})\sim10^{4}M$.

\begin{figure}
\begin{centering}
a)\includegraphics[scale=0.24]{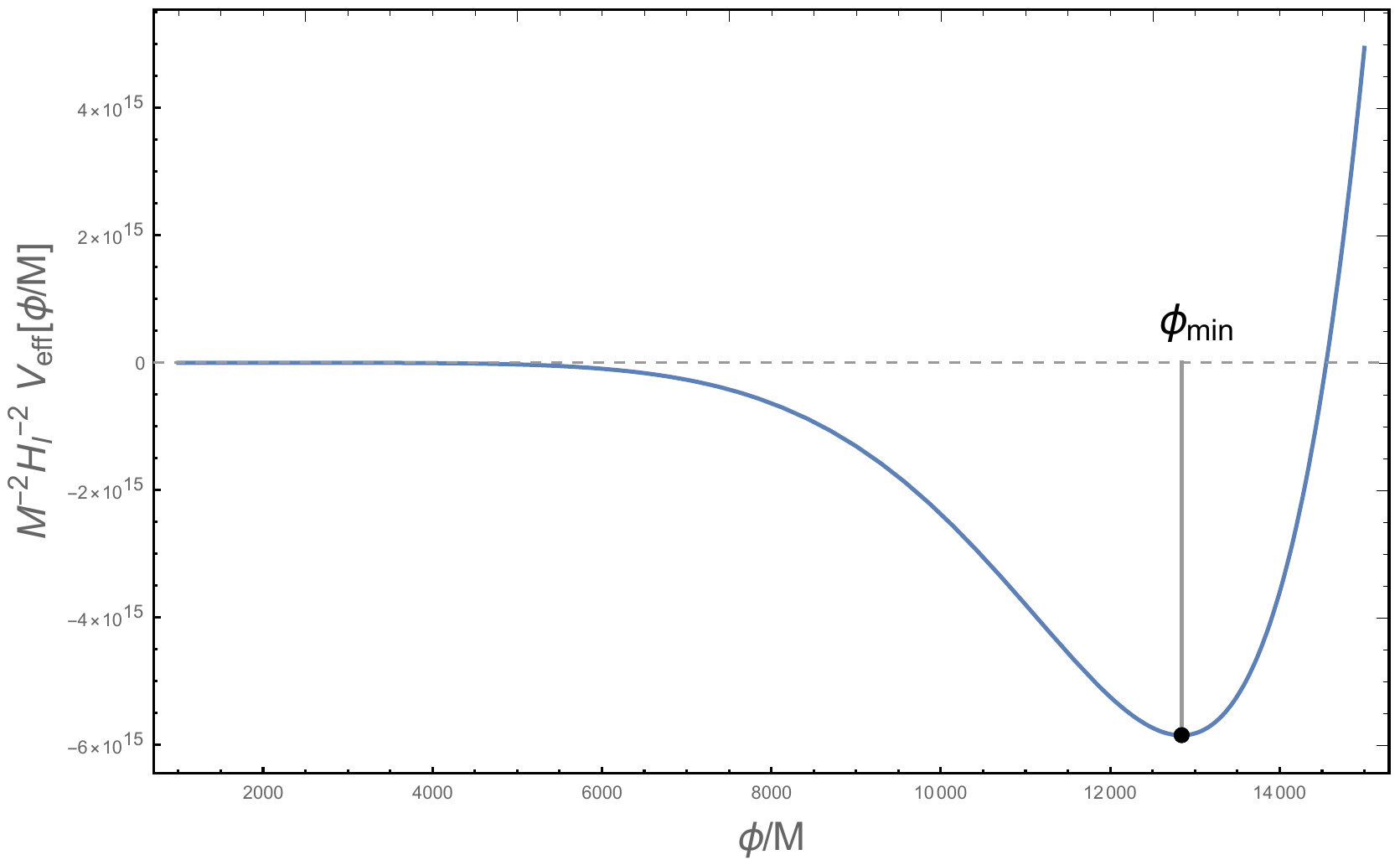}$\quad\quad$b)\includegraphics[scale=0.252]{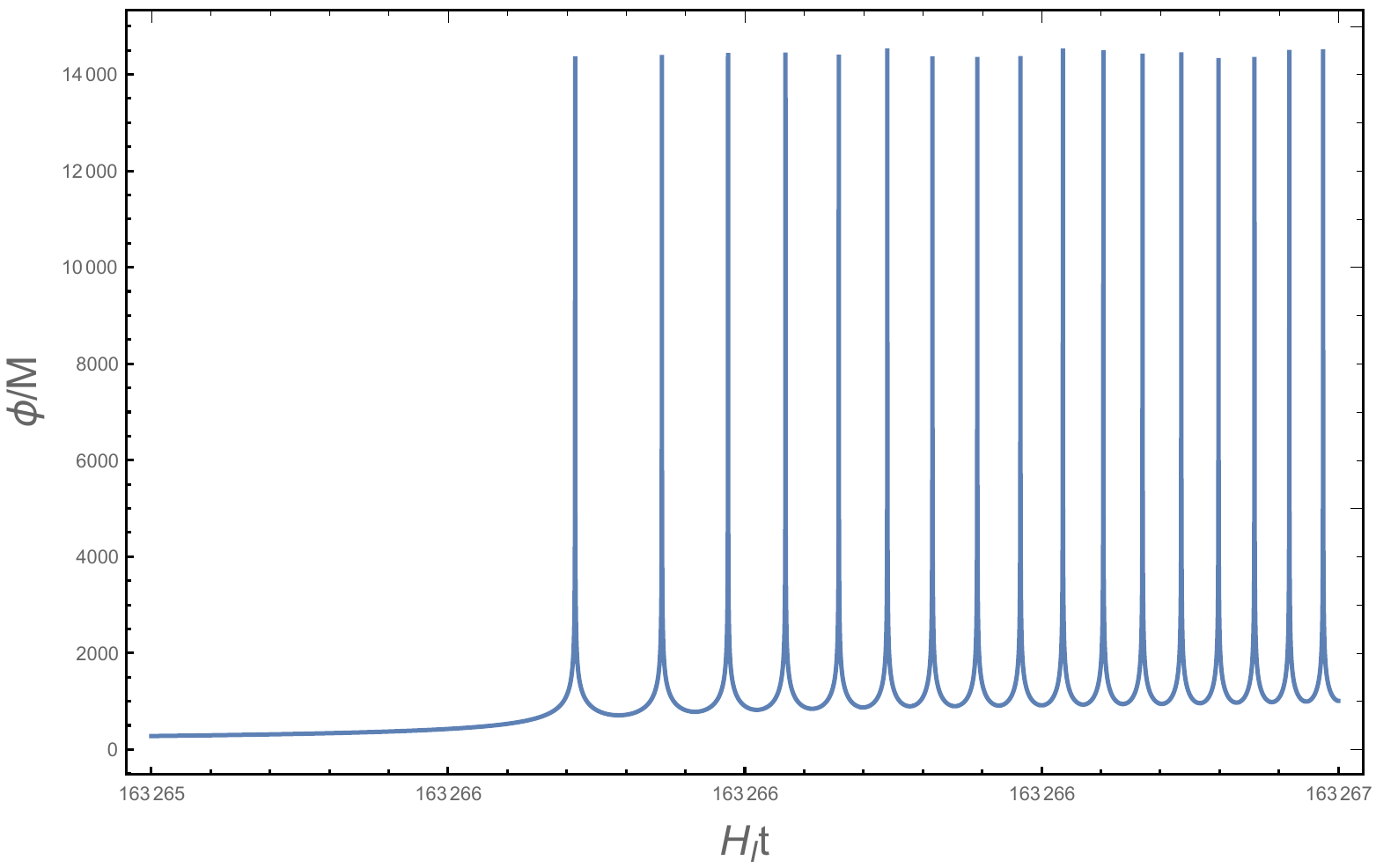}
\par\end{centering}
\caption{\label{fig:1-loop_pot_sol}a) 1-loop corrected potential for $\phi/M\gg1$in
the units $V_{eff}\left(\phi/M\right)/M^{2}H_{I}^{2},$ plotted for
the choice of parameters $gM^{4}/\left(H_{I}^{2}\Lambda^{2}\right)=10^{-2}$and
$\rho_{0}/\left(H_{I}^{2}M^{2}\right)=10^{-6}$. The minimum of the
potential is seen to be at $\phi_{\text{min}}/M\sim12840$. b) Oscillation
of $\phi/M$ around the minimum $\phi_{\text{min}}$, in units of
$H_{I}t$ for the same parameter choices. We see that the field \textquotedblleft falls\textquotedblright{}
into the potential for $H_{I}t_{*}\sim1.6\times10^{5}.$ (The differences
in amplitudes of the oscillations is an artifact of low precision
in the numerical computation.)}
\end{figure}

To compute the evolution of $\phi$ on the corrected potential, note
that the third term in Eq.$\,$(\ref{eq:1-loop-1}) has a negligible
effect and may be ignored. This is because for $\phi/M\gg1$, the
third term is exponentially suppressed relative to the other two whereas
for $\phi/M\lesssim1$, the third term is suppressed by $\bar{\rho}\left(\frac{H_{I}}{M}\right)^{2}\ll1$
and $\bar{\rho}^{2}/\bar{g}^{2}\ll1$ w.r.t the first two terms respectively.
Hence the potential seen by $\phi$ is 
\begin{equation}
V_{\text{eff}}\left(\phi\right)\approx\rho_{0}\left(1-\tanh\left(\frac{\phi}{M}\right)\right)+\frac{g^{2}}{64\pi^{2}}\left(\ln\left(\frac{g}{\bar{\Lambda}^{2}\Lambda^{2}}\phi^{4}\right)-\frac{3}{2}\right)\frac{\phi^{8}}{\Lambda^{4}}.\label{eq:1-loop-simplifd-1}
\end{equation}
For small $\phi$ values, the coupling of the $\phi^{8}$ interaction
term is negative and gradually grows positive giving the potential
the shape as in figure (\ref{fig:1-loop_pot_sol}a). The logarithmic
dependence of the coupling therefore introduces a minimum in the potential
seen by $\phi$ around which the field is trapped and oscillates as
seen in the numerical computation of figure (\ref{fig:1-loop_pot_sol}b).
Since the first term of the in Eq.$\,$(\ref{eq:1-loop-simplifd-1})
becomes exponentially suppressed for $\phi/M\gg1$, the minimum is
determined by the second term as 
\begin{align}
\frac{\phi_{\text{min}}}{M} & =e^{1/4}\left(\frac{gM^{2}/\Lambda^{2}}{10^{4}}\right)^{-1/4}.
\end{align}

For $H_{I}/M=10^{-5}\left(N/60\right)$ satisfying the resolution
condition in Eq.$\,$(\ref{eq:quant_resolution}) and the coupling
choice $gM^{4}/\left(H_{I}^{2}\Lambda^{2}\right)=10^{-2}$ (from the
isolation of the nonadiabaticty explained in Eq.~(\ref{eq:linearversuswidth})),
the above implies $\phi_{\text{min}}/M\sim1.3\times10^{4}\left(N/60\right)^{-1/2}$.
If the $\phi$ oscillations die down and the field takes values $\phi\left(a_{p}\right)\approx\phi_{\text{min}}$
today, then the density of the $\chi$-dark matter estimate is given
by 
\begin{align}
\Omega_{\chi}h^{2} & =0.27\left(\frac{T_{rh}}{10^{7}GeV}\right)\left(\frac{H_{I}}{10^{2}GeV}\right)\left(\frac{M}{10^{9}GeV}\right)\left(1+0.21\log\left(\frac{\rho_{0}/(M^{2}H_{I}^{2})}{10^{-6}}\right)\right)\label{eq:relic}
\end{align}
where the mass of the dark matter is 
\begin{equation}
\sqrt{\partial_{\chi}^{2}V(\phi_{\mathrm{min}},\chi=0)}=1.6\times10^{11}\left(\frac{M}{10^{9}\mathrm{GeV}}\right)\,\mathrm{GeV}.
\end{equation}
The fiducial $M$ choice of $10^{9}$ GeV can be associated with an
arbitrary association with the intermediate scale. It is easy to slide
this number higher without upsetting the conditions required for the
validity of Eq.~(\ref{eq:relic}). For every choice of $M$ and $H_{I}$,
there exists a range of $g/\Lambda^{2}$ scale bounded by the isolation
of the nonadiabaticty explained in Eq.~(\ref{eq:linearversuswidth}),
and for the fiducial values of $M\sim10^{9}$GeV and $H_{I}\sim10^{2}$GeV,
we have 
\begin{equation}
\frac{\Lambda}{\sqrt{g}}\sim10^{17}\mathrm{GeV}
\end{equation}
which implies that the designer coupling of $\phi$ to the dark matter
may come from a UV model construction near the GUT scale.

Note, if the oscillations in the field $\phi$ do not die down sufficiently,
then it could make a significant contribution to the dark matter density
today. To estimate a bound on $\Omega_{\phi}h^{2}$, we numerically
solve for the evolution of the averaged energy density of the field
$\langle\rho_{\phi}\rangle$, where the average is taken over the
time duration $\tau$ of a few oscillations but such that $\tau\ll H^{-1}.$
For systems like the example here, where the time period of coherent
oscillations $T\ll H^{-1}$, averaging over the duration of several
oscillations defines an approximate equation of state for $\rho_{\phi}$
- dependent on the potential $V_{eff}\left(\phi\right)$ and given
by 
\begin{equation}
w\left(\langle\rho_{\phi}\rangle\right)=\frac{\left\langle \frac{\dot{\phi}^{2}}{2}-\left(V_{eff}\left(\phi\right)-V_{eff}\left(\phi_{\text{min}}\right)\right)\right\rangle }{\left\langle \frac{\dot{\phi}^{2}}{2}+\left(V_{eff}\left(\phi\right)-V_{eff}\left(\phi_{\text{min}}\right)\right)\right\rangle }
\end{equation}
where the average in the above is also taken w.r.t $\tau$ and the
potential has been shifted by a constant such that the shifted potential
is positive definite.\footnote{For our scenario with parameters, $gM^{4}=10^{-2}H_{I}^{2}\Lambda^{2}$
, $\rho_{o}=10^{-6}M^{2}H_{I}^{2}$ and $H_{I}/M=10^{-5},$ the ratio
of the oscillation frequency to the Hubble constant may be estimated
as 
\begin{equation}
\frac{\left(V_{eff}''\left(\phi_{\text{min}}\right)\right)^{1/2}}{H(t_{m})}\approx10^{9}\gg1
\end{equation}
where the Hubble constant has been evaluated using a numerical solution
at the earliest times when the oscillations begin.} The equation of state $w\left(\langle\rho_{\phi}\rangle\right)$
as defined above is a function of the energy density $\langle\rho_{\phi}\rangle$
and is only approximately constant for $\Delta t\ll H^{-1}$.

\begin{figure}
\begin{centering}
\hspace{-3mm}a)\includegraphics[scale=0.31]{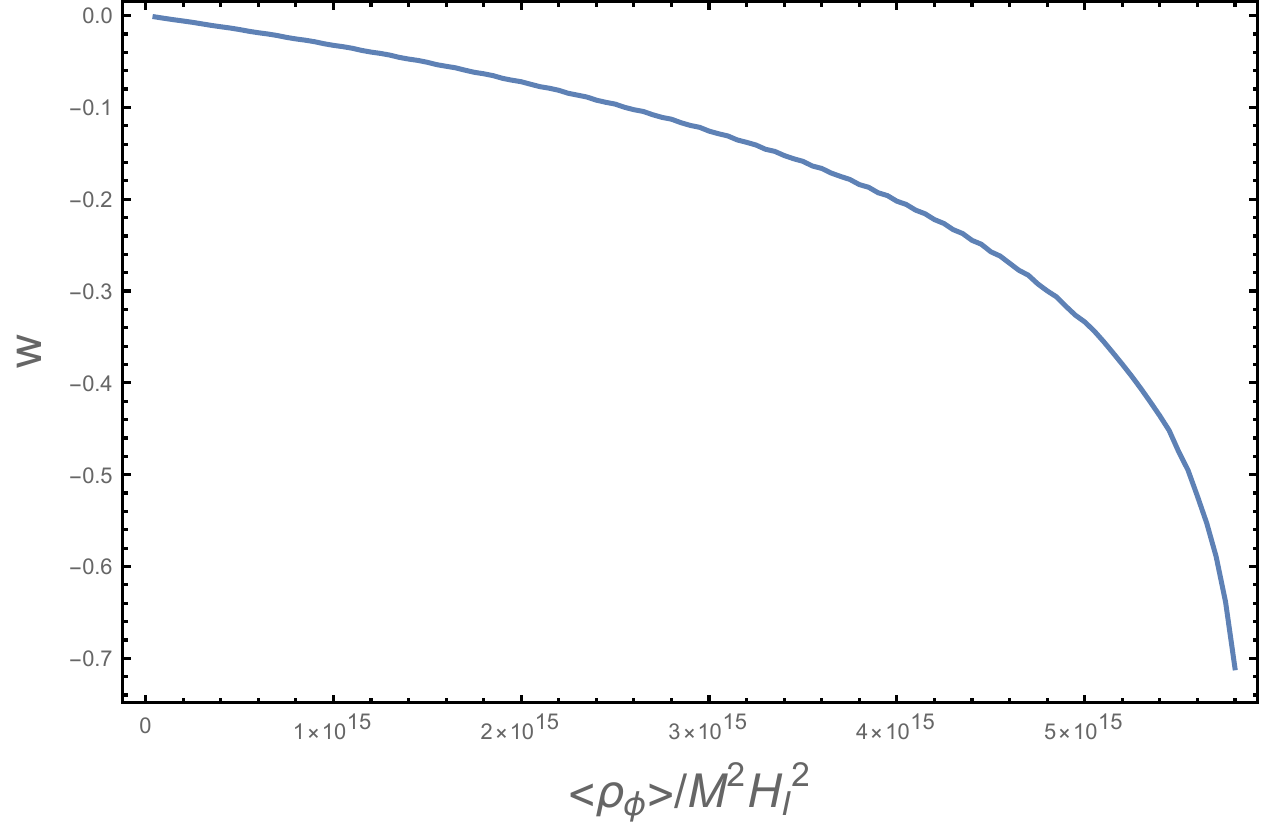}$\hspace{1cm}$b)\includegraphics[scale=0.3]{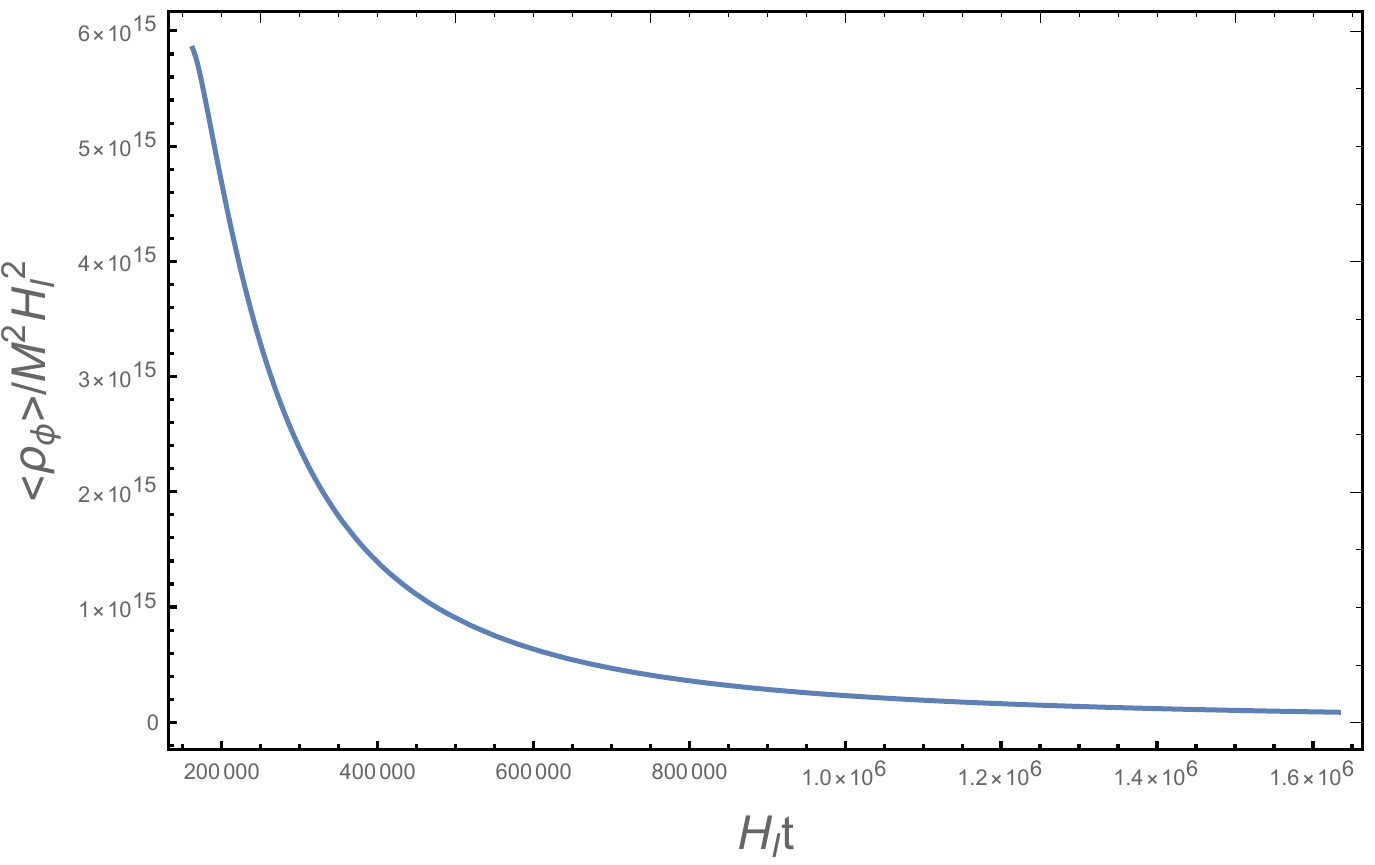}
\par\end{centering}
\caption{\label{fig:Num_estimates}a) Numerically estimated average equation
of state. The average is over $n\gg1$ oscillations and for a duration
$\tau\ll H^{-1}.$ b) The evolution of the average energy density
(in units of $M^{2}H_{I}^{2}$) using the numerically estimated equation
of state $w(\langle\rho_{\rho}\rangle).$ The initially negative equation
of state reflects the flatness and concave down nature of the potential
sampled during the initial set of oscillations.}
\end{figure}

The expression for $w\left(\langle\rho_{\phi}\rangle\right)$ may
be further simplified using the equation of motion (EOM) of $\phi$.
For the duration $\tau$ for which the Hubble friction can be neglected,
the EOM implies 
\begin{equation}
\ddot{\phi}+V'\left(\phi\right)\approx0.
\end{equation}
In our situation, it is easy to estimate that the number of oscillations
in $1/H$ time period is large. In such situations, we can derive
the following virial theorem 
\begin{equation}
\langle\dot{\phi}^{2}\rangle\approx\langle\phi V'\left(\phi\right)\rangle
\end{equation}
where the average is over several oscillations. This implies 
\begin{align}
w\left(\langle\rho_{\phi}\rangle\right) & \approx\frac{\frac{\langle\phi V'\left(\phi\right)\rangle}{2}-\left\langle V_{\mathrm{eff}}\left(\phi\right)-V_{\mathrm{eff}}\left(\phi_{\mathrm{min}}\right)\right\rangle }{\frac{\langle\phi V'\left(\phi\right)\rangle}{2}+\left\langle V_{\mathrm{eff}}\left(\phi\right)+V_{\mathrm{eff}}\left(\phi_{\mathrm{min}}\right)\right\rangle }
\end{align}
and the energy conservation equation 
\begin{align}
\frac{d\langle\rho_{\phi}\rangle}{dt}+3H\langle\rho_{\phi}\rangle\left(1+w\left(\langle\rho_{\phi}\rangle\right)\right) & =0\,.
\end{align}
Numerically estimating $w\left(\langle\rho_{\phi}\rangle\right)$
followed by numerically solving the above differential equation gives
Fig.~\ref{fig:Num_estimates}.

Here we see that as Hubble friction removes energy from the field
oscillations, the potential seen by $\phi$ reduces to a quadratic
around $\phi_{\text{min}}$ and the equation of state approaches that
of a matter dominated universe, i.e. $w\rightarrow0$. For the parameters
used here, $w\approx0$ for $t_{m}\sim10^{6}H_{I}^{-1}.$ Using the
matter-like evolution of the field density, the energy density today
may be estimated from the numerically found energy density around
$t_{m}$ as 
\begin{equation}
\Omega_{\phi}h^{2}\sim1.3\times10^{14}\left(\frac{10^{7}\mathrm{GeV}}{T_{\mathrm{rh}}}\right)^{-1}\left(\frac{H_{I}}{10^{2}\mathrm{GeV}}\right)^{2}\left(\frac{M}{10^{9}\mathrm{GeV}}\right)^{2}\gg1.
\end{equation}
To ensure $\Omega_{\phi}h^{2}\lesssim1$, one can introduce the following
coupling to photons 
\begin{equation}
\mathcal{L_{\phi}}\supset\frac{\phi}{2M_{2}}F_{\mu\nu}F^{\mu\nu}.
\end{equation}
When the oscillations of the scalar field become matter-like, the
field decays dominantly through the $\phi\rightarrow\gamma\gamma$
channel into radiation. The corresponding decay rate is given by 
\begin{equation}
\Gamma=\frac{1}{32\pi}\frac{m^{3}}{M_{2}^{2}}
\end{equation}
where $m=\left(V_{eff}''\left(\phi_{\text{min}}\right)\right)^{1/2}$is
the mass of the non-relativistic dark matter particle. To ensure that
the field does not contribute significantly to the energy density
during BBN, it should ideally decay away before this time: 
\begin{equation}
\Gamma\gtrsim\frac{100\mathrm{MeV}^{2}}{M_{p}}\gg H(t_{\text{BBN}}).
\end{equation}
Including the effects of this decay, the energy density in the coherently
oscillating field now dilutes as 
\begin{equation}
\rho_{\phi}(t)\approx\left[\frac{a(t_{m})}{a(t)}\right]^{3}\rho_{\phi}(t_{m})\exp\left[-\Gamma_{\phi}(t-t_{m})\right].
\end{equation}
Comparing this to the energy density in radiation at the time of BBN,
we find 
\begin{equation}
\frac{\rho_{\phi}(t_{\text{BBN}})}{\rho_{rad}(t_{\text{BBN}})}\sim\exp\left[-\Gamma_{\phi}(t_{\text{BBN}}-t_{m})+42.3\right]\left(\frac{T_{\mathrm{rh}}}{10^{7}\mathrm{GeV}}\right)\left(\frac{M}{10^{9}\mathrm{GeV}}\right)^{2}\ll1
\end{equation}
since for $\Gamma_{\phi}\sim100\mathrm{MeV}^{2}/M_{p}$, the exponent
of the decay factor is 
\begin{equation}
\Gamma_{\phi}(t_{\text{BBN}}-t_{m})\sim3.8\times10^{3}\left(\frac{H_{I}}{10^{2}\mathrm{GeV}}\right)^{7/3}\left(\frac{T_{\mathrm{rh}}}{10^{7}\mathrm{GeV}}\right)^{-8/3}.
\end{equation}

\section{\label{sec:Covariance-of-linear}Covariance of linear differential
operator inducing a symmetry representation}

In this section, we will show how the covariance of a non-first-order
linear differential operator $\mathcal{O}_{x}$ under coordinate change
$\bar{x}=Lx$ will induce a symmetry representation of the propagator
governing a homogeneous differential equation rewritten in a first
order formalism. We will do this in two steps. In the first step,
we show how such covariance leads to a generation of new solutions.
In the second step, we use this solution generation technique together
with a judicious basis choice to find a first order formalism propagator
symmetry. Finally, we apply this to our differential equation of interest.
The key nontriviality will be the transformation property of a judiciously
chosen set of basis functions under the $L$ transform.

\subsection{Generating new solutions}

In this subsection, we show how the covariance of a linear differential
operator $\mathcal{O}_{x}$ under a coordinate transform can generate
new solutions to the homogeneous differential equation governed by
$\mathcal{O}_{x}$.

Let $\mathcal{O}_{x}$ be a linear differential operator and let 
\begin{equation}
\mathcal{O}_{x}\chi(x)=0\label{eq:origeq}
\end{equation}
be a homogeneous linear differential equation. Define a linear transformation
\begin{equation}
\bar{x}=Lx\label{eq:coordtransform}
\end{equation}
which when substituted into Eq.~(\ref{eq:origeq}) gives 
\begin{equation}
\mathcal{O}_{x}=\mathcal{O}_{L^{-1}\bar{x}}.
\end{equation}
We then see Eq.~(\ref{eq:origeq}) by algebraic substitution of the
transformations that 
\begin{equation}
\mathcal{O}_{L^{-1}\bar{x}}\chi(L^{-1}\bar{x})=0.\label{eq:substituted}
\end{equation}
We define there to be a homogeneous differential equation covariance
representation \textbf{if} the operator satisfies 
\begin{equation}
\mathcal{O}_{L^{-1}\bar{x}}=D(L)\mathcal{O}_{\bar{x}}.\label{eq:symmetry}
\end{equation}
where $D(L)$ is invertible and commutes with $\mathcal{O}_{\bar{x}}$.
Eq.~(\ref{eq:symmetry}) turns Eq.~(\ref{eq:substituted}) into
\begin{equation}
D(L)\mathcal{O}_{\bar{x}}\chi(L^{-1}\bar{x})=0.\label{eq:symm}
\end{equation}

Note that if we now drop the bar on $\bar{x}$ in Eq.~(\ref{eq:symm}),
we are considering a different solution: the new solution $\chi(L^{-1}x)$
to 
\begin{equation}
\mathcal{O}_{x}\chi(L^{-1}x)=0\label{eq:newsol}
\end{equation}
satisfies the boundary condition 
\begin{equation}
\chi(L^{-1}x_{P})=\chi_{P}\hspace{1em}\partial_{x}\chi(x)|_{x=L^{-1}x_{P}}=\left(\partial_{x}\chi\right)_{P}\label{eq:BCshift}
\end{equation}
where the right hand side contains the same values that would have
been imposed in the original solution at $x_{P}$ (and not $L^{-1}x_{P}$).
Note that by simply dropping the bar, we have assumed $\chi$ is an
object whose definition does not correspond to components of a coordinate
dependent basis. In contrast, we could have had 
\begin{equation}
D(L)\mathcal{O}_{\bar{x}}\chi^{\mu}(L^{-1}\bar{x})=0
\end{equation}
coming from 
\begin{equation}
D(L)\mathcal{O}_{\bar{x}}\left(\chi^{\mu}(L^{-1}\bar{x})e_{\mu}\right)=0
\end{equation}
which is equivalent to 
\begin{equation}
D(L)\mathcal{O}_{\bar{x}}\left(\chi^{\mu}(L^{-1}\bar{x})\bar{e}_{\nu}\left(R^{-1}(L)\right)_{\phantom{\nu}\mu}^{\nu}\right)=0
\end{equation}
giving rise to 
\begin{equation}
D(L)\mathcal{O}_{\bar{x}}\left(\bar{\chi}^{\nu}(\bar{x})\bar{e}_{\nu}\right)=0\hspace{1em}\bar{\chi}^{\nu}(\bar{x})\equiv\left(R^{-1}(L)\right)_{\phantom{\nu}\mu}^{\nu}\chi^{\mu}(L^{-1}\bar{x})
\end{equation}
where $e_{\mu}$ is a coordinate dependent basis and $R$ is a matrix
that accounts for its coordinate dependence. In the latter case, we
would have the new solution being$\left(R^{-1}(L)\right)_{\phantom{\nu}\mu}^{\nu}\chi^{\mu}(L^{-1}x)$
instead of $\chi^{\nu}(L^{-1}x)$ assuming $\bar{x}$ and $x$ cover
the same set of points.

Summarizing, according to Eq.~(\ref{eq:symm}), because $D^{-1}(L)$
leaves the $0$ on the right hand side of a homogeneous linear differential
equation invariant, $\chi(L^{-1}\bar{x})$ also satisfies the differential
equation (Eq.~(\ref{eq:newsol})). Since $\chi_{P}$ and $\left(\partial_{x}\chi\right)_{P}$
represents an arbitrary boundary condition data, this solution generating
mechanism may be used to generate a propagator symmetry. We will turn
to this next.

\subsection{1st order formalism propagator symmetry}

Here we will apply Eq.~(\ref{eq:newsol}) to find a propagator symmetry
in a first order differential equation rewriting of a second order
complex differential equation.

Suppose we rewrite the second order differential equation 
\begin{equation}
\left[\partial_{z}^{2}+\omega^{2}(z)\right]\chi(z)=0
\end{equation}
where $z$ is a complex number as a first order differential equation
\begin{equation}
\partial_{z}V(z)=M(z)V(z)\label{eq:firstorder}
\end{equation}
where 
\begin{equation}
\chi(z)=F(z)V(z)
\end{equation}
and $F$ is a fixed basis of functions defined to be 
\begin{equation}
F(z)=(\mathcal{F}_{+}(z),\mathcal{F}_{-}(z)).\label{eq:basisdef}
\end{equation}
For the purposes of our induced representation construction, we choose
$F$ to satisfy a particular representation 
\begin{equation}
F(L^{-1}z)=D_{F}(L^{-1})F(z)E_{F}(L^{-1})\label{eq:EFrep}
\end{equation}
where $D_{F}(L^{-1})$ is a complex number and $E_{F}(L^{-1})$ is
a matrix.\footnote{Note that different choices of $D_{F}(L^{-1})$ can lead to different
choices of $E_{F}(L^{-1})$. We will see that the end result depends
on $E_{F}^{-1}(L^{-1})...E_{F}(L^{-1})$ which is invariant under
the choice made for $D_{F}(L^{-1})$.} Whether or not this choice can be made for a nontrivial matrix $E_{F}(L^{-1})$
is the key non-triviality of the construction. We will see that in
our application of this formalism to a particular class of $\omega^{2}$,
the WKB basis for $F$ will generate a nontrivial $E_{F}$ belonging
to a nontrivial $S_{2}$ representation.

With a different boundary condition as discussed in Eqs.~(\ref{eq:newsol})
and (\ref{eq:BCshift}), we generate a new solution by identifying
$\chi$ at $L^{-1}z$ with $\chi_{2}$ at $z$: 
\begin{equation}
\chi_{2}(z)=\chi(L^{-1}z).
\end{equation}
In the $F$ basis, this becomes 
\begin{equation}
F(z)V_{2}(z)=F(L^{-1}z)V(L^{-1}z)\label{eq:step1}
\end{equation}
where $\chi_{2}(z)=F(z)V_{2}(z)$ sharing the same basis function
as $\chi(z)$. Using Eq.~(\ref{eq:EFrep}), Eq.~(\ref{eq:step1})
becomes 
\begin{equation}
F(z)V_{2}(z)=D_{F}(L^{-1})F(z)E_{F}(L^{-1})V(L^{-1}z).\label{eq:FV2}
\end{equation}
Suppose $V_{2}(z)$ corresponds to data propagated from $z_{0}$ denoted
as $V_{2}(z_{0})$: 
\begin{equation}
V_{2}(z)=U(z,z_{0})V_{2}(z_{0})
\end{equation}
where $U$ is the propagator solution to Eq.~(\ref{eq:firstorder}):
\begin{equation}
U(z,z_{0})\equiv P\left[e^{\int_{C(z_{0},z)}dz'M(z')}\right]
\end{equation}
with the path ordering symbol $P$ along the path $C$ starting at
$z_{0}$ and ending on $z$. Putting this into Eq.~(\ref{eq:FV2})
gives 
\begin{equation}
F(z)U(z,z_{0})V_{2}(z_{0})=D_{F}(L^{-1})F(z)E_{F}(L^{-1})V(L^{-1}z).
\end{equation}

Similarly, let $V(L^{-1}z)$ correspond to data $V(L^{-1}z_{0})$
propagated from $L^{-1}z_{0}$: 
\begin{equation}
F(z)U(z,z_{0})V_{2}(z_{0})=D_{F}(L^{-1})F(z)E_{F}(L^{-1})U(L^{-1}z,L^{-1}z_{0})V(L^{-1}z_{0}).\label{eq:prev}
\end{equation}
Setting $z=z_{0}$ in this equation, we find 
\begin{equation}
F(z_{0})V_{2}(z_{0})=D_{F}(L^{-1})F(z_{0})E_{F}(L^{-1})V(L^{-1}z_{0})
\end{equation}
where we used $U(z_{0},z_{0})=1$. The general solution to this equation
is 
\begin{equation}
V_{2}(z_{0})=Z+D_{F}(L^{-1})E_{F}(L^{-1})V(L^{-1}z_{0})\label{eq:solwithz}
\end{equation}
where $Z$ solves the zero mode equation 
\begin{equation}
F(z_{0})Z=0.\label{eq:zmodedef}
\end{equation}
Writing more explicitly 
\begin{equation}
F(z_{0})=(\mathcal{F}_{+},\mathcal{F}_{-})
\end{equation}
we can parameterize the general solution to Eq.~(\ref{eq:zmodedef})
as 
\begin{equation}
Z=f_{s}(z_{0})F_{\perp}(z_{0})
\end{equation}
where $f_{s}$ is an arbitrary scaling function of $z_{0}$ and $F_{\perp}(z_{0})\equiv(\mathcal{F}_{-}(z_{0}),-\mathcal{F}_{+}(z_{0}))$.

Putting Eq.~(\ref{eq:solwithz}) into Eq.~(\ref{eq:prev}) therefore
becomes 
\begin{align}
D_{F}(L^{-1})F(z)\left[E_{F}(L^{-1})U(L^{-1}z,L^{-1}z_{0})-U(z,z_{0})E_{F}(L^{-1})\right]V(L^{-1}z_{0})\nonumber \\
=F(z)U(z,z_{0})F_{\perp}(z_{0})f_{s}(z_{0})
\end{align}
Choosing $f_{s}(z_{0})$ to vary independently of $V(L^{-1}z_{0})$,
we find 
\begin{equation}
D_{F}(L^{-1})F(z)\left[E_{F}(L^{-1})U(L^{-1}z,L^{-1}z_{0})-U(z,z_{0})E_{F}(L^{-1})\right]V(L^{-1}z_{0})=0
\end{equation}
Since $V(L^{-1}z_{0})$ is arbitrary, we find 
\begin{equation}
F(z)\left[E_{F}(L^{-1})U(L^{-1}z,L^{-1}z_{0})-U(z,z_{0})E_{F}(L^{-1})\right]=0.\label{eq:withprojection}
\end{equation}
Hence, up to ambiguities of the projection, we are motivated to define
a symmetry transformation 
\begin{equation}
U(z,z_{0})=E_{F}(L^{-1})U(L^{-1}z,L^{-1}z_{0})E_{F}^{-1}(L^{-1})\label{eq:Umatrixrep}
\end{equation}
which as anticipated before does not depend on different choices of
the phases $D_{F}(L^{-1})$ since $E_{F}...E_{F}^{-1}$ cancels any
such factors. Expanding $U$ to linear order in $M$, this also implies
a differential relationship of 
\begin{equation}
dzM(z)=E_{F}(L^{-1})L^{-1}dzM(L^{-1}z)E_{F}^{-1}(L^{-1}).\label{eq:linearizedversion}
\end{equation}

In summary, we considered situations where the second order ordinary
differential equation governed by the differential operator $\mathcal{O}_{x}$
has a symmetry representation $D(L)$ under the coordinate transform
$\bar{x}=Lx$. This can be written in terms of the first order formalism
of Eq.~(\ref{eq:firstorder}) with a judicious basis choice of $F$,
and one can compute the representation of $L$ acting on $F$ as Eq.~(\ref{eq:EFrep})
which involves the matrix $E_{F}(L^{-1})$. If $E_{F}(L^{-1})$ is
nontrivial, it induces a useful symmetry of the propagator through
Eq.~(\ref{eq:Umatrixrep}).

\subsection{Our model}

Consider 
\begin{equation}
\mathcal{O}_{(z-z_{0})}=\partial_{(z-z_{0})}^{2}+A(z-z_{0})^{n}.
\end{equation}
Under the rotation 
\begin{equation}
\bar{z}-\bar{z}_{0}=L\left(z-z_{0}\right)
\end{equation}
where $L\equiv e^{i\theta}$, the operator transforms as 
\begin{equation}
\mathcal{O}_{(z-z_{0})}=e^{i2\theta}\left[\partial_{\bar{z}-\bar{z}_{0}}^{2}+Ce^{-i(n+2)\theta}(\bar{z}-\bar{z}_{0})^{n}\right]
\end{equation}
Hence, we see if 
\begin{equation}
\theta=\frac{2\pi}{n+2}
\end{equation}
$L$ in Eq.~(\ref{eq:Umatrixrep}) has been constructed.

With a first order formalism written in terms of the WKB basis functions
we choose 
\begin{equation}
\mathcal{F}_{\pm}(z)=f_{\pm}(z)
\end{equation}
in Eq.~(\ref{eq:basisdef}) where $f_{\pm}(z)$ are WKB basis functions
of Eq.~(\ref{eq:Fdef-1}) with ($\eta$ complexified and) the origin
taken as $(z_{*})=0$. Under $L$, the basis vector $F$ transforms
as 
\begin{equation}
F\left(L^{-1}z\right)=iL^{-1/2}F(z)B\label{eq:basistransform}
\end{equation}
where 
\[
B=\left(\begin{array}{cc}
0 & 1\\
1 & 0
\end{array}\right)
\]
which allows us to choose 
\begin{equation}
E_{F}(L^{-1})=B
\end{equation}
in Eq.~(\ref{eq:Umatrixrep}). Hence, Eq.~(\ref{eq:Umatrixrep})
becomes 
\begin{equation}
\bar{U}_{g}(\bar{z},\bar{z}_{0})=BU_{g}(z,z_{0})B^{-1}.\label{eq:propagatorrep}
\end{equation}
where $U_{g}$ is the propagator matrix of Eq.~(\ref{Eq:ConnecMat})
in any gauge $g$.

It is important to recognize that the representation given by Eq.~(\ref{eq:propagatorrep})
is not generated by a coordinate transformation Eq\@.~(\ref{eq:coordtransform})
acting on Eq.~(\ref{Eq:ConnecMat}) in every gauge. This is stemming
from the ambiguities of the projection effects in going from Eq.~(\ref{eq:withprojection})
and (\ref{eq:propagatorrep}). The representation given by Eq.~(\ref{eq:propagatorrep})
is generated by the coordinate transformation in the 0-gauge and the
$F^{2}$-gauge.

The linearized version corresponding to Eq.~(\ref{eq:linearizedversion})
becomes 
\begin{equation}
d\bar{z}M(\bar{z})=BdzM(z)B^{-1}.
\end{equation}

\section{\label{sec:An-asymptotic-property}An asymptotic property of off
diagonal propagator}

In this section, we present an argument for the vanishing of $\mu^{p>0}U_{21}$
in the limit $\mu\rightarrow0$ (also applicable to $\mu^{p>0}U_{12}$)
where $U_{21}$ and $U_{12}$ are off-diagonal propagators connecting
adjacent anti-Stokes lines.

Start with 
\begin{equation}
\chi(z)=F(z)V(z)=F(z)U(z,z_{0})V(z_{0})\label{eq:maineq-1}
\end{equation}
which is postulated to be a solution to the mode equation (where we
have suppressed the wave vector $k$ to reduce notational clutter).
Now, choose $V(z_{0})$ to be 
\begin{equation}
V(z_{0})=\left(\begin{array}{c}
1\\
0
\end{array}\right)
\end{equation}
such that the right hand side of Eq.~(\ref{eq:maineq-1}) becomes
\begin{equation}
F(z)U(z,z_{0})V(z_{0})=F(z)\left(\begin{array}{c}
U(z,z_{0})_{11}\\
U(z,z_{0})_{21}
\end{array}\right).\label{eq:intermed1}
\end{equation}
The left hand side of Eq.~(\ref{eq:maineq-1}) on the annulus is
known to have an asymptotic expansion of 
\begin{equation}
\chi(z)\sim F(z)V_{r}\label{eq:existence}
\end{equation}
where 
\begin{equation}
V_{r}=O(\mu^{0})\label{eq:const}
\end{equation}
as long as $z$ is either in a single Stokes sector (defined to be
a region on the annulus bounded by two anti-Stokes lines with at least
one Stokes line in between) or is on an (anti-)Stokes line as the
asymptotic expansion is taken. In other words, \emph{all} solutions
including Eq.~(\ref{eq:intermed1}) can be matched with an asymptotic
expansion satisfying Eq.~(\ref{eq:const}) 
\begin{equation}
F(z)O(\mu^{0})\sim F(z)\left(\begin{array}{c}
U(z,z_{0})_{11}\\
U(z,z_{0})_{21}
\end{array}\right)
\end{equation}
as long as $z$ is restricted to a particular region in the complex
plane.

Let $z_{0}$ be an anti-Stokes line and let $z=z_{0}\gamma$ where
$\gamma=\exp\left(\frac{2\pi i}{n+2}\right)$: 
\begin{equation}
F(z_{0}\gamma)O(\mu^{0})\sim F(z_{0}\gamma)\left(\begin{array}{c}
U(z_{0}\gamma,z_{0})_{11}\\
U(z_{0}\gamma,z_{0})_{21}
\end{array}\right)
\end{equation}
which makes 
\begin{equation}
U(z_{0}\gamma,z_{0})_{21}=O(\mu^{0}).
\end{equation}
This implies 
\begin{equation}
\lim_{\mu\rightarrow0}\mu^{n>0}U_{21}(\mbox{adjacent anti-Stokes line propagation)}=0.
\end{equation}
Note that although this conclusion is also implied in \citep{FF:1965},
our line of reasoning is distinct from what is presented there in
that \citep{FF:1965} uses the properties of the exact power series
solution and in our language utilizes a particular gauge.

\section{Mismatch of analytic properties of mode and basis functions}

Stokes phenomena occurs in situations where the basis of the asymptotic
expansion has an analytic property that is mismatched from the analytic
property of the function that is being resolved. In such cases, the
asymptotic expansion of the coefficients acquire discreteness. In
this appendix, we use three examples to illustrate this.

A trivial example is 
\begin{equation}
f(z)=e^{-1/z}
\end{equation}
which one can ask to be expanded about $z=0$ in terms of the $\{z^{n}\}$
basis: 
\begin{equation}
f(z)\sim\sum_{n=0}^{\infty}c_{n}(z)z^{n}
\end{equation}
where $c_{n}(z)$ will have a discontinuous behavior across the anti-Stokes
rays at $\arg z=\{\frac{\pi}{2},\frac{3\pi}{2}\}$: 
\begin{equation}
c_{n}(z)\sim\begin{cases}
0 & \Re z>0\\
\mbox{undefined} & \mbox{otherwise}
\end{cases}.
\end{equation}
The mismatch in the analytic behavior of $z^{n}$ and $e^{-1/z}$
about $z=0$ leads to an asymptotic expansion coefficient function
$c_{n}(z)$ which attains a discrete character on the complex plane.

For an example of a situation where the asymptotic expansion is not
divergent, consider the resolution of 
\begin{equation}
f(z)=e^{z}\label{eq:orig-1}
\end{equation}
in the basis $\{e^{\sqrt{z^{2}+1}},e^{-\sqrt{z^{2}+1}}\}$ as $|z|\rightarrow\infty$:
\begin{equation}
f(z)\sim a(z)e^{\sqrt{z^{2}+1}}+b(z)e^{-\sqrt{z^{2}+1}}\label{eq:asymp}
\end{equation}
where $\{a(z),b(z)\}$ are defined to not contain essential singularities
at $|z|\rightarrow\infty$. To fix the branch cut of the basis, consider
$z^{2}+1=(z-i)(z+i)$ and take the branch cut to be from $\pm i$
to $\pm i\infty$. This means that $\sqrt{z^{2}+1}$ take the same
sign on the positive and negative real axis. This endows the essential
singularity at $z=\infty$ with a branch cut. In this way, the basis
$\{e^{\sqrt{z^{2}+1}},e^{-\sqrt{z^{2}+1}}\}$ has a different analytic
structure at $\infty$ than $e^{z}$. The Stokes (anti-Stokes) line
is then on the real (imaginary) axis as $\left|z\right|\rightarrow\infty$.
For specificity of the sign on the branch cuts, we will define the
analytic function $\sqrt{z^{2}+1}$ to be continuous in the region
$\arg z\in(-\pi/2,\pi/2]$ and $\arg z\in(\pi/2,3\pi/2]$ and to be
positive on the positive real axis.

If Eq.~(\ref{eq:asymp}) were promoted to an equality as 
\begin{equation}
f(z)=a_{e}(z)e^{\sqrt{z^{2}+1}}+b_{e}(z)e^{-\sqrt{z^{2}+1}},
\end{equation}
then there will be an exact solution of the form 
\begin{equation}
a_{e}(z)=e^{z-\sqrt{z^{2}+1}}-b_{e}(z)e^{-2\sqrt{z^{2}+1}}
\end{equation}
where $b_{e}(z)$ an arbitrary function. Now, even though we want
$\{a(z),b(z)\}$ not to have any essential singularities as they are
expanded asymptotically, $\{a_{e}(z),b_{e}(z)\}$ may contain essential
singularities. In other words, the functions $\{a(z),b(z)\}$ are
not defined by expanding $\{a_{e}(z),b_{e}(z)\}$ but by asymptotically
expanding the two sides of Eq.~(\ref{eq:asymp}).

If $b_{e}(z)$ is chosen to not contain any essential singularities
as $|z|\rightarrow\infty$, then 
\begin{equation}
a_{e}(z)\sim1-\frac{1}{2z}+\frac{1}{8z^{2}}+O(z^{-3})=a(z)\label{eq:az1}
\end{equation}
in the region excluding the anti-Stokes lines (which we happened to
have placed on the branch cut in this example). On the anti-Stokes
lines in the region $z=ri\rightarrow i\infty$, our definitions give
\begin{align}
\sqrt{z^{2}+1} & =\sqrt{1-r^{2}}\\
 & =i\left(r-\frac{1}{2r}-\frac{1}{8r^{3}}+O(r^{-5})\right).
\end{align}
This implies 
\begin{equation}
a_{e}(z)=1+\frac{i}{2r}+O(r^{-2})-b_{e}(ri)e^{-2ir+i/r+O(r^{-2})}.
\end{equation}
Since $b_{e}(ri)$ was chosen to not have any essential singularities,
if we similarly choose $a_{e}(z)$ to not have any essential singularities,
we need $b_{e}(ri)=0$. Furthermore, matching Eqs.~(\ref{eq:orig-1})
and (\ref{eq:asymp}) (and not just expanding $b_{e}(ri)=0$) gives
\begin{equation}
b(ri)\sim0
\end{equation}
even though $b(z)$ is unconstrained in the region $\arg z\in(-\pi/2,\pi/2)$.

When $z=ri-\epsilon$, as $\epsilon\rightarrow0^{+}$ and $r\rightarrow\infty$,
we have 
\begin{equation}
\sqrt{z^{2}+1}=-i\left(r-\frac{1}{2r}-\frac{1}{8r^{3}}+O(r^{-5})\right)
\end{equation}
where the minus sign is a signature of our current choice of branch
cut. Now, we continuously connect this in the $\arg z\in(\pi/2,3\pi/2]$
region to 
\begin{align}
\sqrt{r^{2}e^{2i\theta}+1} & =e^{i\theta}\sqrt{r^{2}+e^{-2i\theta}}\\
 & \sim-re^{i\theta}
\end{align}
which gives along the $z=-ri\rightarrow-i\infty$ region 
\begin{equation}
e^{\sqrt{z^{2}+1}}\sim e^{-re^{i3\pi/2}}=e^{ri}.
\end{equation}
This makes 
\begin{equation}
a(-ri)\sim0
\end{equation}
even though a smooth $a(z)$ is unconstrained in the region $\arg z\in(\pi/2,3\pi/2)$.

The fact that a smooth $b(z)$ can vary arbitrarily in $\arg z\in(-\pi/2,\pi/2)$
without changing the asymptotic expansion (owing to $e^{-\sqrt{z^{2}+1}}\sim0$)
means if the asymptotic expansion in this region has further mathematical
specification that the leading coefficients are constants (as in the
WKB cases), that condition can be satisfied: i.e. 
\begin{equation}
f(z)\sim a(z)e^{\sqrt{z^{2}+1}}
\end{equation}
independently of varying $b(z)$ which goes to zero at $\arg z=\pi/2$.
In the $\Re z<0$ Stokes region, we have 
\begin{equation}
a(z)\sim0\hspace{1em}\Re z<0\label{eq:az0}
\end{equation}
region with $b(z)$ still arbitrary owing again to $e^{-\sqrt{z^{2}+1}}\sim0$
in this region. Hence, the basis coefficient function has again become
discrete.

As a third example, suppose with the same $f(z)=e^{z}$ and basis
$\{e^{\sqrt{z^{2}+1}},e^{-\sqrt{z^{2}+1}}\}$ for asymptotic expansion,
suppose we take the branch cut for the basis to be instead between
$-i$ and $i$ such that the essential singularity of the basis no
longer has a branch cut and there is no mismatch in the analytic property
of the basis functions and the function $e^{z}$ that is being resolved.
In this situation, $\sqrt{z^{2}+1}$ changes sign on the negative
real axis compared to the sign on the positive real axis. This means
although we still have Eq.~(\ref{eq:az1}) in the $\Re z>0$ region,
there does not need to be a change to Eq.~(\ref{eq:az0}) in the
$\Re z<0$ region since $e^{\sqrt{z^{2}+1}}\sim0$ there. Similarly,
$b(z)$ coefficient need not change. Hence, we can have the leading
asymptotic expansion being 
\begin{equation}
a(z)\sim1\hspace{1em}b(z)\sim0
\end{equation}
everywhere. This illustrates the fact that having the basis $\{e^{\sqrt{z^{2}+1}},e^{-\sqrt{z^{2}+1}}\}$
match the analytic property of $e^{z}$ at $|z|=\infty$ does not
lead to Stokes phenomena, unlike the situation of Eqs.~(\ref{eq:az1})
and (\ref{eq:az0}).

\bibliographystyle{JHEP}
\phantomsection\addcontentsline{toc}{section}{\refname}\bibliography{topologicalbogo}

\end{document}